\documentclass[11pt]{article}
\usepackage{epsf,amsmath,amssymb,axodraw,graphicx,scalefnt}
\usepackage{subfigure}
\usepackage{cite}
\usepackage{color}
\usepackage{rotating}
\newcommand{\mhmax}[1]{$m_h^{\rm max}(#1)$}
\newcommand{\gluophobic}[1]{gluophobic$(#1)$}
\newcommand{\nomixing}[1]{no-mixing$(#1)$}
\newcommand{\smallalpha}[1]{small-$\alpha_{\rm eff}(#1)$}
\newcommand{\note}[1]{}
\newcommand{\fourth}{$4^{\rm th}$}
\newcommand{\dred}{{\abbrev DRED}}

\newcommand{\bare}{{\rm B}}
\newcommand{\qcd}{{\abbrev QCD}}
\newcommand{\ew}{{\abbrev EW}}
\newcommand{\pt}{p_T}

\newcommand{\lrq}{l_{rq}}
\newcommand{\lts}{l_{ts}}

\newcommand{\lqs}{l_{qs}}
\newcommand{\lbs}{l_{bs}}
\newcommand{\Lhb}{L_{hb}}
\newcommand{\Lhq}{L_{hq}}
\newcommand{\lhc}{{\abbrev LHC}}
\newcommand{\sm}{{\abbrev SM}}
\newcommand{\mssm}{{\abbrev MSSM}}
\newcommand{\susy}{{\abbrev SUSY}}

\newcommand{\lo}{{\abbrev LO}}
\newcommand{\nlo}{{\abbrev NLO}}
\newcommand{\nnlo}{{\abbrev NNLO}}
\newcommand{\msbar}{\overline{\mbox{\abbrev MS}}}
\newcommand{\drbar}{\overline{\mbox{\abbrev DR}}}

\newcommand{\abbrev}{\rm\scalefont{.9}}
\newcommand{\muF}{\mu_{\rm f}}
\newcommand{\muR}{\mu_{\rm r}}
\newcommand{\higgs}{h}
\newcommand{\mhiggs}{M_\higgs}
\newcommand{\mquark}{M_q}
\newcommand{\mtop}{M_{t}}
\newcommand{\mbottom}{M_{b}}
\newcommand{\muSUSY}{\mu_{\rm SUSY}}
\newcommand{\msusy}[1]{\tilde{M}_{#1}}
\newcommand{\mstop}[1]{\tilde{M}_{t#1}}
\newcommand{\msbottom}[1]{\tilde{M}_{b#1}}
\newcommand{\msquark}[1]{\tilde{M}_{q#1}}
\newcommand{\mgluino}{\tilde{M}_{g}}
\newcommand{\pdf}{{\abbrev PDF}}
\newcommand{\ep}{\epsilon}
\newcommand{\api}{\frac{\alpha_s}{\pi}}
\newcommand{\eqn}[1]{Eq.\,(\ref{#1})}
\newcommand{\neqn}[1]{(\ref{#1})}
\newcommand{\fig}[1]{Fig.\,\ref{#1}}

\newcommand{\sct}[1]{Sect.\,\ref{#1}}
\newcommand{\dd}{{\rm d}}

\newcommand{\order}[1]{{\cal O}(#1)}
\newcommand{\bld}[1]{\boldmath{$#1$}}

\renewcommand{\Re}{{\rm Re}}

\date{}
\title{\vspace*{-6em}
  \begin{flushright}
    {\sf\small December 2010 --- WUB/10-35}\\[-.5em]
    {\sf\small (v3) December 2011}
  \end{flushright}
  \vspace*{2em} Supersymmetric Higgs production in gluon fusion}
\author{%
  Robert V. Harlander, Franziska Hofmann,
  Hendrik Mantler\\[2em] 
  {\it Fachbereich C, Bergische Universit\"at
    Wuppertal}\\{\it 42097 Wuppertal, Germany}
  }

\begin{document}
\maketitle

\begin{abstract}
The cross section through gluon fusion is calculated for the production
of the light neutral Higgs boson through next-to-leading order \qcd{}
within the Minimal Supersymmetric Standard Model. The quark-mediated
contributions are taken into account exactly, while for the genuinely
supersymmetric terms we use expressions obtained in the limit of large
squark, gluino and top quark masses. We present numerical results for the
total inclusive cross section as well as for kinematical distributions
of the Higgs boson. We also consider the effect of an \mssm{}-like
\fourth{} generation on the total Higgs production cross section.
\end{abstract}

\section{Introduction}

With the first sensitivity of the Tevatron to a Higgs
signal~\cite{tevatron:2010ar}\footnote{See {\tt
    http://tevnphwg.fnal.gov/} for updates.} and the first data taken at
the Large Hadron Collider (\lhc{}), Higgs phenomenology has received an
additional significant boost in the last few years. It is pleasing to
see that the efforts towards precise predictions for Higgs production
(for reviews, see
Refs.\,\cite{Djouadi:2005gi,Djouadi:2005gj,moriond2010,Harlander:2007zz})
at the \lhc{} are paying off, currently leading to significant bounds on
the Standard Model (\sm{}) Higgs boson mass.

Another interesting application of the on-going Higgs searches are the
restrictions on parameters of theories that go beyond the \sm{}, such as
the Minimal Supersymmetric Standard Model
(\mssm{})~\cite{Benjamin:2010xb} (see also Ref.\,\cite{Baglio:2010bn})
or the \sm{} with a fourth fermion
generation~\cite{Aaltonen:2010sv}.\footnotemark[1]

One of the most important production mechanisms for these searches is
gluon fusion which is known in the \sm{} through next-to-next-to-leading
order (\nnlo{})
\qcd{}~\cite{Georgi:1977gs,Djouadi:1991tka,Dawson:1990zj,Spira:1995rr,
  Harlander:2002wh,Anastasiou:2002yz,Ravindran:2003um,Marzani:2008az,
  Harlander:2009mq,Pak:2009dg,Harlander:2010my}. Effects beyond that
order have been investigated as
well~\cite{Catani:2003zt,Idilbi:2005ni,Idilbi:2006dg,Ravindran:2006cg,
  Moch:2005ky,Ahrens:2008nc}, with the reassuring conclusion that the
theory uncertainty derived from the renormalization and factorization
scale variation of the fixed order \nnlo{} result seems to be reliable.
Electro-weak (\ew{}) corrections have been found to be below around 6\%
relative to the leading order (\lo{})
result~\cite{Aglietti:2004nj,Actis:2008ug}, and a calculation of the
mixed \qcd{}/\ew{} effects in the unphysical limit $\mhiggs\ll M_W$
supports the assumption of a factorization of \qcd{} and \ew{}
corrections~\cite{Anastasiou:2008tj}.

Concerning the \mssm{}, at \lo{} squark loop effects have to be taken
into account. Due to supersymmetry (\susy{}), their couplings to the
Higgs bosons are typically suppressed by powers of $\mquark/\msquark{}$,
which is why they only contribute significantly to the cross section if
the squark mass $\msquark{}$ is not too large. In that case, however,
there are certain regions of the \susy{} parameter space where a quite
drastic cancellation among the quark and the squark induced amplitudes
may occur~\cite{Djouadi:1998az,Carena:2002qg}.

Apart from the \sm{}-like diagrams involving only quarks and gluons,
\qcd{} corrections to gluon fusion in the \mssm{} require the
calculation of the corresponding corrections to the \lo{} squark
diagrams, plus diagrams involving quarks, squarks, and gluinos
simultaneously. Meanwhile, all the ingredients for the next-to-leading
order (\nlo{}) \qcd{} corrections to gluon fusion in the \mssm{} have
been calculated: The top/stop/gluino effects were taken into account
through an effective Lagrangian
approach~\cite{Harlander:2003bb,Harlander:2004tp,Harlander:2003kf,
  Degrassi:2008zj}, a result for scalar \qcd{} (scalar quarks with
arbitrary mass) was calculated both numerically and
analytically~\cite{Anastasiou:2006hc,Aglietti:2006tp,Muhlleitner:2006wx},
and for the bottom/sbottom/gluino diagrams recently a result in terms of
an expansion in the inverse \susy{} masses was
presented~\cite{Degrassi:2010eu}. The quark/squark/gluino effects have
also been calculated fully numerically both for the top and the bottom
sector~\cite{Anastasiou:2008rm}. Very recently, even the \nnlo{} effects
in the heavy top/stop limit have been evaluated~\cite{Pak:2010cu}.

In this paper, we present a calculation of the full \nlo{} \qcd{}
effects for hadronic Higgs production in the \mssm{} with real
parameters and combine it with the \nnlo{} \qcd{} effects from purely
top-quark induced contributions known from the \sm{}
calculations. Although most of our results should be valid for Higgs
masses up to $\mhiggs \lesssim 2\min(\mtop{},\msusy{})$ ($\msusy{}$ is the
typical scale of the \susy{} partner masses), we present results only
for the phenomenologically most relevant case of the light \mssm{} Higgs
boson, for which $\mhiggs\lesssim 135$\,GeV~(see, e.g.\
Refs.\,\cite{Heinemeyer:2004ms,Allanach:2004rh} for reviews, and
Ref.\,\cite{Kant:2010tf} for the most up-to-date result).  We have
re-calculated all of the real corrections due to quarks and squarks,
while for the virtual corrections to the pure quark terms the analytic
result of Ref.\,\cite{Harlander:2005rq} is employed. These results are
valid for general quark/squark masses and can therefore be used for the
top as well as the bottom sector. Concerning the virtual squark effects,
we use the effective Lagrangian approach of
Ref.\,\cite{Harlander:2004tp} for the top sector, evaluating the Wilson
coefficient with the help of {\tt evalcsusy.f}~\cite{evalcsusy}. For the
bottom sector, we follow the approach of
Ref.\,\cite{Degrassi:2010eu,Hofmann:diss} in order to evaluate pure
sbottom and the bottom/sbottom/gluino diagrams in terms for $1/\msusy{b}$,
where $\msusy{b}$ denotes the average sbottom and gluino mass.

Apart from the inclusive total cross section, we also present
kinematical distributions for the Higgs boson: while the transverse
momentum ($\pt$) and pseudo-rapidity ($\eta$) distributions at
$\order{\alpha_s^3}$ have to be considered as \lo{} (at
$\order{\alpha_s^2}$, it is $\pt\equiv 0$ and $\eta=\infty$) and have
been evaluated before~\cite{Brein:2003df,Field:2003yy,Brein:2007da}, to
our knowledge this is the first time that rapidity ($y$) distributions
are being presented in the full \mssm{} at this order. Our calculations
are based on a numerical routine which will be made publicly available
and can be obtained from the authors upon request.

The remainder of this paper is organized as follows:
Section~\ref{sec::calc} outlines our notation, describes our calculation
and methods, and quotes some of the most important
formulas. Section~\ref{sec::total} defines the scenarios for our
numerical analyses, describes how we obtain our best prediction of the
cross section, and provides numerical results for the total inclusive
cross section. Section~\ref{sec::distrib} presents results for the \lo{}
transverse momentum and the \nlo{} rapidity distribution of the Higgs
boson.  Section~\ref{sec::4th} considers the effect of an \mssm{}-like
\fourth{} generation on the total inclusive cross section through \nlo{}
\qcd{}, and Section~\ref{sec::conclusions} contains our conclusions.  In
the Appendix, we collect the Feynman rules for the Higgs couplings used
in this paper, give some more analytical formulas, and provide numerical
results for other \susy{} scenarios.

\section{Calculation of the cross section}\label{sec::calc}

\subsection{General outline}
For the sake of clarity, we collect some of the most important notation
at this point. Concerning the \susy{} parameters, we follow the usual
notation (as mentioned before, we work in the \mssm{} with real parameters):
\begin{itemize}
\item The angle $\alpha$ rotates the {\abbrev CP} even neutral
  components of the two Higgs doublets $H_1$ and $H_2$ into their mass
  eigenstates $h,H$, where by definition $\mhiggs<M_H$. In all our
  formulas, $\alpha$ actually denotes the loop-corrected effective
  mixing angle $\alpha_{\rm eff}$.
\item The ratio of the two Higgs vacuum expectation values is denoted by
  $\tan\beta = v_2/v_1$.
\item The coefficient of the bilinear term $\sim H_1\cdot H_2$ in the
  superpotential is called $\muSUSY$. At tree-level, it is related to
  other \susy{} parameters as in \eqn{eq::ab} of
  Appendix~\ref{app::couplings}.
\end{itemize}
Furthermore, the following abbreviations will be applied:
\begin{equation}
\begin{split}
&\lqs = \ln\frac{\mquark^2}{\msusy{q}^2}\,,\qquad
\lrq = \ln\frac{\muR^2}{\mquark^2}\,,\qquad
\Lhq = \ln\frac{\mhiggs^2}{\mquark^2} - i\pi\,,\\
&\tau_q = \frac{4\mquark^2}{\mhiggs^2}\,,\qquad
\tilde\tau_{q,i} = \frac{4\msquark{i}^2}{\mhiggs^2}\,,\qquad
\label{eq::defs}
\end{split}
\end{equation}
where $\mquark$ and $\msquark{i}$ denote the on-shell quark and squark
mass ($q\in\{b,t\}$), while $\msusy{q} =
(\msquark{1}+\msquark{2}+\mgluino)/3$ is the average of the $q$-squark
and gluino ($\mgluino$) masses.  The mass of the light {\abbrev CP}-even
Higgs boson is called $\mhiggs$, and $\muR$ is the renormalization
scale. The factorization scale will be denoted by $\muF$.

Unless stated otherwise, the strong coupling
constant will always be defined in the framework of standard five-flavor
\qcd{}, renormalized in the $\msbar$ scheme:
\begin{equation}
\begin{split}
\alpha_s \equiv \alpha_s^{(5),\scriptsize{\msbar}}\,.
\end{split}
\end{equation}

Following the notation of Ref.\,\cite{Spira:1995rr}, we write the
hadronic cross section at the hadronic center-of-mass energy $s$ through
\nlo{} \qcd{} as
\begin{equation}
\begin{split}
\sigma(pp\to H+X) = \sigma_0\left[ 1 + C\,\api\right]\tau_h
\frac{\dd{\cal L}^{gg}}{\dd\tau_h}
 + \Delta\sigma_{gg}
 + \Delta\sigma_{gq}
 + \Delta\sigma_{q\bar q}\,,
\label{eq::hadsigma}
\end{split}
\end{equation}
where
\begin{equation}
\begin{split}
\frac{\dd{\cal L}^{gg}}{\dd \tau} = \int_\tau^1\frac{\dd x}{x} 
g(x)\,g(\tau/x)\,,
\end{split}
\end{equation}
with the gluon density $g(x)$ and $\tau_h=\mhiggs^2/s$.  The
normalization factor $\sigma_0$ determines the \lo{} cross section. In
the framework of the \mssm{}, we write it as
\begin{equation}
\begin{split}
\sigma_0 = \frac{G_{\rm
    F}\alpha^2_s(\muR^2)}{288\sqrt{2}\pi}|{\cal A}|^2\,,\qquad
{\cal A} = \sum_{q\in\{t,b\}}\left(
a^{(0)}_q + \tilde a^{(0)}_q\right)\,,
\label{eq::siglo}
\end{split}
\end{equation}
where
\begin{equation}
\begin{split}
&a^{(0)}_q = g_q\frac{3\tau_q}{2}\left(1 +
  (1-\tau_q)f(\tau_q)\right)\,,\qquad \tilde a^{(0)}_q =
  -\frac{3\tau_q}{8}\sum_{i=1}^2g^h_{q,ii}\left(1 -\tilde\tau_{q,i}
  f(\tilde \tau_{q,i})\right)\,,
\label{eq::alo}
\end{split}
\end{equation}
\begin{equation}
\begin{split}
f(\tau) = \left\{\begin{array}{ll}
\arcsin^2\frac{1}{\sqrt{\tau}}\,,&\tau \geq 1\,,\\
\displaystyle - \frac{1}{4}\left(\log\frac{1+\sqrt{1-\tau}}{1-\sqrt{1-\tau}} -
i\pi\right)^2\,,&\tau < 1\,.
\end{array}
\right.
\label{eq::ftau}
\end{split}
\end{equation}

\begin{figure}
\begin{center}
\begin{tabular}{ccc}
\raisebox{-2.6em}{\includegraphics[bb=200 510 370 630,width=.25\textwidth]{%
    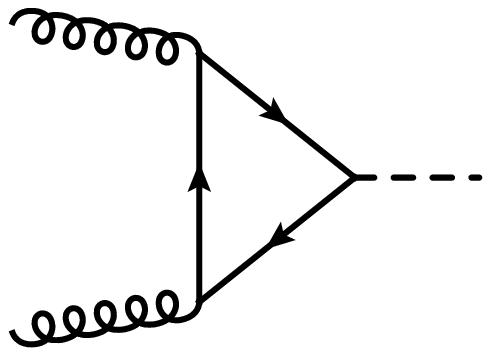}} &
\raisebox{-2.6em}{\includegraphics[bb=200 510 370 630,width=.25\textwidth]{%
    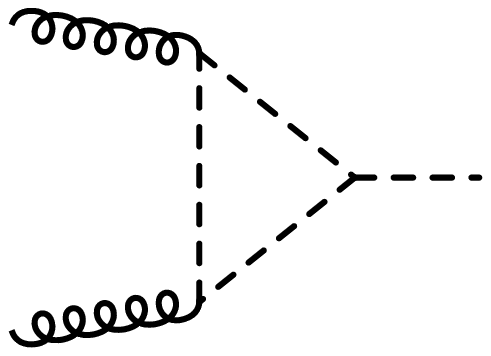}} &
\raisebox{-2.6em}{\includegraphics[bb=200 510 370 630,width=.25\textwidth]{%
    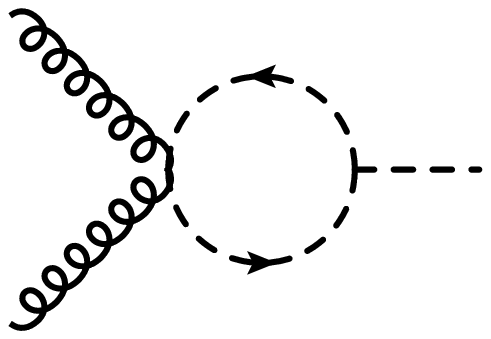}}\\
(a)&(b)&(c)
\end{tabular}
\caption[]{\label{fig::lodias}Feynman diagrams contributing to Higgs
  production in gluon fusion at \lo{}.}
\end{center}
\end{figure}

The $a^{(0)}(\tau_q)$ and the $\tilde a^{(0)}(\tilde\tau_{q,i})$ terms
are due to quark and squark diagrams, respectively, as the ones
displayed in \fig{fig::lodias}. It may be useful to quote the relevant
limits of these functions:
\begin{equation}
\begin{split}
a_q^{(0)} &\stackrel{\tau_q\gg 1}{=} g^h_q\left[1 + \frac{7}{30\tau_q} 
+ \cdots\right]
\,,\qquad
\tilde a_q^{(0)} \stackrel{\tilde \tau_{q,i}\gg 1}{=} 
\frac{\tau_q}{8}\,\sum_{i=1}^2 \frac{g^h_{q,ii}}{\tilde{\tau}_{q,i}}
\left[1 + \frac{8}{15\tilde
    \tau_{q,i}} 
 + \cdots\right]\,,\\
a_q^{(0)} &\stackrel{\tau_q\ll 1}{=} \frac{3}{2}g^h_q\tau_q\left[
  1-\frac{\Lhq^2}{4}
  +\frac{\tau_q\Lhq}{4}\left(1+\Lhq\right)
+\cdots
\right]\,.
\end{split}
\end{equation}
The coupling constants $g^h_q$ and $g^h_{q,ij}$ are given in
Appendix~\ref{app::couplings}; those for $q=t$ can also be found in
Ref.\,\cite{Harlander:2004tp}.

In principle, the sum in \eqn{eq::siglo} should run over all quark
flavors. However, only the top and the bottom quark account for a
relevant contribution to the cross section, while all others are
suppressed by their Yukawa coupling.

\begin{figure}
\begin{center}
\begin{tabular}{ccc}
\raisebox{-2.6em}{\includegraphics[bb=200 510 370 630,width=.25\textwidth]{%
    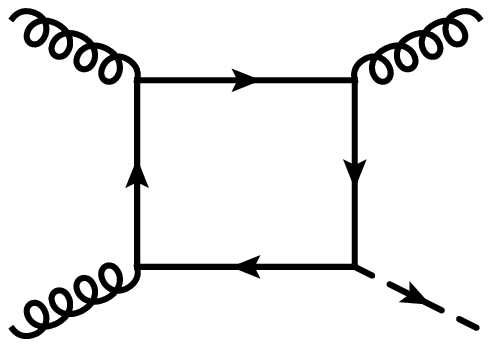}} &
\raisebox{-2.6em}{\includegraphics[bb=200 510 370 630,width=.25\textwidth]{%
    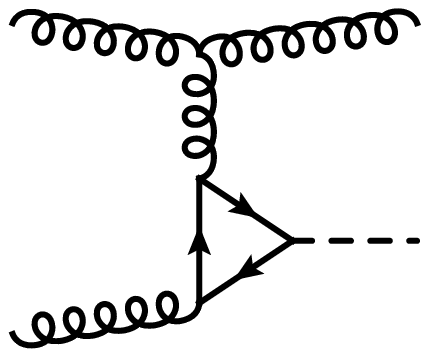}} &
\raisebox{-2.6em}{\includegraphics[bb=200 510 370 630,width=.25\textwidth]{%
    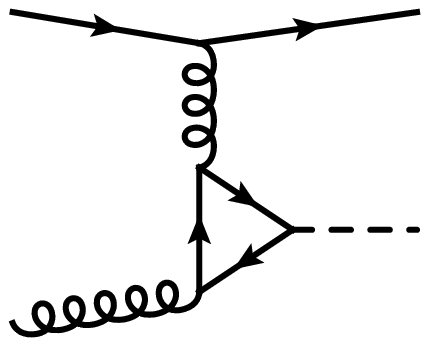}}\\
(a)&(b)&(c)
\end{tabular}
\caption[]{\label{fig::realdias}Sample Feynman diagrams contributing to
  the real corrections at \nlo{} to Higgs production in gluon
  fusion. The particle in the loop can be a quark or a squark of bottom
  or top flavor.}
\end{center}
\end{figure}

The quantities $\Delta\sigma_{gg}$, $\Delta\sigma_{gq}$,
$\Delta\sigma_{q\bar q}$ in \eqn{eq::hadsigma} denote the non-singular
terms of the cross section arising from $gg$, $gq$ and $q\bar q$
scattering. Typical diagrams are shown in \fig{fig::realdias}; the
$q\bar q$ contribution is obtained through crossing from
\fig{fig::realdias}\,(c).  They can be evaluated using well-known
techniques. We have expressed them in terms of Passarino-Veltman
functions~\cite{Passarino:1978jh} and checked our result against the
literature (see, e.g.\
Ref.\,\cite{Ellis:1987xu,Baur:1989cm,Keung:2009bs,Bonciani:2007ex}).

\subsection{Virtual corrections}

The coefficient $C$ in \eqn{eq::hadsigma} denotes the virtual
corrections to the $gg$ initiated process, regularized by the infrared
singular part of the cross section for real gluon emission. We write
\begin{equation}
\begin{split}
C &=2\,\Re \left[{\cal A}_\infty^{-1}
  \sum_{q}\left(a_{q}^{(1)} + {\tilde a}_{q}^{(1)}\right)\right] + \pi^2
+ \beta_0\,\ln\frac{\muR^2}{\muF^2}\,,
\end{split}
\end{equation}
where $\beta_0 = 11/2 - n_l/3$ with $n_l=5$, and ${\cal A}_\infty$ is
the \lo{} amplitude in the limit of large stop and sbottom masses, i.e.,
\begin{equation}
\begin{split}
{\cal A}_\infty = \sum_q \left(a_q^{(0)} + \frac{\tau_q}{8}\sum_{i=1}^2
\frac{g^h_{q,ii}}{\tilde{\tau}_{q,i}}\right)\,.
\end{split}
\end{equation}
The \nlo{} quark-loop contribution $a_{q}^{(1)}$, corresponding to the
\sm{} part, has first been evaluated numerically a long time
ago~\cite{Spira:1995rr} and was later expressed in terms of analytic
functions~\cite{Harlander:2005rq,Anastasiou:2006hc,Aglietti:2006tp}.  We
provide a few terms of its small- and large-mass expansions in
Appendix~\ref{app::a1q}.

\begin{figure}
\begin{center}
\begin{tabular}{ccc}
\raisebox{-2.6em}{\includegraphics[bb=200 510 370 630,width=.25\textwidth]{%
    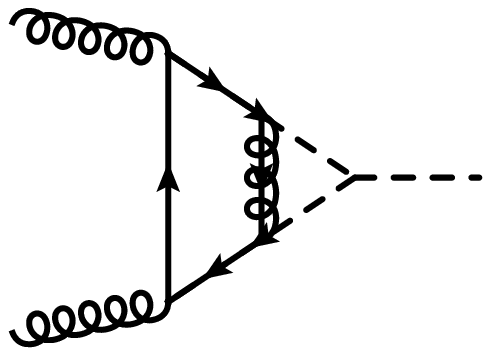}} &
\raisebox{-2.6em}{\includegraphics[bb=200 510 370 630,width=.25\textwidth]{%
    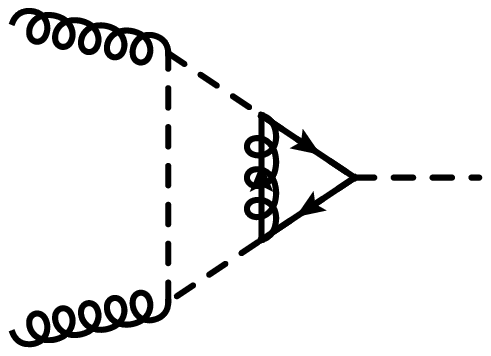}} &
\raisebox{-2.6em}{\includegraphics[bb=200 510 370 630,width=.25\textwidth]{%
    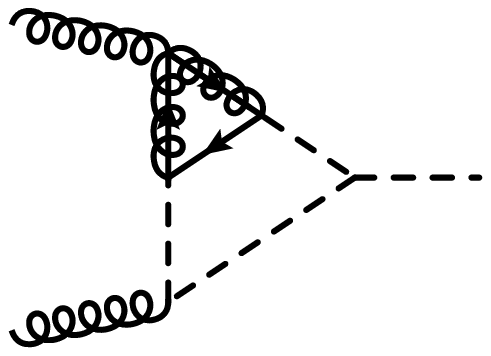}}\\
(a)&(b)&(c)
\end{tabular}
\caption[]{\label{fig::gluino}Sample diagrams for the mixed
  quark/squark/gluino contribution to gluon fusion at \nlo{}.}
\end{center}
\end{figure}

One class of diagrams contributing to the \susy{} part $\tilde
a_q^{(1)}$ is obtained by attaching a virtual gluon to
\fig{fig::lodias}\,(b) and (c).  However, both squark and gluino effects
need to be considered in order to preserve
supersymmetry~\cite{Harlander:2003bb,Harlander:2004tp}; sample diagrams
containing a gluino are shown in \fig{fig::gluino}.  A fully numerical
result for $\tilde a_q^{(1)}$ for general quark/squark/gluino masses was
obtained in Ref.\,\cite{Anastasiou:2008rm}, but the corresponding code
is not publicly available. For the pure squark diagrams, there exists
both an analytic and a numerical
result~\cite{Anastasiou:2006hc,Aglietti:2006tp,Muhlleitner:2006wx}.  In
Ref.\,\cite{Harlander:2003bb,Harlander:2004tp,Degrassi:2008zj}, $\tilde
a_q^{(1)}$ was evaluated for the top sector (i.e., $q=t$) in terms of an
effective Lagrangian, and one can write
\begin{equation}
\begin{split}
\tilde a_{t}^{(1)} = c_{1}^{(1)}-g_t^h\frac{11}{4} +
\order{\mhiggs^2}\,,
\end{split}
\end{equation}
where $c_{1}^{(1)}$ is the \nlo{} term of the Wilson coefficient,
defined in Eq.\,(2.5) of Ref.\,\cite{Harlander:2004tp}. It can be
evaluated with the help of the publicly available program {\tt
  evalcsusy.f}~\cite{evalcsusy}.

Concerning the \susy{} bottom quark/squark sector, a fully numerical
result for general masses was presented in
Ref.\,\cite{Anastasiou:2008rm}. Recently, this contribution was
calculated with the help of asymptotic expansions in the limit of large
\susy{} masses~\cite{Degrassi:2010eu}. We present an independent result,
following the same strategy. However, due to the \susy{} mass spectrum
emerging from the most popular \susy{} breaking scenarios, we decided to
take the limit $\msusy{b}\equiv\msbottom{1}=\msbottom{2}=\mgluino$ which
leads to an extremely simple result ($\mgluino$ is the gluino mass).
Let us briefly describe the calculation.

For the pure sbottom diagrams (i.e., without
gluinos), the procedure is very similar to the one applied in
Ref.\,\cite{Harlander:2009bw,Harlander:2009mq,%
  Harlander:2010my,Pak:2009bx,Pak:2009dg}, where top quark mass
suppressed terms for the \sm{} cross section $\sigma(gg\to H+X)$ were
evaluated (mind you, through \nnlo{}; here, we consider \nlo{} only).
The expansion of the mixed bottom/sbottom/gluino diagrams is a little
more involved, but still rather straightforward due to the
algorithmically defined method of asymptotic expansions (see, e.g.\
Ref.\,\cite{Smirnovbook}). Let us consider an example:
\begin{equation}
\begin{split}
\raisebox{-2.6em}{\includegraphics[bb=180 550 415 720,width=.2\textwidth]{%
    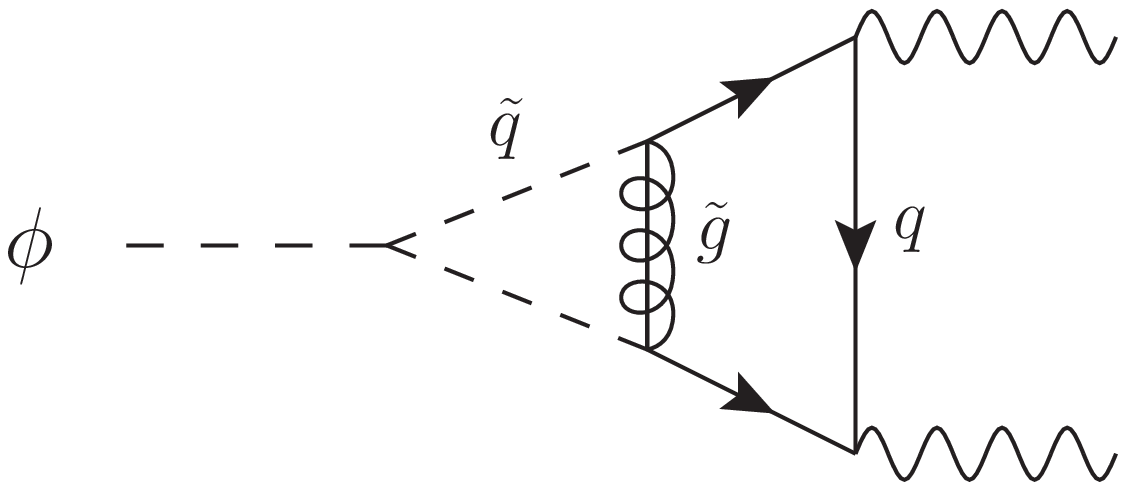}} \qquad
&\stackrel{\msusy{}\to\infty}{\to }
\raisebox{-3.3em}{\includegraphics[bb=180 550 415 720,width=.2\textwidth]{%
    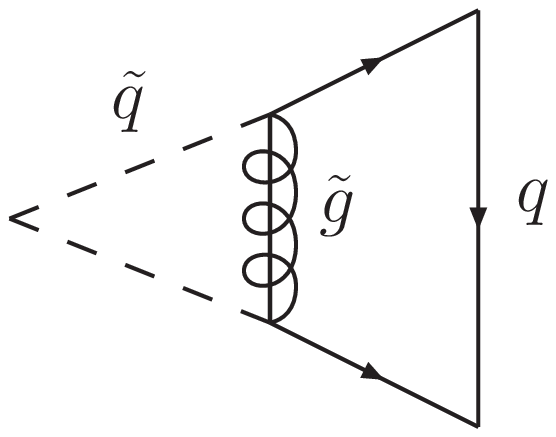}} \otimes
\raisebox{-3.6em}{\includegraphics[bb=180 550 415 720,width=.2\textwidth]{%
    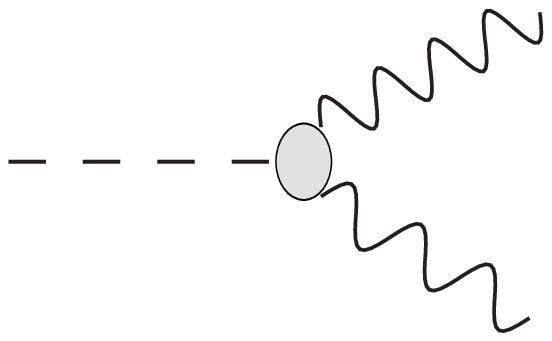}}\\[-1.5em]
&\quad +
\raisebox{-3.8em}{\includegraphics[bb=180 550 415 720,width=.2\textwidth]{%
    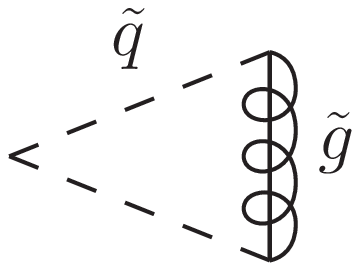}} \otimes\qquad
\raisebox{-3.2em}{\includegraphics[bb=180 550 415 720,width=.2\textwidth]{%
    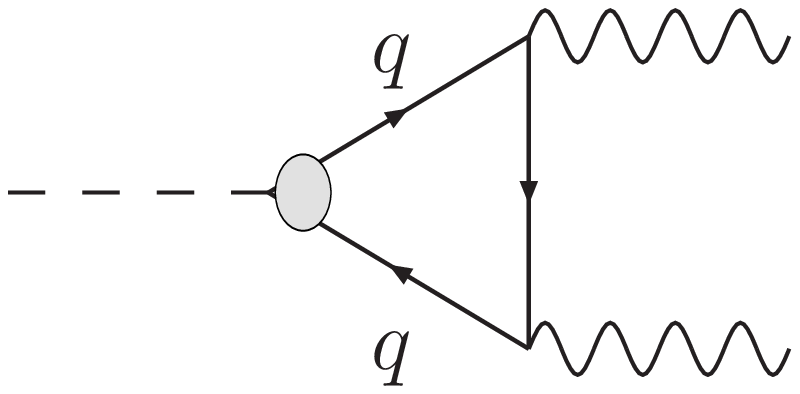}}\\[-2em]
\label{eq:asympt1}
\end{split}
\end{equation}
The notation is as follows: in the original diagram, Taylor-expand all
the propagators of the sub-diagram $\gamma$ left of $\otimes$ in terms
of $p/\msusy{}$ {\it before integration}. Here, $p$ denotes any
dimensional quantity (mass or momentum) of $\gamma$ except its loop
momentum or $\msusy{}$. The resulting Feynman integrals to evaluate in
the above case are therefore: (i) one- and two-loop tadpole integrals
(i.e., vanishing external momenta) depending on $\msusy{}$, but not on
$\mbottom$; (ii) one-loop vertex integrals with external momenta
$q_1^2=q_2^2=0$ and $2q_1\cdot q_2 = \mhiggs^2$, and mass $\mbottom$ (no
dependence on $\msusy{}$). Both types of integrals can be calculated
analytically: type (i) with the help of {\tt
  MATAD}~\cite{Steinhauser:2000ry}, type (ii) by standard
Passarino-Veltman reduction~\cite{Passarino:1978jh}, for example.

For clarity, let us consider another example:
\begin{equation}
\begin{split}
\raisebox{-2.6em}{\includegraphics[bb=180 550 415 720,width=.2\textwidth]{%
    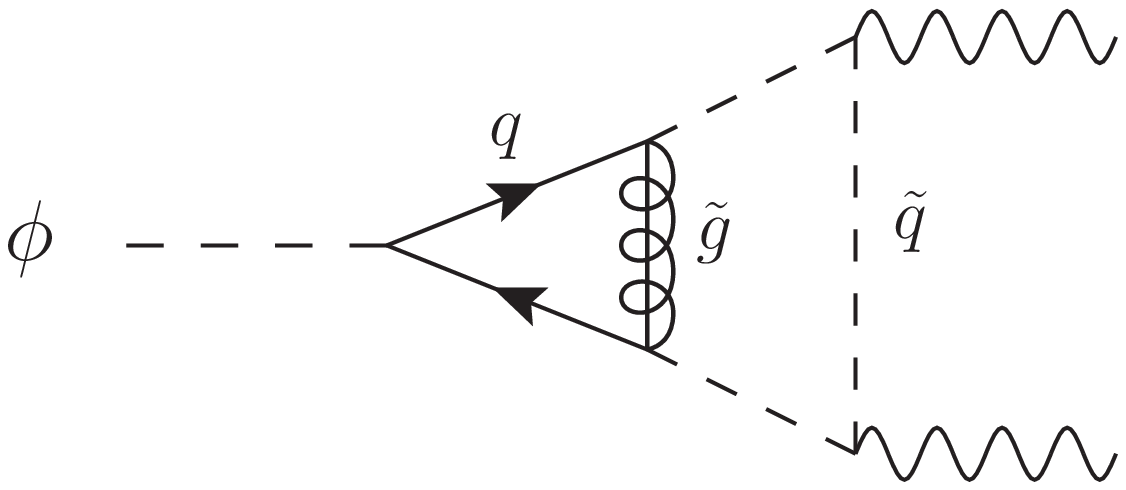}} \qquad
&\stackrel{M_s\to\infty}{\to }
\raisebox{-3.3em}{\includegraphics[bb=180 550 415 720,width=.2\textwidth]{%
    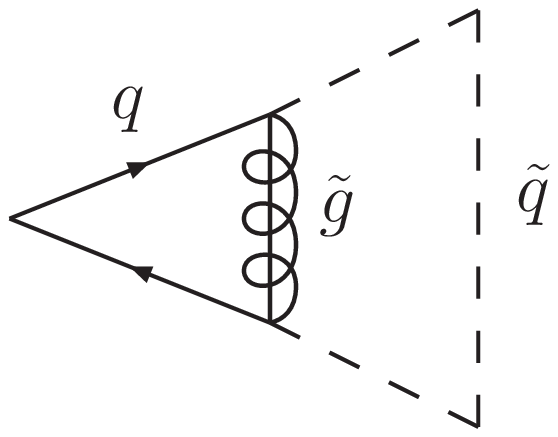}} \otimes
\raisebox{-3.6em}{\includegraphics[bb=180 550 415 720,width=.2\textwidth]{%
    figs/dias/nlosusybasexp5.eps}}\\[-1.5em]
&\quad +
\raisebox{-3.3em}{\includegraphics[bb=180 550 415 720,width=.2\textwidth]{%
    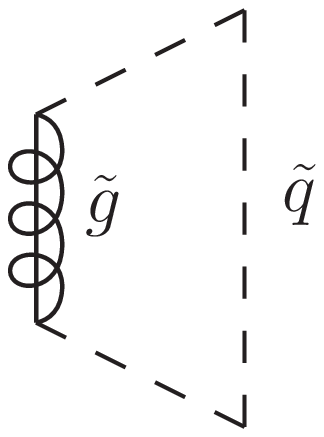}} \otimes\qquad
\raisebox{-4em}{\includegraphics[bb=180 550 415 720,width=.2\textwidth]{%
    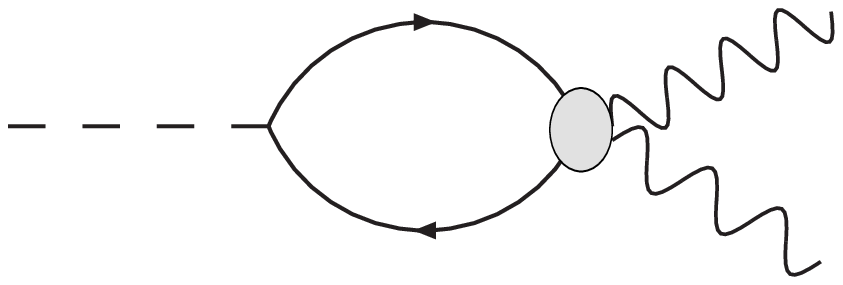}}\\[-2em]
\label{eq:asympt2}
\end{split}
\end{equation}
Using this procedure, the virtual corrections are valid for arbitrary
values of $\mhiggs$ and $\mbottom$, as long as they are both smaller
than\footnote{The factor of two can be deduced from the analytic
  structure of the amplitude.} $2\msusy{b}$.

We perform the calculation in Dimensional Reduction (\dred{}) by
explicitely taking into account $\ep$-scalars as propagating
particles. In order to avoid infra-red singularities in the
corresponding Feynman integrals, we assign a large mass $M_\ep$ to the
$\ep$-scalars for which we can assume $M_\ep\to \infty$ at the end of
the calculation. Practical details about the implementation of
$\ep$-scalars, in particular their Feynman rules, have been given
in~Refs.\,\cite{Harlander:2006rj,Harlander:2006xq,Harlander:2009mn}.

Concerning renormalization, we use on-shell conditions for the bottom
quark mass, one of the sbottom masses, and the sbottom quark mixing
angle. The other sbottom mass is then fixed by {\abbrev SU(2)}
symmetry. The gluino does not require renormalization at this
order. This top/stop-like renormalization scheme may cause perturbative
problems for large $\muSUSY$ and
$\tan\beta$~\cite{Brignole:2002bz}. However, due to our limit of
degenerate sbottom masses we will restrict our analysis to moderate
values of $\tan\beta$ anyway.  The comparison of various other
renormalization schemes is beyond the scope of the current paper and
will be deferred to a forthcoming publication.

The external gluon wave functions $G_\mu$ are renormalized on-shell
using
\begin{equation}
\begin{split}
G_\mu^\bare &= Z_3^{1/2}\,G_\mu\,,\\ Z_3 &= 1 -
\frac{\alpha_s^{\scriptsize{\drbar}}}{\pi}\,
\frac{1}{\ep}\,\left(\frac{T}{6}\,( n_s + 2) + \frac{C_A}{6}\right) +
\delta z\,,\\ \delta z &=
\frac{\alpha_s^{\scriptsize{\drbar}}}{\pi}\left(
\frac{T}{12}\sum_{\substack{q\in\{b,t\}\\ i=1,2}} \ln\frac{\muR^2}{\msquark{i}^2}
+ \frac{T}{3} \ln\frac{\muR^2}{\mtop^2} + \frac{C_A}{6}
\ln\frac{\muR^2}{\mgluino^2}\right) - \api\,\frac{C_A}{6}\,,\\
\end{split}
\end{equation}
where $n_s=6$ is the number of squark flavors, $C_A=3$ and $T=1/2$ are
color factors, and the last term in $\delta z$ is due to the
$\ep$-scalars. In fact, at this order of $\alpha_s$, all of the finite
part of $Z_3$ cancels against the conversion~(see, e.g.\
Ref.\,\cite{Harlander:2005wm})
\[
\alpha_s^{\scriptsize{\drbar}} = (1-\delta
z)\alpha_s^{\scriptsize{\msbar},(5)}\,,
\]
where $\alpha_s^{\scriptsize{\drbar}}$ denotes the strong coupling
constant renormalized in the full \mssm{} according to $\drbar$
subtraction.

The bottom mass $\mbottom$, the sbottom mass $\msbottom{1}$, and the
sbottom mixing angle $\theta_b$ are renormalized on-shell in analogy to
the top quark sector, see, e.g.\ Ref.\,\cite{Harlander:2004tp}. The
sbottom mass $\msbottom{2}$ is determined from the {\abbrev SU(2)}
relation from the other (top and bottom) on-shell quark and squark
masses and mixing angles, and its on-shell value $\msbottom{2}^{\rm
  OS}$. The two mass definitions are related by a finite shift~(see,
e.g.\ Ref.\,\cite{Heinemeyer:2004xw}\footnote{Ref.\,\cite{Heinemeyer:2004xw}
  fixes $\msbottom{2}$ on-shell and evaluates $\msbottom{1}$ from the
  {\abbrev SU(2)} condition; we found better numerical stability by
  fixing $\msbottom{1}$ instead.}):
\begin{equation}
\begin{split}
\msbottom{2} = \msbottom{2}^{\rm OS}\left(1 + \api\Delta^{b,(1)}_{{\rm
    OS}}\right)\,.
\end{split}
\end{equation}

The sbottom sector contribution to the Higgs production cross section
can now be written as ($s_{nb} = \sin(n\theta_b)$, $c_{nb} =
\cos(n\theta_b)$)
\begin{equation}
\begin{split}
\tilde a_b^{(1)} &=
-\frac{\mbottom^2}{\msusy{b}^2}\frac{\sin\alpha}{\cos\beta} \left({\cal
  K}^b_1 - \frac{1}{2}\Delta^{b,(1)}_{{\rm OS}}\right) -
\frac{\muSUSY}{\msusy{b}}\frac{\cos(\alpha-\beta)}{\cos^2\beta}
\left({\cal K}^b_2+
\frac{1}{4}\sqrt{\frac{\tau_q}{\tilde{\tau}_q}}s_{2b}\Delta^{b,(1)}_{{\rm
    OS}}\right) \\&\quad
+ \frac{M_Z^2}{\msusy{b}^2}\sin(\alpha+\beta)\left[{\cal
  K}^b_3 - \frac{1}{4}\left(s_b^2 +
  \frac{2}{3}c_{2b}\sin^2\theta_W\right)\Delta^{b,(1)}_{{\rm
    OS}}\right]
 + \order{\msusy{b}^{-4}}\,,
\end{split}
\end{equation}
with the weak mixing angle $\theta_W$, cf.\,\eqn{eq::cew}.
Our result for the coefficients ${\cal K}^q_i$ is (cf.~\eqn{eq::defs})
\begin{equation}
\begin{split}
{\cal K}^q_1 &= \frac{25}{24} + \frac{17}{12}\,\lqs -
\frac{5}{24}\left(B_0^{\rm fin}(\tau_q) - \lrq\right) + \frac{\tau_q}{2}
+ \tau_q\left(\frac{17}{24} - \frac{\tau_q}{2}\right)\,f(\tau_q)\,,
\\ {\cal K}^q_2 &= -\frac{\tau_q}{2}\left[1 + (1-\tau_q)f(\tau_q)\right]
+ \frac{\tau_q}{36\tilde{\tau}_q}\left[ -37 + \frac{15}{2}\lqs - 3\tau_q
  - 3( 2 - \tau_q - \tau_q^2)f(\tau_q)\right]\,, \\ {\cal K}^q_3 &=
\frac{19}{32} + \frac{\tau_q}{6} + \frac{\tau_q}{6}(1-\tau_q)f(\tau_q)\,.
\label{eq::kq}
\end{split}
\end{equation}
Note that the ${\cal K}_i^q$ are given for a general quark flavor $q$,
so we will be able to apply them also for the top contribution further
below.  The only difference arises from the functions $f(\tau)$, defined
in \eqn{eq::ftau}, and $B_0^{\rm fin}(\tau)$, given by
\begin{equation}
\begin{split}
B_0^{\rm fin}(\tau_q) -\lrq = \left\{
\begin{array}{ll}
\displaystyle 2 - 
2\sqrt{\tau_q-1}\arctan\left(\frac{1}{\sqrt{\tau_q-1}}\right)\,,
& \tau_q > 1\,,\\[1.3em]
\displaystyle 2 -
\sqrt{1-\tau_q}\left(\log\frac{1+\sqrt{1-\tau_q}}{1-\sqrt{1-\tau_q}}
- i\pi\right)\,, & \tau_q \leq 1\,.
\end{array}
\right.
\end{split}
\end{equation}
Due to our limit of a degenerate \susy{} mass spectrum, all the
dependence on the squark mixing angle $\theta_q$ drops out in
\eqn{eq::kq}. This also removes all terms $\sim \mgluino/\mquark$ found
in Ref.\,\cite{Degrassi:2010eu}. In fact, many of the terms evaluated in
that reference vanish in our approximation, while many of the terms of
\eqn{eq::kq} vanish in the approximation of Ref.\,\cite{Degrassi:2010eu}
since they are of higher orders in $\mbottom$ or $1/\msusy{b}$. Needless
to say that for the terms that are non-zero in both approximations we
find complete agreement.\footnote{Thanks to P.~Slavich for
  confirmation.}

Since the bottom quark mass is very small, the following expansion in
$\mbottom/\mhiggs$, which leads to particularly simple expressions,
approximates \eqn{eq::kq} at the order of $1\%$ for a reasonable range
of masses~\cite{Hofmann:diss}:
\begin{equation}
\begin{split}
{\cal K}_1^b &= \frac{5}{8} + \frac{17}{12}\lbs + \frac{5}{24}\Lhb
+\order{\tau_b}\,,
\\
{\cal K}_2^b &=
  -\frac{\tau_b}{2}\left(1 - \frac{\Lhb^2}{4}\right)
  + \frac{\tau_b}{\tilde\tau_b}\left(
  -\frac{37}{36} + \frac{5}{24}\lbs + \frac{\Lhb^2}{24}\right)
+\order{\tau_b^2}\,,
\\
{\cal K}_3^b &=
\frac{19}{32} + \frac{\tau_b}{6}\left(
    1 - \frac{\Lhb^2}{4} \right)
+\order{\tau_b^2}\,.
\end{split}
\end{equation}

A particularly useful check of \eqn{eq::kq} is to evaluate it for $q=t$
and expand it in the limit $\mtop\gg\mhiggs$:
\begin{equation}
\begin{split}
\tilde a_t^{(1)} =
\frac{\mtop^2}{\msusy{t}^2}\frac{\cos\alpha}{\sin\beta} {\cal K}^t_1 +
\frac{\muSUSY}{\msusy{t}}\frac{\cos(\alpha-\beta)}{\sin^2\beta} {\cal K}^t_2
- \frac{M_Z^2}{\msusy{t}^2}\sin(\alpha+\beta){\cal K}^t_3 +
\order{\msusy{t}^{-4}}
\end{split}
\end{equation}
\begin{equation}
\begin{split}
{\cal K}_1^t &= \frac{19}{12} + \frac{17}{12}\lts + \frac{1}{120\tau_t}
+ \order{1/\tau_t^{2}}\,,
\\
{\cal K}_2^t &= -\frac{1}{3} - \frac{7}{90\tau_t}
+ \frac{\tau_t}{\tilde \tau_t}\left[ -\frac{11}{12} + \frac{5}{24}\lts
  +\order{1/\tau_t}\right]
+ \order{1/\tau_t^{2}}\,,
\\
{\cal K}_3^t &= \frac{203}{288} + \frac{7}{270\tau_t}
+ \order{1/\tau_t^{2}}\,.
\label{eq::Kt}
\end{split}
\end{equation}
This result can also be directly obtained by asymptotic expansions of
the corresponding Feynman diagrams along the lines of
Refs.\,\cite{Harlander:2003bb,Harlander:2004tp,Harlander:2003kf,
  Degrassi:2008zj}. Of course, we find complete agreement, therefore
validating the result of \eqn{eq::kq}. Let us stress that we do not use
\eqn{eq::Kt} in our numerical analysis, but rather the more general
results obtained with the help of {\tt
  evalcsusy.f}~\cite{Harlander:2004tp}.

\section{Total inclusive cross section}\label{sec::total}

\subsection{Scenarios}\label{sec::scenarios}

In this section we present numerical results for the hadronic cross
section for Higgs production in gluon fusion in the \mssm{}. Through
$\order{\alpha_s^3}$, we include the effects from quark and squark loops
as described above, i.e., the full quark mass dependence is kept for all
leading order ($a_q^{(0)}$) and real radiation contributions
($\Delta\sigma_{ij}$), as well as for the virtual quark loop terms
($a_q^{(1)}$). The virtual top squark and top/stop/gluino contribution
is taken into account in the limit
$\{\mtop,\mstop{1},\mstop{2},\mgluino\}\gg \mhiggs$, without further
restrictions on the masses or \susy{} parameters, with the help of {\tt
  evalcsusy.f}. For the 2-loop virtual bottom squark and bottom/sbottom/gluino
contribution, and only there, we apply the limit
$\msbottom{1}=\msbottom{2}=\mgluino\equiv \msusy{b}\gg \mhiggs,
\mbottom$, where in numerical evaluations we set $\msusy{b}\equiv
(\msbottom{1} + \msbottom{2} + \mgluino)/3$. These approximations should
be well-justified for the most popular \susy{} benchmark scenarios~(see,
e.g.\ Refs.\,\cite{Allanach:2002nj,Carena:2002qg}), in particular since
they are only applied to the virtual effects where the partonic
center-of-mass energy is fixed to $\hat s\equiv \mhiggs^2$.

In order to obtain numerical results for the cross section, one needs to
insert numbers for the unknown \susy{} parameters. In this paper, we
will consider the following \susy{}
scenarios~\cite{Carena:2002qg,Carena:2005ek}:
\begin{description}
\item[\underline{gluophobic\bld{(\pm)}:}]
  \begin{equation}
    \begin{split}
      &M_{\rm SUSY} = 350\,{\rm GeV}\,,\qquad \muSUSY = \pm300\,{\rm
        GeV}\,,\\ &M_2 = 300\,{\rm GeV}\,,\qquad M_3 = 500\,{\rm
        GeV}\,,\qquad X_t = -750\,{\rm GeV}\\[1em]
      \stackrel{\tan\beta=10}{\Rightarrow}\qquad &\mstop{1} \approx
      150\,{\rm GeV} \,,\qquad \mstop{2} \approx 520\,{\rm
        GeV}\,,\\ &\msbottom{1} \approx 340\,{\rm GeV}\,,\qquad
      \msbottom{2} \approx 370\,{\rm GeV}\,,\qquad \mgluino = 500\,{\rm
        GeV}
      \label{eq::gluophobic}
    \end{split}
  \end{equation}
\item[\underline{\bld{m_{h}^{\rm max}(\pm)}:}]
  \begin{equation}
    \begin{split}
      &M_{\rm SUSY} = 1\,{\rm TeV}\,,\qquad \muSUSY = \pm 200\,{\rm
        GeV}\,,\\ &M_2 = 200\,{\rm GeV}\,,\qquad M_3 = 800\,{\rm
        GeV}\,,\qquad X_t = 2\,{\rm TeV}\\[1em]
      \stackrel{\tan\beta=10}{\Rightarrow}\qquad &\mstop{1} \approx
      830\,{\rm GeV} \,,\qquad \mstop{2} \approx 1170\,{\rm
        GeV}\,,\\ &\msbottom{1} \approx \msbottom{2} \approx 1\,{\rm
        TeV}\,,\qquad \mgluino = 800\,{\rm GeV}
      \label{eq::mhmax}
    \end{split}
  \end{equation}
\item[\underline{no-mixing\bld{(\pm)}:}]
  \begin{equation}
    \begin{split}
      &M_{\rm SUSY} = 2\,{\rm TeV}\,,\qquad
      \muSUSY = \pm 200\,{\rm GeV}\,,\\
      &M_2 = 200\,{\rm GeV}\,,\qquad
      M_3 = 1600\,{\rm GeV}\,,\qquad
      X_t = 0\\[1em]
      \stackrel{\tan\beta=10}{\Rightarrow}\qquad &\mstop{1} \approx \mstop{2}
      \approx\msbottom{1} \approx \msbottom{2} \approx 2\,{\rm TeV}\,,\qquad
      \mgluino = 1600\,{\rm GeV}
      \label{eq::nomix}
    \end{split}
  \end{equation}
\item[\underline{small \bld{\alpha_{\rm eff}(\pm)}:}]
  \begin{equation}
    \begin{split}
      &M_{\rm SUSY} = 800\,{\rm GeV}\,,\qquad
      \muSUSY = \pm 2\,{\rm TeV}\,,\\
      &M_2 = 500\,{\rm GeV}\,,\qquad
      M_3 = 500\,{\rm GeV}\,,\qquad
      X_t = -1.1\,{\rm TeV}\\[1em]
      \stackrel{\tan\beta=10}{\Rightarrow}\qquad &\mstop{1} \approx 690\,{\rm GeV}\,,\qquad
      \mstop{2} \approx 920\,{\rm GeV}\,,\\
      &\msbottom{1} = 760\,{\rm GeV}\,,\qquad
      \msbottom{2} = 840\,{\rm GeV}\,,\qquad
      \mgluino = 500\,{\rm GeV}
      \label{eq::smalla}
    \end{split}
  \end{equation}
\end{description}
The input parameters are the squark mass scale $M_{\rm SUSY}$, the
bilinear Higgs coupling $\muSUSY$, the gaugino mass parameters $M_2$ and
$M_3$, and the off-diagonal term in the stop mixing matrix
$\mtop{}X_t$. Furthermore, it is always assumed that $A_b=A_t\equiv
X_t+\muSUSY/\tan\beta$, where $A_b$ and $A_t$ are trilinear couplings of
the \susy{} potential. For a more detailed description of these input
parameters, let us refer to the documentation and references of the
program {\tt
  FeynHiggs}~\cite{Frank:2006yh,Degrassi:2002fi,Heinemeyer:1998np,
  Heinemeyer:1998yj} which we use to determine the corresponding Higgs,
sbottom and stop masses and mixing angles.\footnote{Once the theoretical
  accuracy of the Higgs cross section increases further, the three-loop
  result for the Higgs mass has to be taken into
  account~\cite{Martin:2007pg,Harlander:2008ju,Kant:2010tf}.} The
numbers for the squark masses quoted above are obtained for our default
setting $\tan\beta=10$. For the quark masses, we use the input values
\begin{equation}
\begin{split}
\mtop = 173.3\,{\rm GeV}\,,\qquad m_b(m_b) = 4.2\,{\rm GeV}\quad
\Rightarrow\quad \mbottom = 4.79\,{\rm GeV}\,,
\end{split}
\end{equation}
where $\mtop$ and $\mbottom$ are on-shell masses, and $m_b$ is the
$\msbar{}$ running mass.

In the main part of this paper, we will restrict ourselves to the
\gluophobic{+} and/or the \mhmax{+} scenario.  The results for negative
sign, including the other two scenarios, are deferred to
Appendix~\ref{app::res}.

\subsection{Numerical results}

\begin{figure}
  \begin{center}
    \begin{tabular}{cc}
      \includegraphics[bb=60 130 560
        770,width=.36\textwidth,angle=-90]{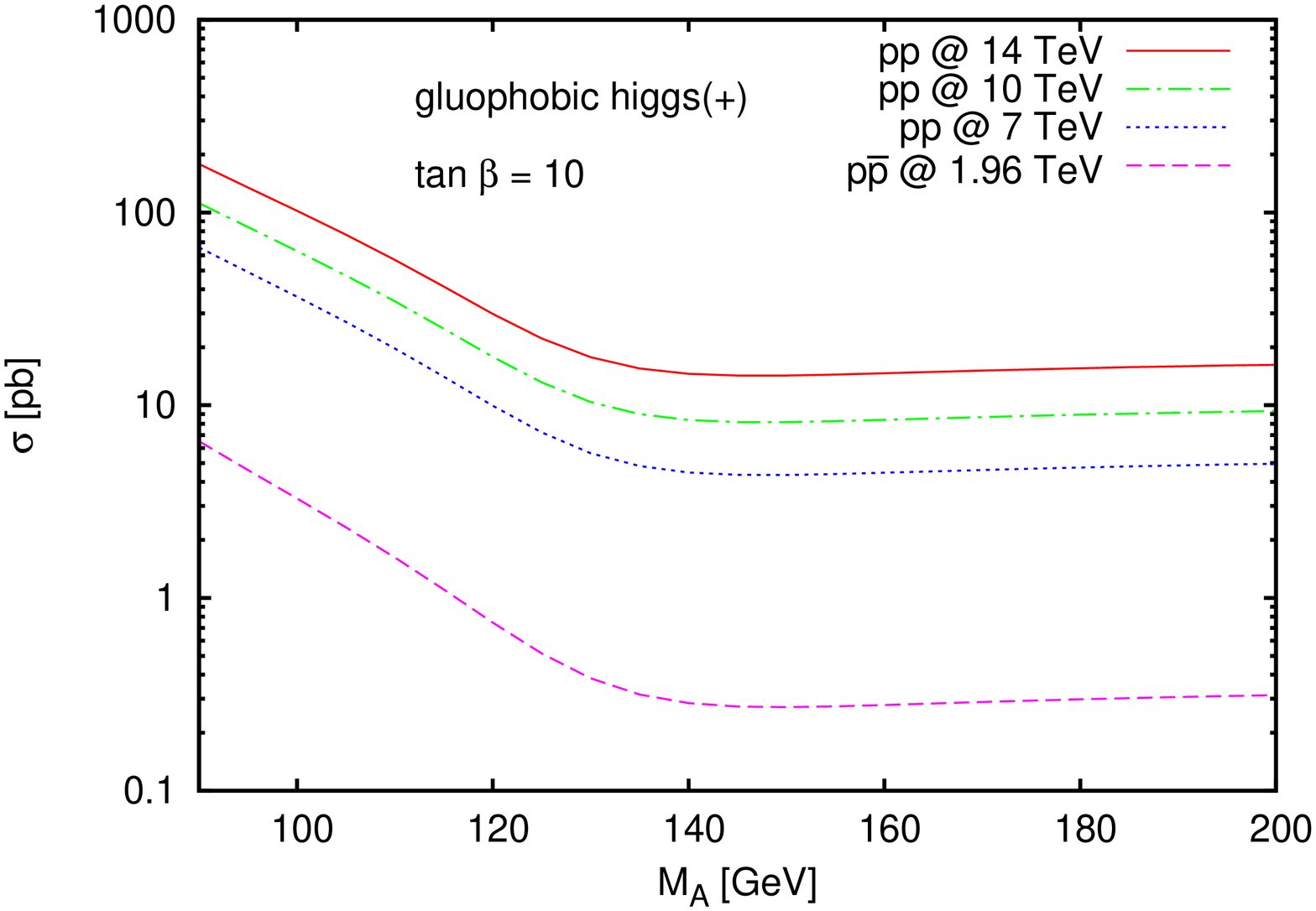} &
      \includegraphics[bb=60 130 560
        770,width=.36\textwidth,angle=-90]{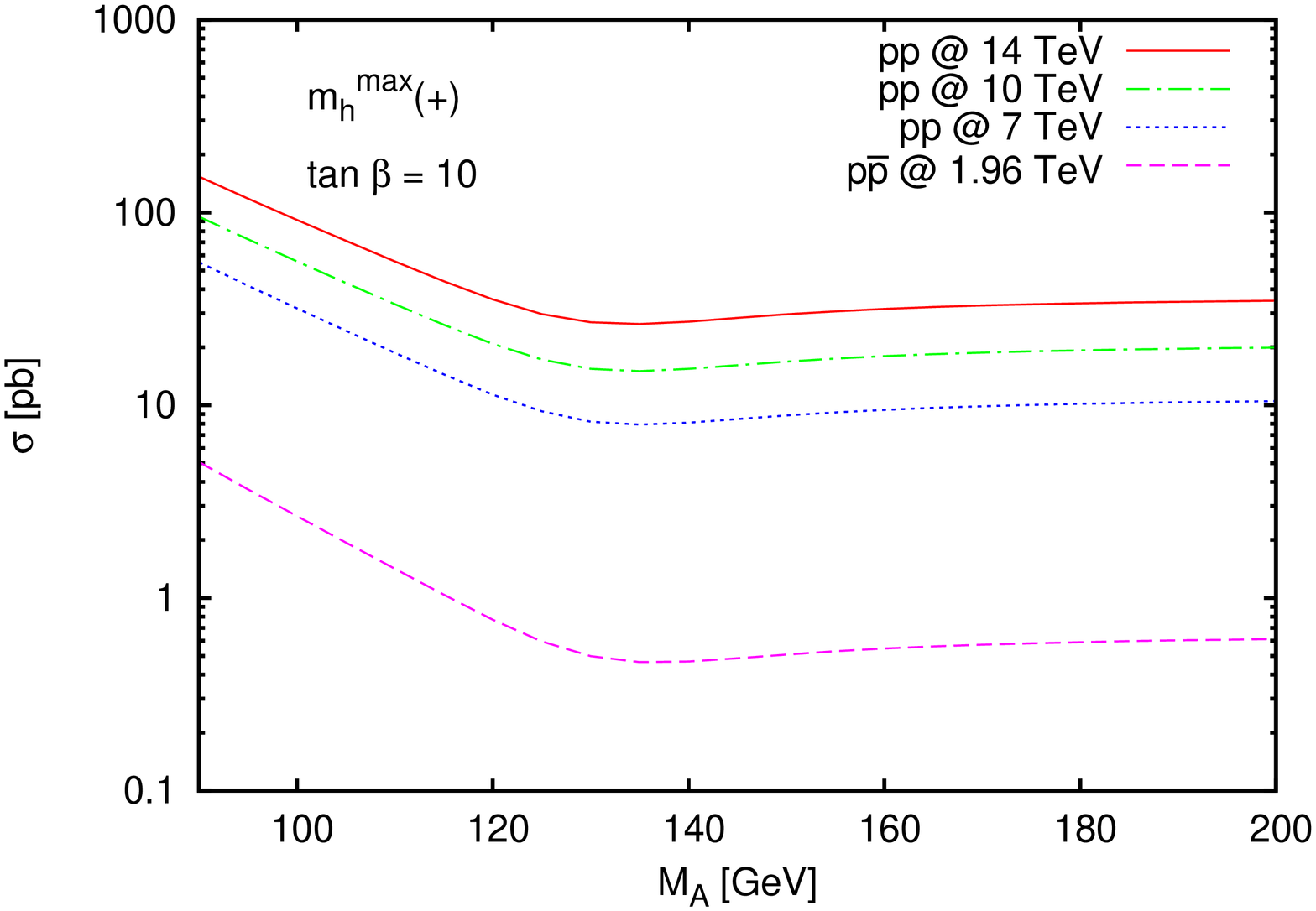} \\
      (a) & (b) \\
      \includegraphics[bb=60 130 560
        770,width=.36\textwidth,angle=-90]{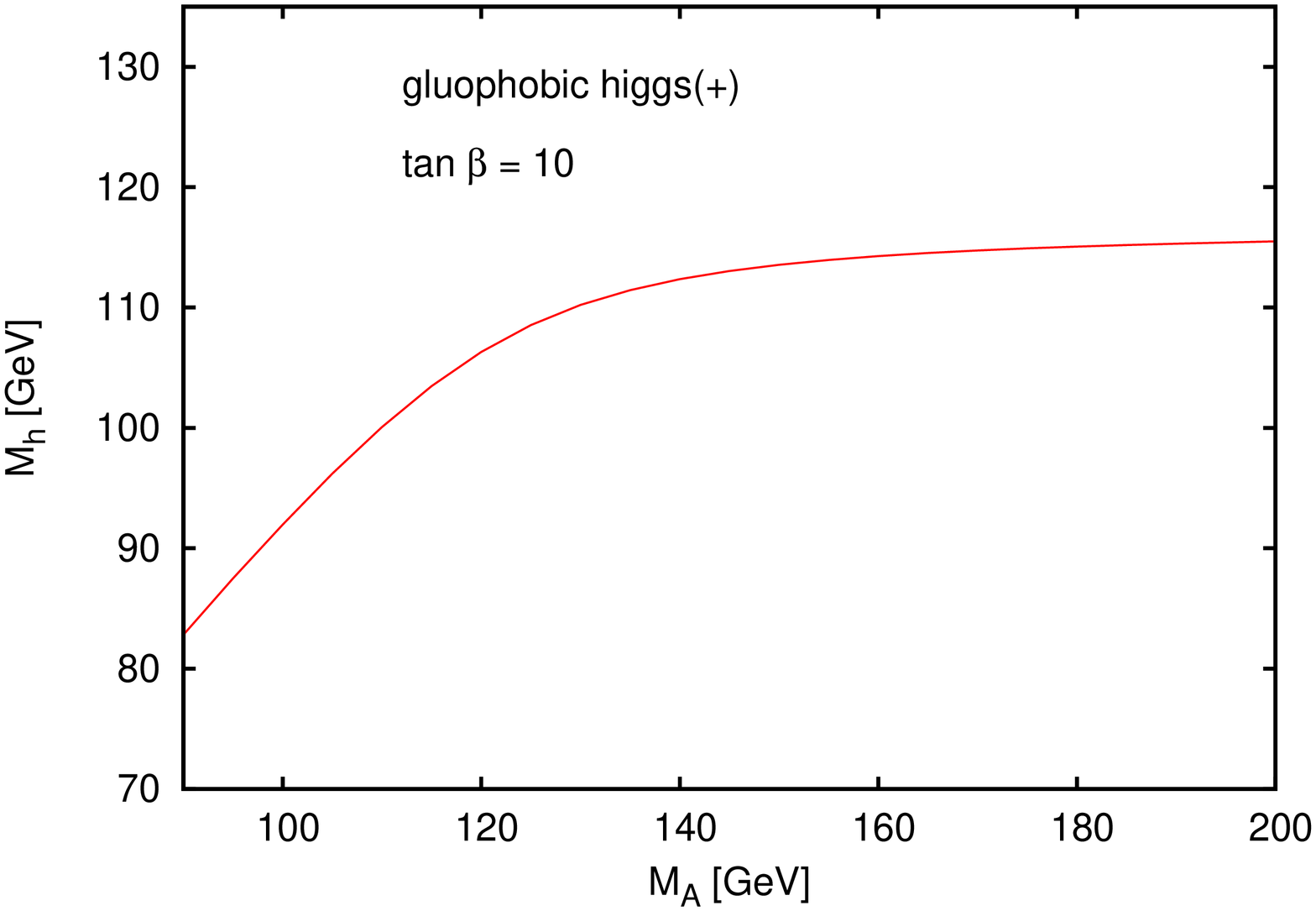} &
      \includegraphics[bb=60 130 560
        770,width=.36\textwidth,angle=-90]{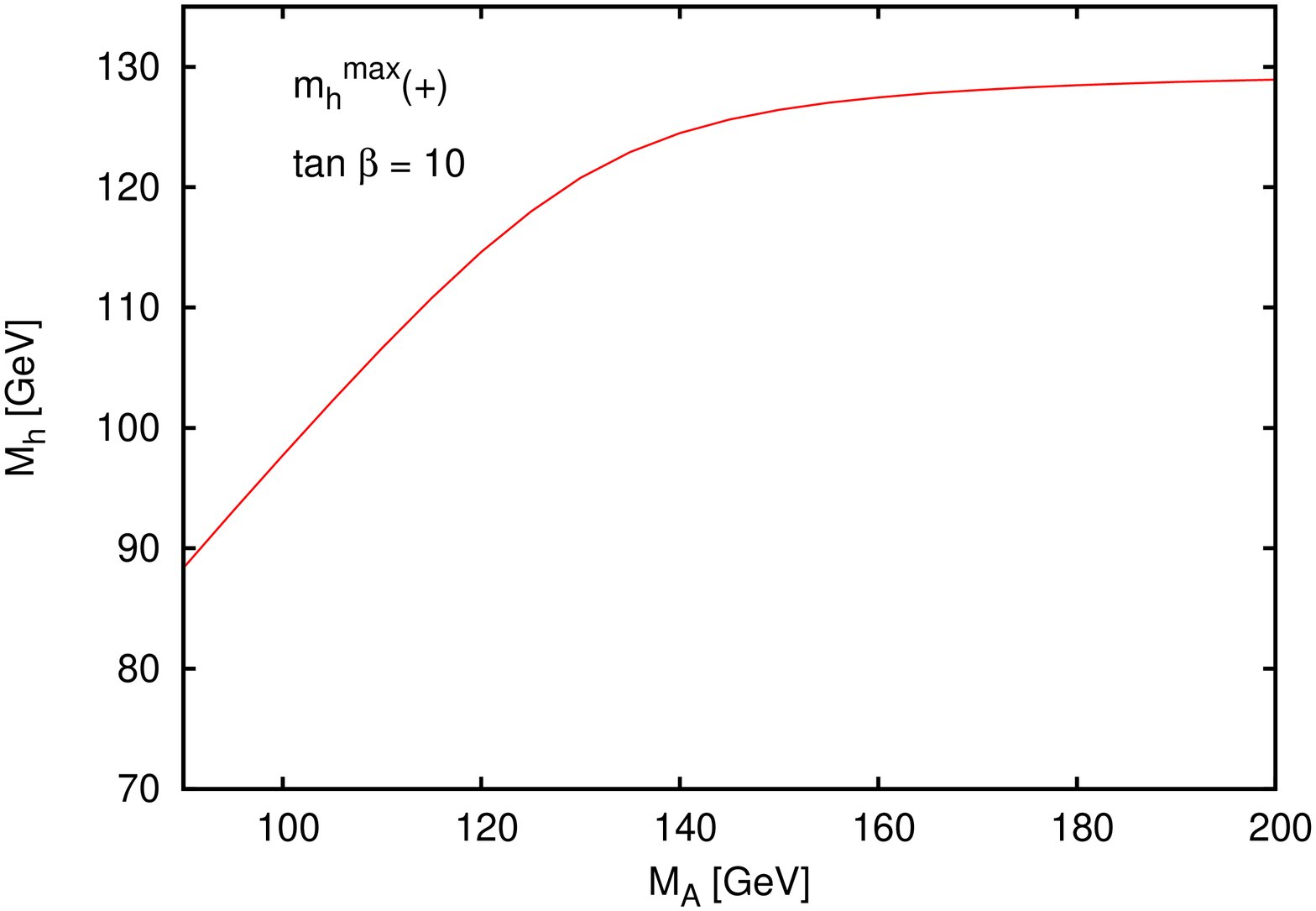} \\
      (c) & (d)
    \end{tabular}
    \parbox{.9\textwidth}{
      \caption[]{\label{fig::gluophobic.mhmax+}\sloppy Inclusive total
        cross section for gluon fusion in the \mssm{} for the scenarios
        defined in Eqs.\,(\ref{eq::gluophobic}) and (\ref{eq::mhmax}) at
        the Tevatron and the \lhc{} for various energies.  (a)
        gluophobic~$(+)$;\quad (b) $m_{h}^{\rm max}(+)$.  Panels (c) and
        (d) show the corresponding light Higgs boson mass.  }}
  \end{center}
\end{figure}

The total inclusive Higgs production cross section within the \mssm{}
through \nlo{} \qcd{} is shown for the \gluophobic{+} and the \mhmax{+}
scenario in \fig{fig::gluophobic.mhmax+}\,(a) and (b), respectively. The
figures include results for the Tevatron ($p\bar p@ 1.96$\,TeV) as well
as for the \lhc{} at various energies, where $\tan\beta=10$. Here and in
the following, the renormalization and factorization scales are set to
$\mu\equiv\muF=\muR=\mhiggs/2$~\cite{Anastasiou:2008tj}, unless
indicated otherwise.  Also shown is the corresponding light Higgs mass
$\mhiggs$, see \fig{fig::gluophobic.mhmax+}\,(c) and (d). The results
for the other scenarios can be found in Appendix~\ref{app::res}.

In the \sm{}, where the cross section has been studied in great
detail~\cite{Georgi:1977gs,Djouadi:1991tka,Dawson:1990zj,Spira:1995rr,%
  Harlander:2002wh,Anastasiou:2002yz,Ravindran:2003um,Marzani:2008az,%
  Harlander:2009mq,Pak:2009dg,Harlander:2010my,Catani:2003zt,Idilbi:2005ni,%
  Idilbi:2006dg,Moch:2005ky,Ahrens:2008nc,Aglietti:2004nj,Actis:2008ug,
  Anastasiou:2008tj,deFlorian:2009hc,Baglio:2010ae}, it was found that the \nnlo{} \qcd{} corrections
are essential in order to reduce the scale uncertainty to an acceptable
level. In addition, several calculations of beyond-\nnlo{} effects lead
to the conclusion that the fixed-order \nnlo{} result provides a fairly
precise prediction of the inclusive rate.

Since we expect a similar behavior in the \mssm{}, we need to transfer
the available information from the \sm{} result to the \mssm{} case. We
therefore define our best prediction of the total inclusive cross
section as
\begin{equation}
\begin{split}
\sigma^{\rm \mssm} &= \sigma_{\rm NLO}^{\rm \mssm} + (g_t^h)^2\left[
  (1+\delta_{\rm EW})\sigma_{\rm NNLO}^{{\rm \sm},t} - \sigma_{\rm
    NLO}^{{\rm \sm},t}\right]\,,
\label{eq::best}
\end{split}
\end{equation}
where $\sigma_{\rm \nlo{}}^{\rm \mssm}$ is our result for the total
inclusive cross section though $\order{\alpha_s^3}$ within the \mssm{}
as described in the preceding sections. We consistently evaluate it with
\nlo{} parton density functions (\pdf{}s).\footnote{In this paper, we
  use the central set of {\abbrev MSTW2008}~\cite{Martin:2009iq}
  throughout. Detailed studies of the \pdf{} dependence will be
  considered in a forthcoming paper.}  The quantity $\sigma_{\rm
  (N)NLO}^{{\rm \sm},t}$ is the top-quark induced \sm{} cross section
evaluated at {\abbrev (N)NLO} \qcd{} (i.e., with {\abbrev (N)NLO}
\pdf{}s), and $g_t^h = \cos\alpha/\sin\beta$.  In addition, the
electro-weak correction factor within the \sm{}, $\delta_{\rm
  EW}$~\cite{Actis:2008ug}, is included by assuming complete
factorization, as it was indicated to be a reasonable
assumption~\cite{Anastasiou:2008tj}.

In this paper we restrict most of our analysis to moderate and small
$\tan\beta$. Therefore, we do not actually expect the resummation of the
dominant $\tan\beta$ terms along the lines of
Ref.\,\cite{Carena:1999py,Carena:2000uj} to be important. Nevertheless,
we implement it, not least as a useful check: upon expansion of the
resummed expression in terms of $\alpha_s$, we recover the coefficient
${\cal K}_2^b$ (through $\order{1/\msusy{b}^0}$) which gives the leading
term in $\tan\beta$ at \nlo{}. The numerical effect will be studied in
more detail below.  The renormalization and factorization scales are
again set to $\muF=\muR=\mhiggs/2$.

\newcommand{\tablehead}{$ M_A $ & $ M_h $ & $(g_t^h)^2$ & $1+\delta_{\rm
    EW}$ & $\sigma_{\rm NLO}^{{\rm SM},t}$ & $\sigma_{\rm NNLO}^{{\rm
      SM},t}$ & $\sigma_{\rm NLO}^{{\rm MSSM}}$ & $\sigma^{{\rm MSSM}}$}
\newcommand{\tablecaption}[3]{Data for the #1 scenario with
  $\tan\beta=10$ at the #2 with #3\,TeV. (Masses are given in GeV, cross
  sections in pb.)}

\begin{table}
\begin{center}
\begin{tabular}{cccccccc}
\tablehead\\\hline
$90$&$82.8$&$0.063$&$1.037$&$31.49$&$37.28$&$65.90$&$66.35$\\
$95$&$87.5$&$0.085$&$1.038$&$28.29$&$33.41$&$49.03$&$49.57$\\
$100$&$92.0$&$0.119$&$1.039$&$25.63$&$30.20$&$36.53$&$37.21$\\
$105$&$96.2$&$0.169$&$1.041$&$23.44$&$27.55$&$27.03$&$27.91$\\
$110$&$100.1$&$0.240$&$1.042$&$21.65$&$25.40$&$19.69$&$20.85$\\
$115$&$103.5$&$0.337$&$1.043$&$20.23$&$23.71$&$14.05$&$15.57$\\
$120$&$106.3$&$0.452$&$1.044$&$19.14$&$22.41$&$9.93$&$11.86$\\
$125$&$108.5$&$0.570$&$1.045$&$18.35$&$21.46$&$7.22$&$9.54$\\
$130$&$110.2$&$0.672$&$1.046$&$17.78$&$20.78$&$5.64$&$8.30$\\
$135$&$111.4$&$0.751$&$1.046$&$17.38$&$20.30$&$4.84$&$7.74$\\
$140$&$112.4$&$0.809$&$1.046$&$17.09$&$19.96$&$4.48$&$7.55$\\
$145$&$113.0$&$0.850$&$1.047$&$16.88$&$19.71$&$4.36$&$7.55$\\
$150$&$113.5$&$0.880$&$1.047$&$16.72$&$19.52$&$4.35$&$7.62$\\
$155$&$113.9$&$0.902$&$1.047$&$16.60$&$19.38$&$4.39$&$7.73$\\
$160$&$114.3$&$0.919$&$1.047$&$16.50$&$19.26$&$4.46$&$7.84$\\
$165$&$114.5$&$0.931$&$1.047$&$16.42$&$19.17$&$4.54$&$7.94$\\
$170$&$114.7$&$0.941$&$1.047$&$16.36$&$19.09$&$4.61$&$8.04$\\
$175$&$114.9$&$0.949$&$1.047$&$16.31$&$19.03$&$4.69$&$8.13$\\
$180$&$115.1$&$0.955$&$1.047$&$16.26$&$18.98$&$4.75$&$8.21$\\
$185$&$115.2$&$0.960$&$1.048$&$16.23$&$18.94$&$4.81$&$8.28$\\
$190$&$115.3$&$0.964$&$1.048$&$16.19$&$18.90$&$4.87$&$8.35$\\
$195$&$115.4$&$0.968$&$1.048$&$16.17$&$18.86$&$4.92$&$8.40$\\
$200$&$115.5$&$0.971$&$1.048$&$16.14$&$18.84$&$4.97$&$8.46$\\

\end{tabular}
\end{center}
\caption[]{\label{tab::gluophobic+pp7}%
  \tablecaption{\gluophobic{+}}{\lhc{}}{7}}
\end{table}

\begin{table}
\begin{center}
\begin{tabular}{cccccccc}
\tablehead\\\hline
$90$&$88.4$&$0.030$&$1.038$&$27.74$&$32.74$&$55.20$&$55.38$\\
$95$&$93.1$&$0.039$&$1.040$&$25.03$&$29.47$&$41.77$&$41.99$\\
$100$&$97.7$&$0.054$&$1.041$&$22.72$&$26.68$&$31.82$&$32.09$\\
$105$&$102.2$&$0.075$&$1.043$&$20.73$&$24.30$&$24.36$&$24.70$\\
$110$&$106.6$&$0.106$&$1.044$&$19.03$&$22.28$&$18.72$&$19.17$\\
$115$&$110.8$&$0.155$&$1.046$&$17.59$&$20.56$&$14.47$&$15.07$\\
$120$&$114.6$&$0.227$&$1.047$&$16.40$&$19.14$&$11.36$&$12.19$\\
$125$&$118.0$&$0.329$&$1.049$&$15.44$&$18.00$&$9.30$&$10.43$\\
$130$&$120.8$&$0.453$&$1.050$&$14.70$&$17.13$&$8.22$&$9.70$\\
$135$&$122.9$&$0.579$&$1.051$&$14.17$&$16.50$&$7.93$&$9.76$\\
$140$&$124.5$&$0.687$&$1.051$&$13.80$&$16.06$&$8.11$&$10.22$\\
$145$&$125.6$&$0.767$&$1.052$&$13.54$&$15.75$&$8.47$&$10.79$\\
$150$&$126.4$&$0.824$&$1.052$&$13.36$&$15.54$&$8.85$&$11.31$\\
$155$&$127.0$&$0.864$&$1.052$&$13.23$&$15.38$&$9.18$&$11.74$\\
$160$&$127.5$&$0.892$&$1.052$&$13.13$&$15.27$&$9.46$&$12.08$\\
$165$&$127.8$&$0.913$&$1.053$&$13.06$&$15.18$&$9.69$&$12.36$\\
$170$&$128.1$&$0.928$&$1.053$&$13.00$&$15.11$&$9.88$&$12.58$\\
$175$&$128.3$&$0.939$&$1.053$&$12.95$&$15.06$&$10.03$&$12.76$\\
$180$&$128.5$&$0.948$&$1.053$&$12.91$&$15.01$&$10.16$&$12.90$\\
$185$&$128.6$&$0.955$&$1.053$&$12.88$&$14.98$&$10.27$&$13.03$\\
$190$&$128.7$&$0.961$&$1.053$&$12.86$&$14.94$&$10.36$&$13.13$\\
$195$&$128.8$&$0.965$&$1.053$&$12.83$&$14.92$&$10.44$&$13.22$\\
$200$&$128.9$&$0.969$&$1.053$&$12.81$&$14.90$&$10.51$&$13.29$\\

\end{tabular}
\end{center}
\caption[]{\label{tab::mhmax+pp7}%
  \tablecaption{\mhmax{+}}{\lhc{}}{7}}
\end{table}

We provide the numerical results for $\sigma^{\mssm{}}$ as well as the
individual contributions from \eqn{eq::best} for the \gluophobic{+} and
the \mhmax{+} scenario in Table~\ref{tab::gluophobic+pp7} and
\ref{tab::mhmax+pp7}. As expected, in the gluophobic scenario the cross
section is typically much smaller than the \sm{} value for the same
Higgs mass, at least for Higgs masses above $\sim 100$\,GeV.  In
particular for the phenomenologically most relevant region
$\mhiggs\gtrsim 114$\,GeV, the ratio of the \nlo{} \mssm{} to \sm{}
result is only 25-30\%. Inclusion of the top quark induced \nnlo{} terms
almost doubles the \mssm{} result, so that the \mssm{} to \sm{} ratio
increases to roughly 40-45\%. For $\mhiggs<100$\,GeV, on the other hand,
the $\mssm{}$ cross section can become significantly larger than the
$\sm{}$ result.

This qualitative feature of a suppression of the cross section due to
\susy{} effects for $\mhiggs\gtrsim 114$\,GeV applies also for the
$m_h^{\rm max}$ scenario, but much less pronounced. And, also here, as
one lowers the Higgs mass, the \mssm{} result surpasses the \sm{}
one.

\begin{figure}
  \begin{center}
    \begin{tabular}{cc}
      \includegraphics[bb=60 130 560
        770,width=.36\textwidth,angle=-90]{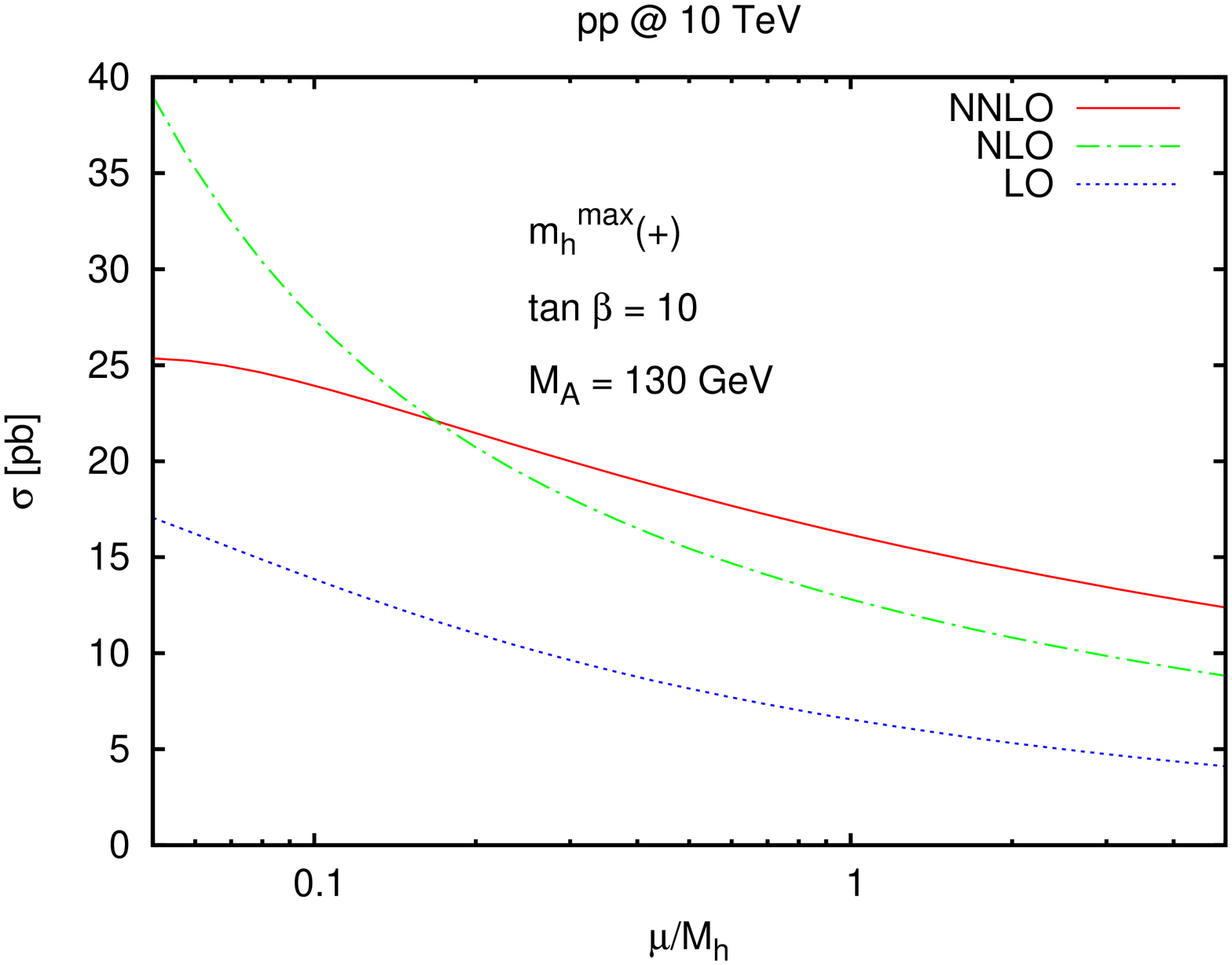} &
      \includegraphics[bb=60 130 560
        770,width=.36\textwidth,angle=-90]{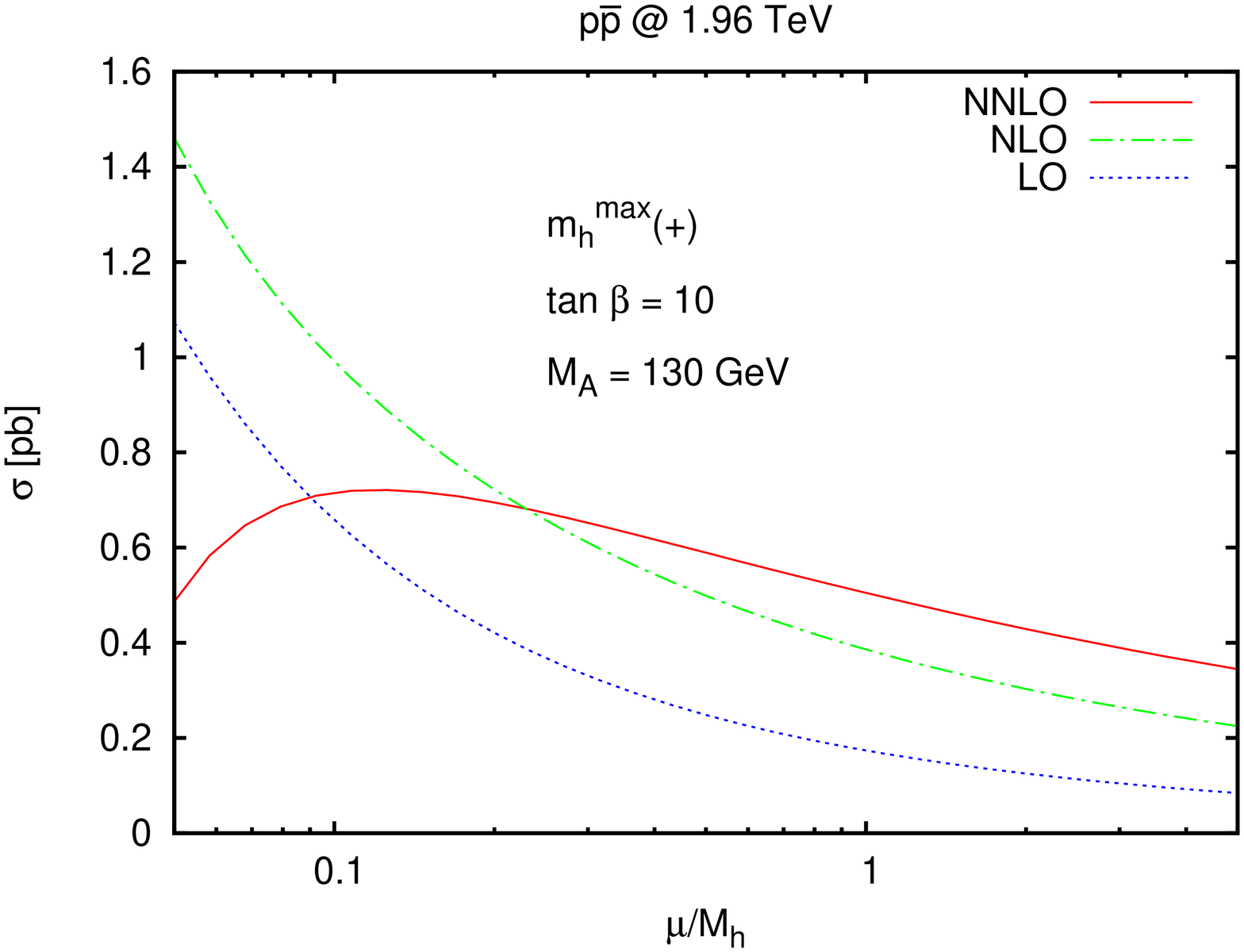} \\
      (a) & (b)
    \end{tabular}
    \parbox{.9\textwidth}{
      \caption[]{\label{fig::mudep}\sloppy Renormalization/factorization
        scale dependence ($\mu=\muF=\muR$) of the inclusive total cross
        section for gluon fusion in the \mssm{}. (a) \lhc{} at 10\,TeV;
        (b) Tevatron.  }}
  \end{center}
\end{figure}

Fig.\,\ref{fig::mudep} shows the scale variation of the \mssm{} cross
section in the \mhmax{+} scenario at \lo{} and \nlo{}, as well as for
our best prediction $\sigma^\mssm{}$ (labelled
``\nnlo{}''). Renormalization and factorization scale are identified in
this plot ($\mu=\muF=\muR$), and varied by a factor 10 around
$\mhiggs/2$. The lower order results exhibit a similar behavior as known
from the \sm{}, and, as expected, the inclusion of the \nnlo{} top quark
terms leads to a considerable stabilization against the scale
variation. The error due to scale variations estimated from this plot by
considering the variation within $\mhiggs/4\leq \mu\leq
\mhiggs$~\cite{Anastasiou:2008tj} is of the order of 15\%.

\begin{figure}
  \begin{center}
    \begin{tabular}{cc}
      \includegraphics[bb=60 130 560
        770,width=.36\textwidth,angle=-90]{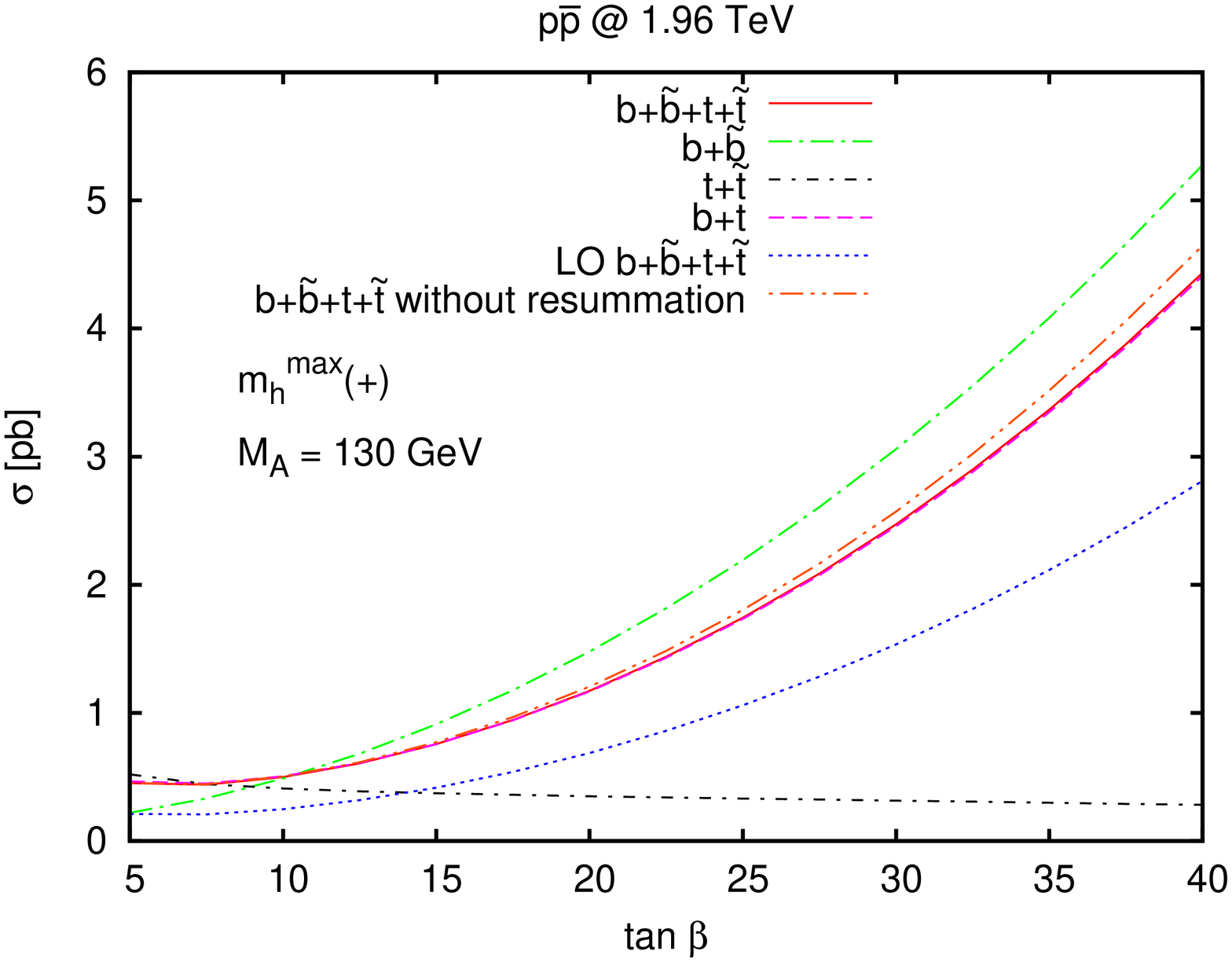} &
      \includegraphics[bb=60 130 560
        770,width=.36\textwidth,angle=-90]{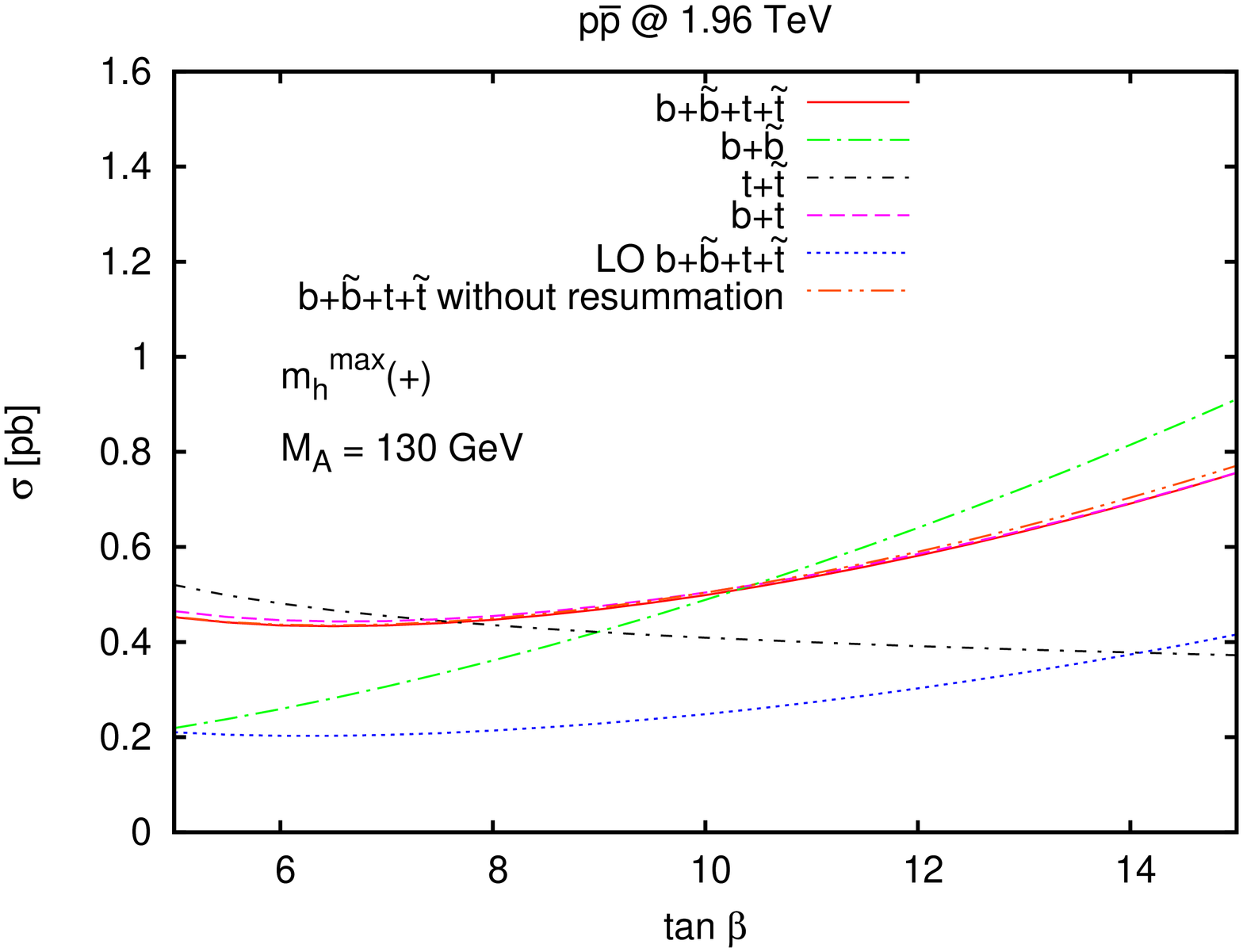}
\\[1em]      (a) & (b)
    \end{tabular}
    \parbox{.9\textwidth}{
      \caption[]{\label{fig::tanb}\sloppy Dependence of (various
        contributions to) the cross section on $\tan\beta$ for Tevatron
        conditions. Panel (b) is an zoom of the low- to
        intermediate-$\tan\beta$ region of panel (a).}}
  \end{center}
\end{figure}

\begin{figure}
  \begin{center}
    \begin{tabular}{cc}
      \includegraphics[bb=60 130 560
        770,width=.36\textwidth,angle=-90]{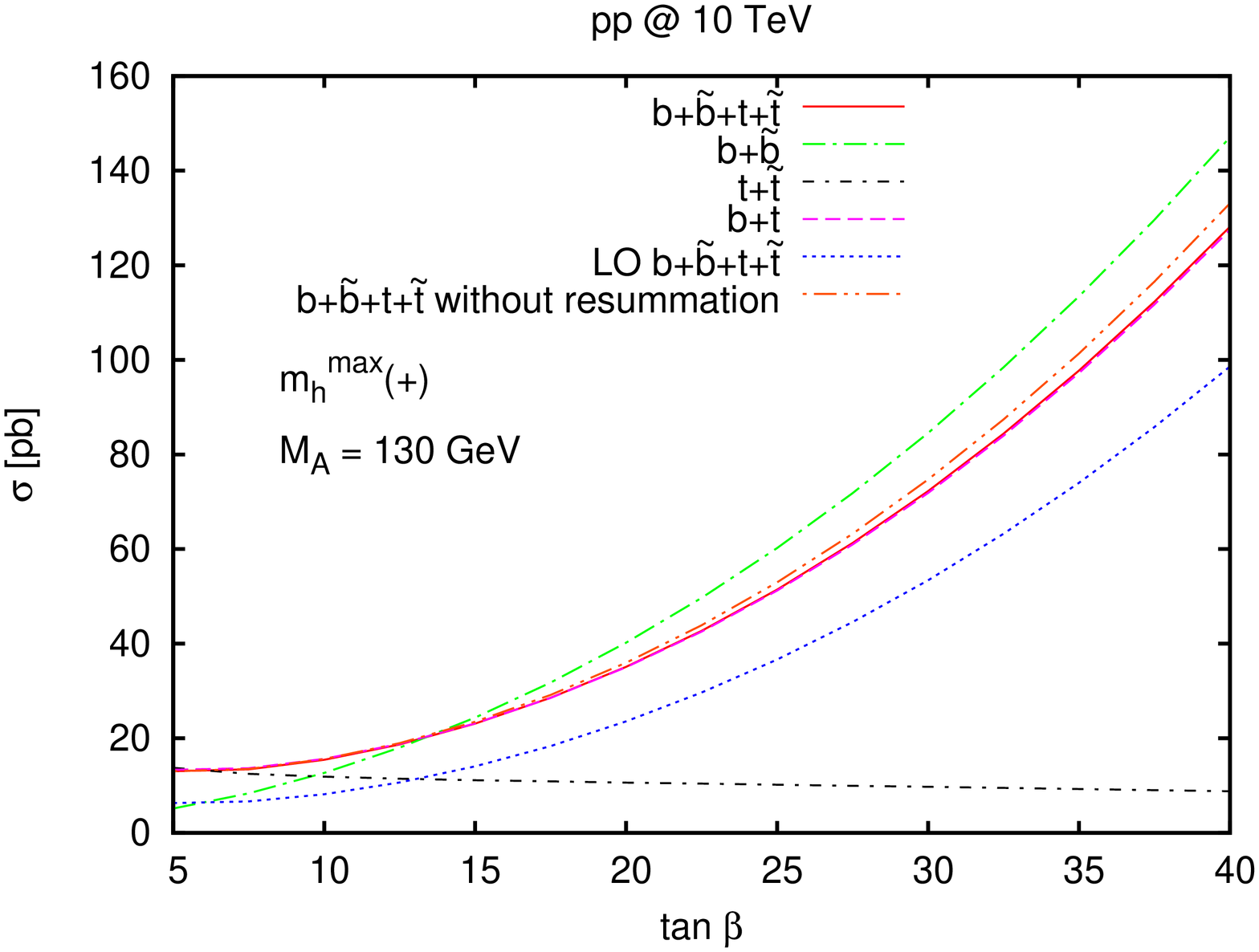} &
      \includegraphics[bb=60 130 560
        770,width=.36\textwidth,angle=-90]{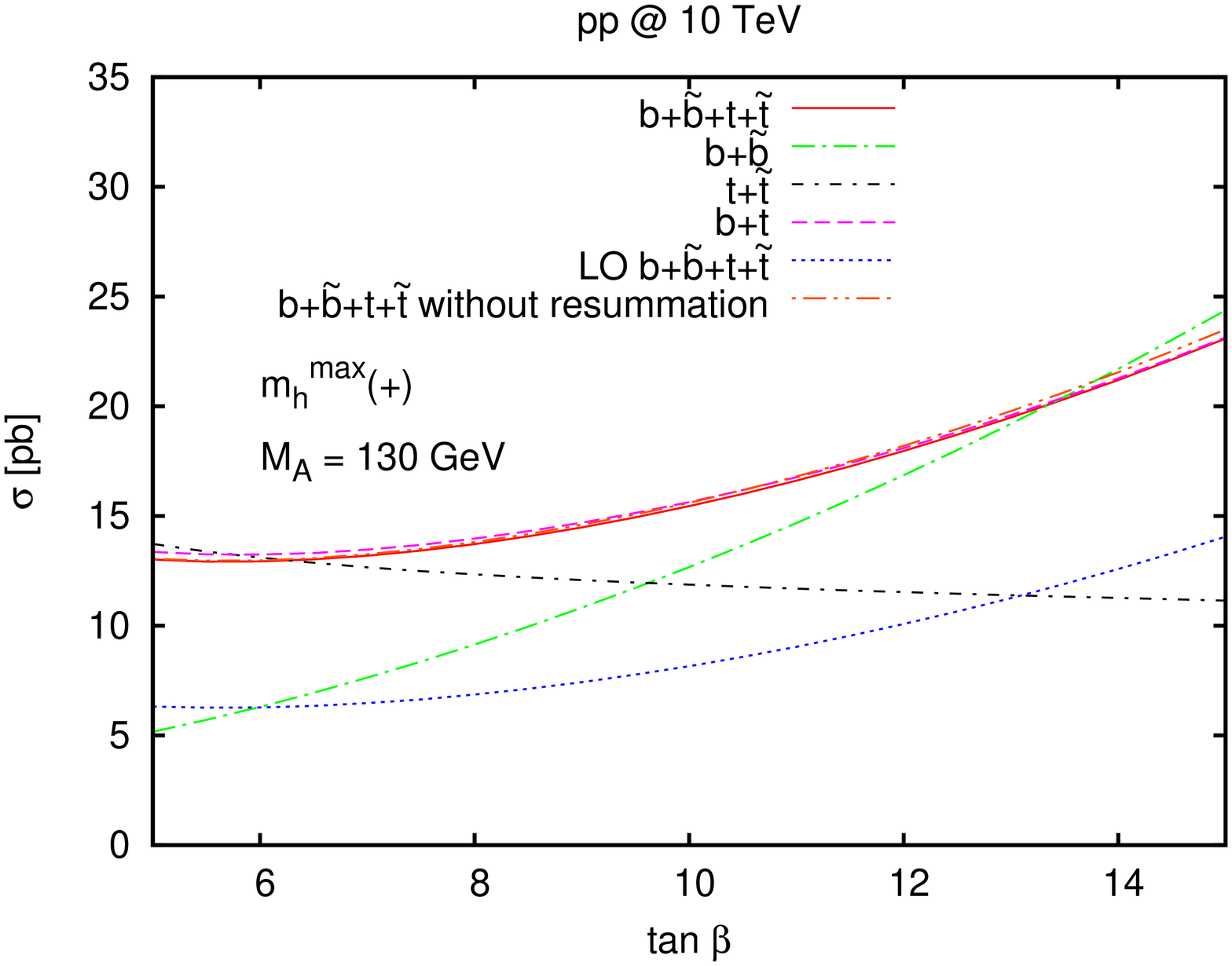}\\[1em]
      (a) & (b)
    \end{tabular}
    \parbox{.9\textwidth}{
      \caption[]{\label{fig::tanblhc}\sloppy Same as \fig{fig::tanb},
        but for the \lhc{} at 10\,TeV.  }}
  \end{center}
\end{figure}

Fig.\,\ref{fig::tanb} shows the dependence of the cross section
(Tevatron, \mhmax{+}) on the choice of $\tan\beta$, where various
contributions are displayed seperately. For example, the curve denoted
$b+\tilde b$ is obtained by setting the top- and stop-Higgs couplings to
zero, $g^h_t=\tilde g^h_{t,ij} = 0$, etc. The figure illustrates that
already for $\tan\beta \approx 10$, the bottom effects outweigh the top
effects (green/long vs.\ black/short dash-dotted). Furthermore, squark
effects have a significant effect even at these large squark masses of
$\order{1{\rm TeV}}$ if the dominant $\tan\beta$ terms are not resummed
in the Higgs-bottom Yukawa coupling (orange/dash-double-dotted
vs.\ purple/dashed). Once this resummation is taken into account, the
squark effects become negligible, however (purple/dashed
vs.\ red/solid). Note that this plot depends strongly on the choice of
$M_A$, of course. Another well-known feature that is shown is the effect
of the \nlo{} corrections which are of the order of 100\% (blue/dotted
vs.\ red/solid). A very similar behavior is observed for \lhc{}
conditions, see \fig{fig::tanblhc}.

\section{Differential distributions}\label{sec::distrib}

\begin{figure}
  \begin{center}
    \begin{tabular}{cc}
      \includegraphics[bb=60 100 560
        750,width=.36\textwidth,angle=-90]{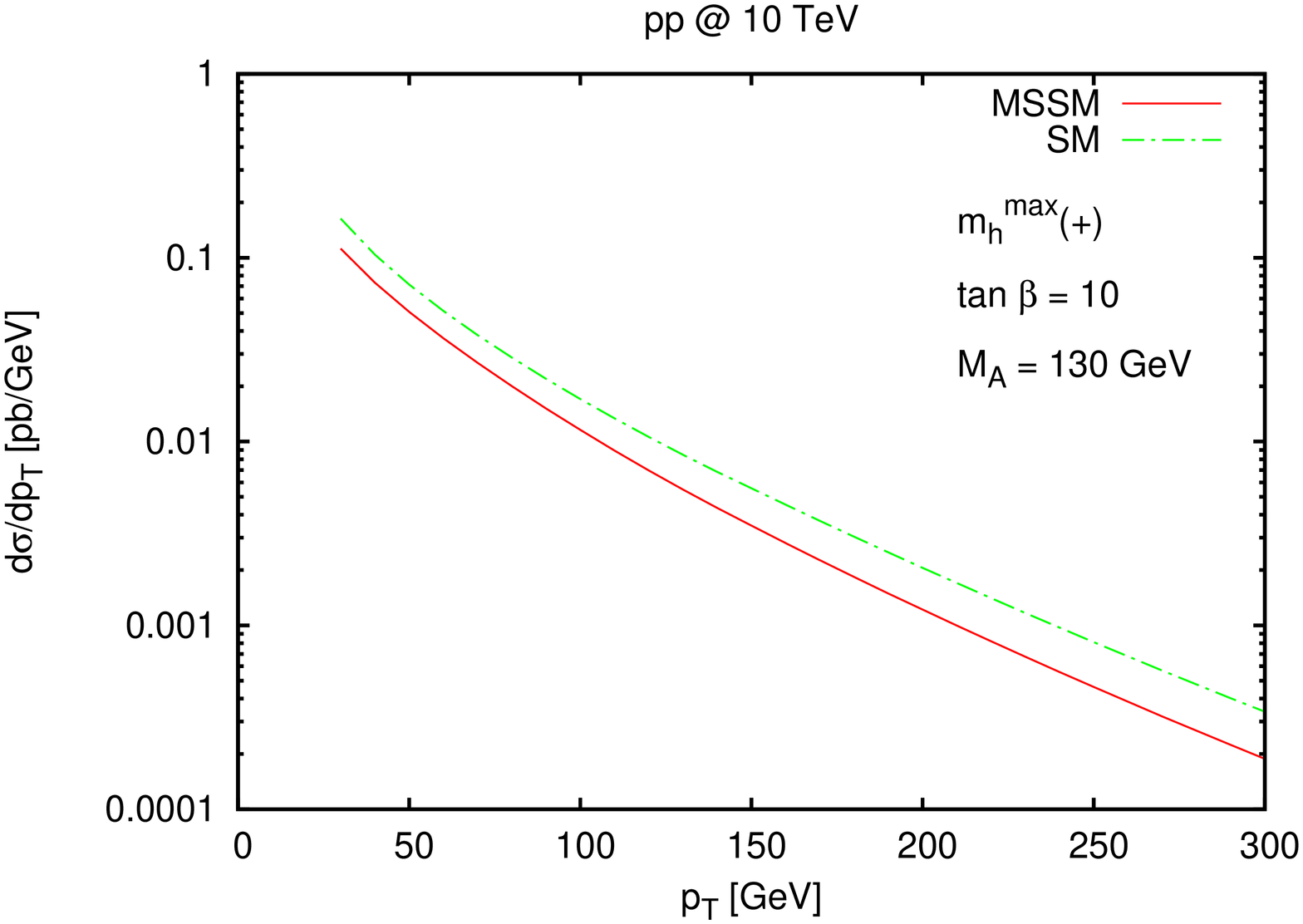} &
      \includegraphics[bb=60 100 560
        750,width=.36\textwidth,angle=-90]{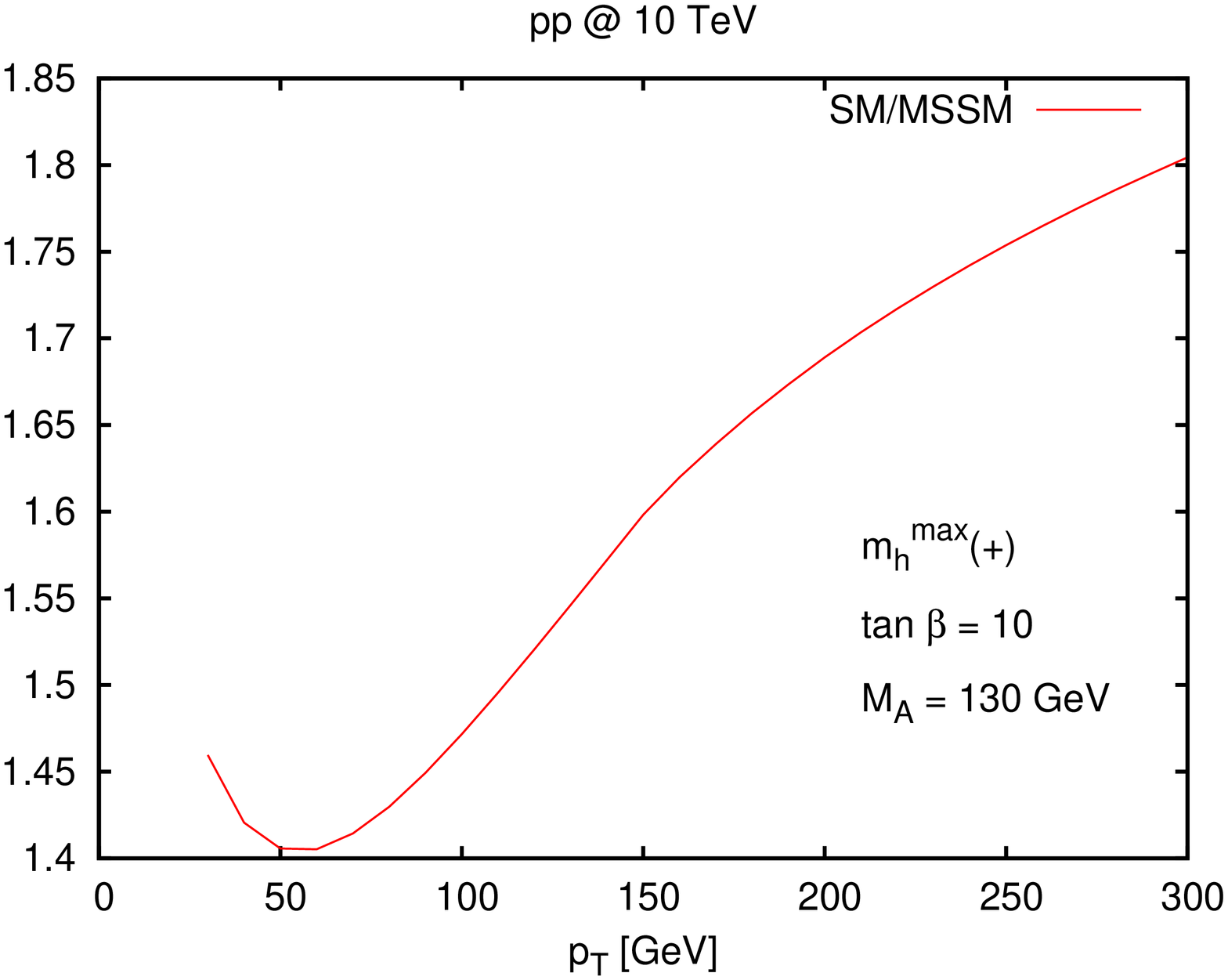} \\
      (a) & (b)
    \end{tabular}
    \parbox{.9\textwidth}{
      \caption[]{\label{fig::ptlhc}\sloppy (a) Transverse momentum
        distribution of the Higgs boson at \lo{} in the \sm{} and the
        \mssm{}, using the \mhmax{+} scenario at the \lhc{} for 10\,TeV.
        (b) Ratio of the \sm{} and the \mssm{} distribution.}}
  \end{center}
\end{figure}

\begin{figure}
  \begin{center}
    \begin{tabular}{cc}
      \includegraphics[bb=60 100 560
        750,width=.36\textwidth,angle=-90]{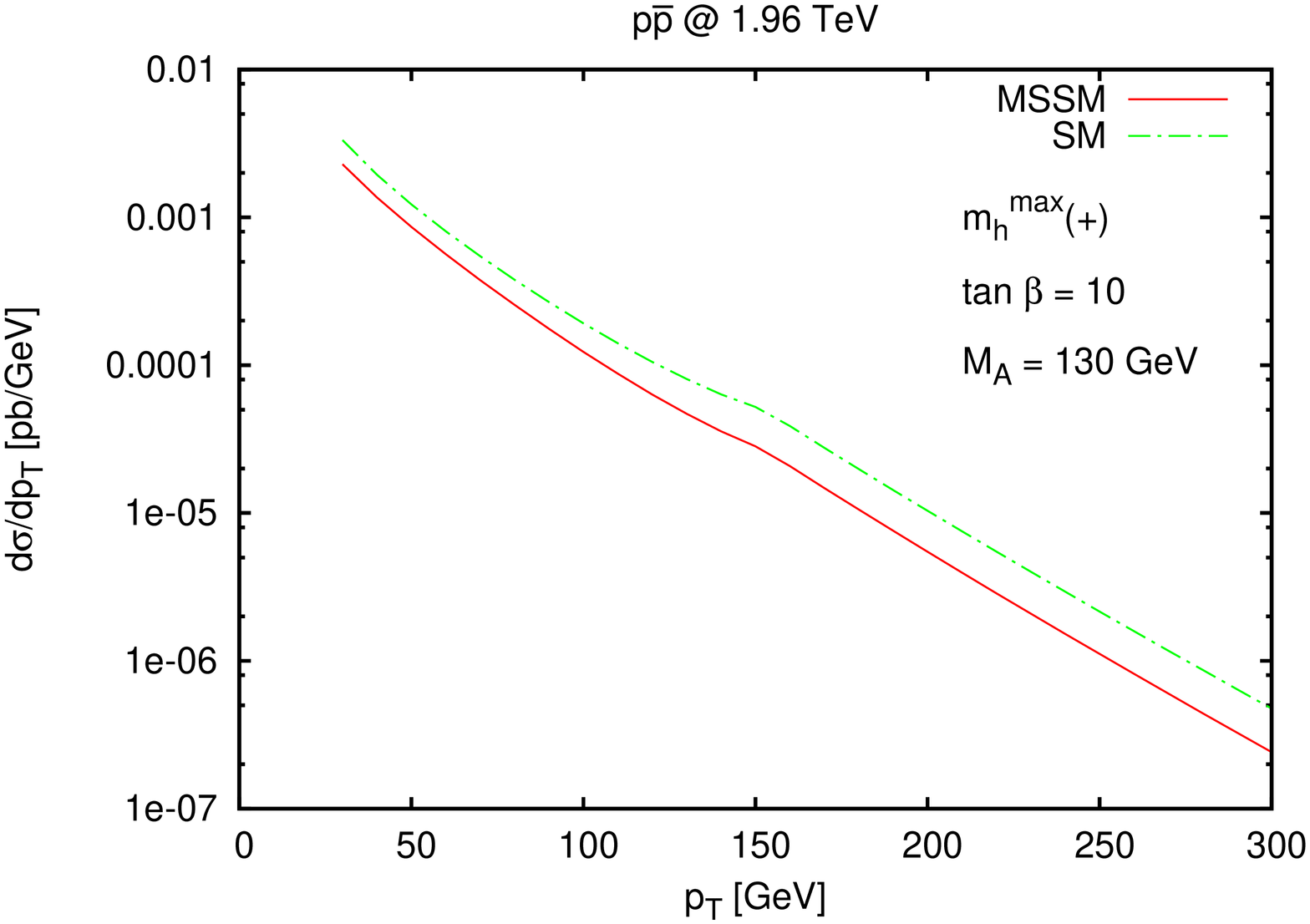} &
      \includegraphics[bb=60 100 560
        750,width=.36\textwidth,angle=-90]{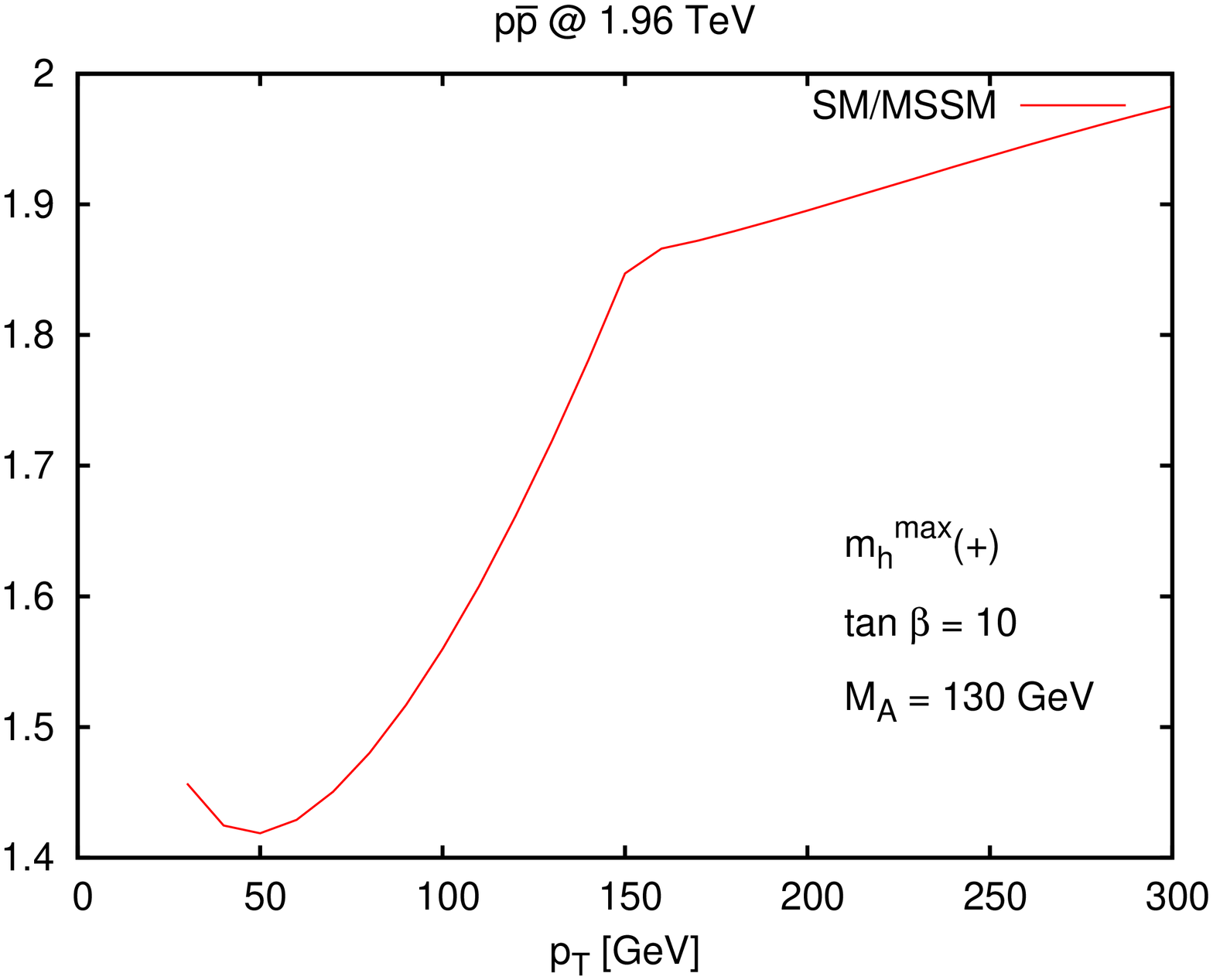} 
\\      (a) & (b)
    \end{tabular}
    \parbox{.9\textwidth}{
      \caption[]{\label{fig::pttev}\sloppy (a) Transverse momentum
        distribution of the Higgs boson at \lo{} in the \sm{} and the
        \mssm{}, using the \mhmax{+} scenario at the Tevatron.
        (b) Ratio of the \sm{} and the \mssm{} distribution.}}
  \end{center}
\end{figure}

In our perturbative partonic approach, the Higgs boson can only have
non-zero transverse momentum $\pt$ when at least one parton is produced
in association. Therefore, the purely virtual corrections do not
contribute to the $\pt$ distribution $\dd\sigma/\dd \pt$. In fact, the
\lo{} $\pt$ distribution in the \mssm{} has been considered
before~\cite{Spira:1995mt,Brein:2003df,Brein:2007da,Field:2003yy,
  Langenegger:2006wu}, and we include it here only for the sake of
completeness.  In Fig.\,\ref{fig::ptlhc}\,(a), the $\pt$ distribution at
the \lhc{} with 10\,TeV is displayed for both the \sm{} and the \mssm{}
in the \mhmax{+} scenario.\footnote{Only the value of $\mhiggs$
  influences the \sm{} prediction when changing the \susy{} parameters.}
Fig.\,\ref{fig::ptlhc}\,(b) diplays the ratio of the two curves. We
observe that the shape of the $\pt$ distribution depends non-trivially
on the model.  The corresponding results for the Tevatron are shown in
Fig.\,\ref{fig::pttev}.  The kink at $\pt\approx 150$\,GeV which is more
pronounced in the \sm{}, originates predominantly from the kinematical
cut at $\sqrt{\hat s}=\sqrt{\pt^2+\mhiggs^2}+\pt$ (see also
Ref.\,\cite{Keung:2009bs}).

\begin{figure}
  \begin{center}
    \begin{tabular}{cc}
      \includegraphics[bb=60 100 560
        750,width=.36\textwidth,angle=-90]{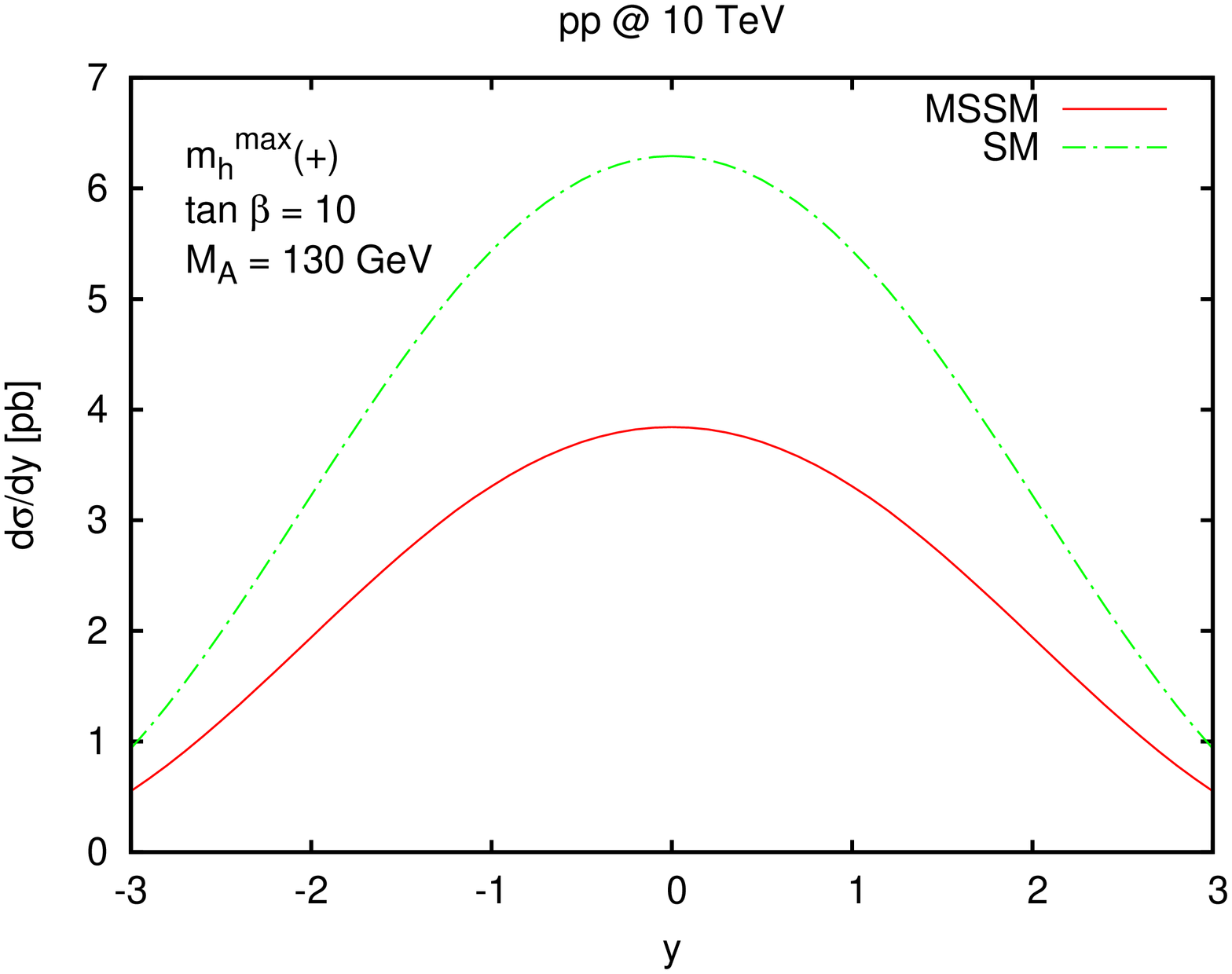} &
      \includegraphics[bb=60 100 560
        750,width=.36\textwidth,angle=-90]{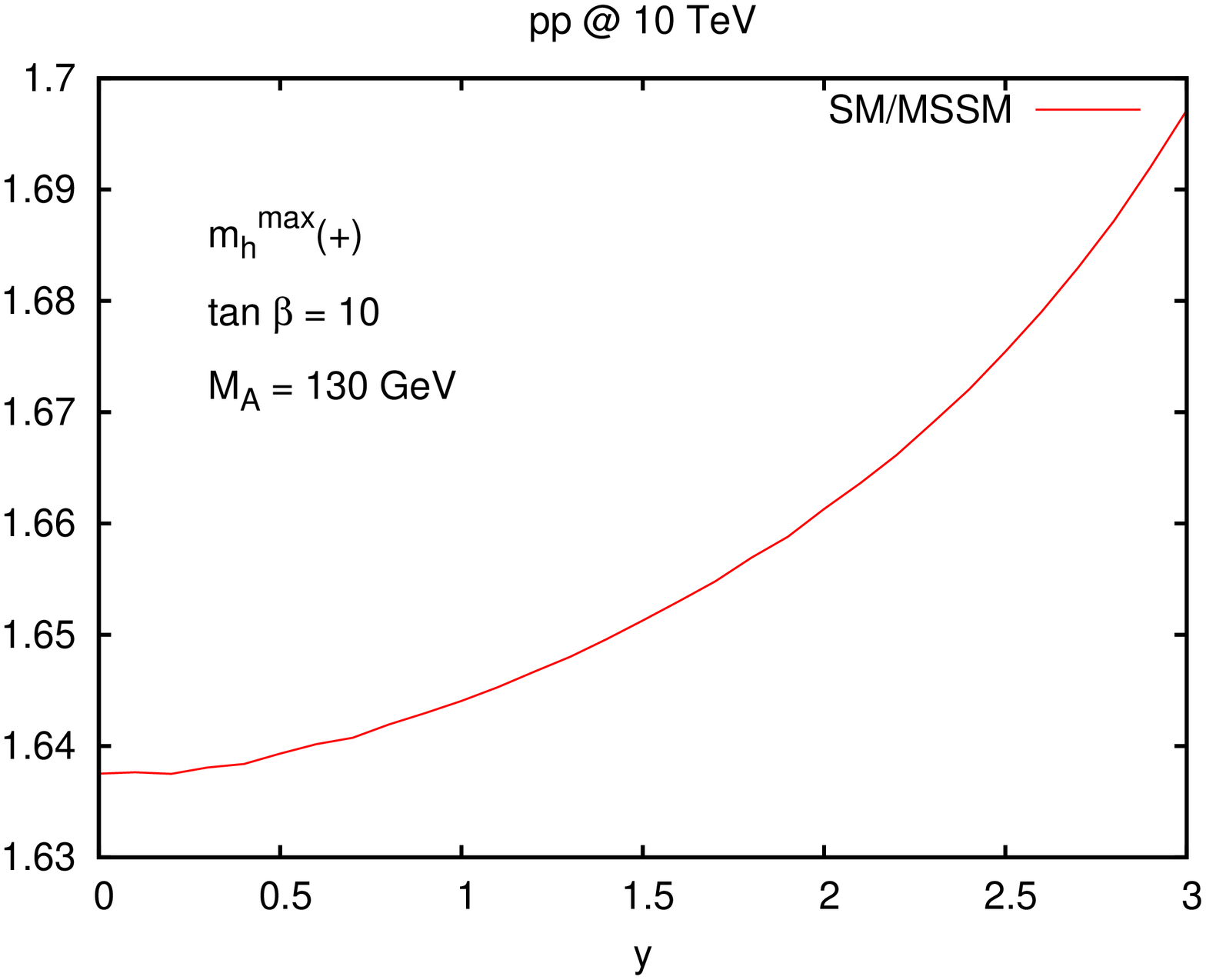} 
\\      (a) & (b)
    \end{tabular}
    \parbox{.9\textwidth}{
      \caption[]{\label{fig::ylhc}\sloppy (a) Rapidity distribution of
        the Higgs boson at \nlo{} in the \sm{} and the \mssm{}, using the
        \mhmax{+} scenario at the \lhc{} for 10\,TeV.  (b) Ratio of the
        \mssm{} and the \sm{} distribution.  }}
  \end{center}
\end{figure}

Concerning the rapidity distribution $\dd\sigma/\dd y$, where
\begin{equation}
\begin{split}
y \equiv \frac{1}{2}\ln\frac{E+p_z}{E-p_z}\,,
\end{split}
\end{equation}
with $E$ and $p_z$ the energy and the longitudinal momentum of the Higgs
boson in the lab frame, there clearly is a non-trivial distribution
already at \lo{}, given by
\begin{equation}
\begin{split}
\frac{\dd\sigma}{\dd y}\bigg|_{\rm \lo} \sim
g(\sqrt{\tau}e^y)g(\sqrt{\tau}e^{-y})\,,\qquad
\tau = \frac{\mhiggs^2}{s}\,,
\end{split}
\end{equation}
where $s$ is the hadronic center-of-mass energy. Therefore, at \nlo{},
also the virtual corrections need to be taken into account. We present
here the first truly \nlo{} results for this quantity in the \mssm{}.
Fig.\,\ref{fig::ylhc}\,(a) shows the rapidity distribution at the \lhc{}
with 10\,TeV both for the \sm{} and the \mssm{} in the \mhmax{+}
scenario, while Fig.\,\ref{fig::ylhc}\,(b) shows again the ratio of the
two curves. The model dependence of the shape is much smaller for the
$y$- than for the $\pt$-distribution. This is due to the fact that in
the former case, a difference can arise only at \nlo{}.

\begin{figure}
  \begin{center}
    \begin{tabular}{cc}
      \includegraphics[bb=60 100 560
        750,width=.36\textwidth,angle=-90]{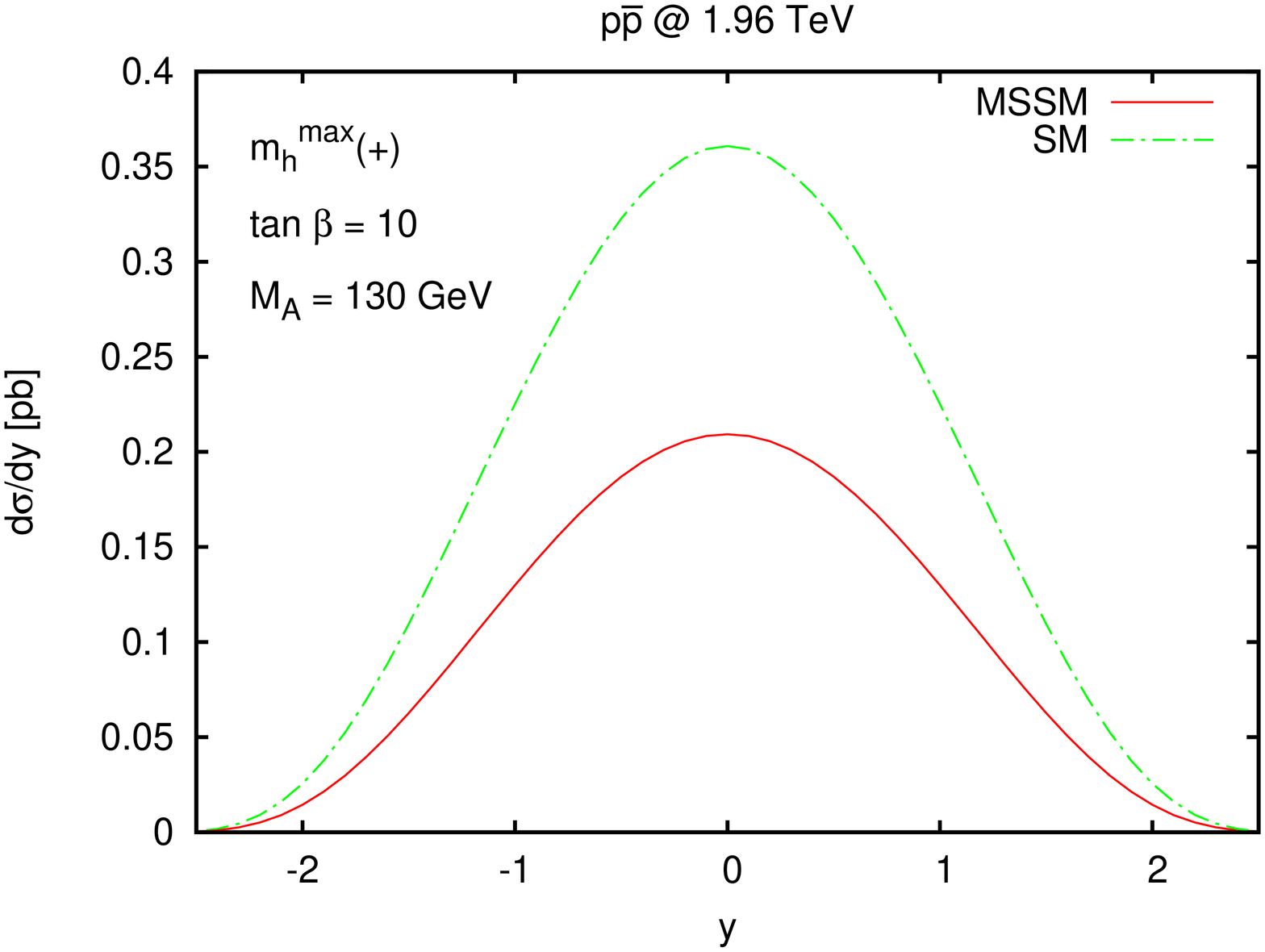} &
      \includegraphics[bb=60 100 560
        750,width=.36\textwidth,angle=-90]{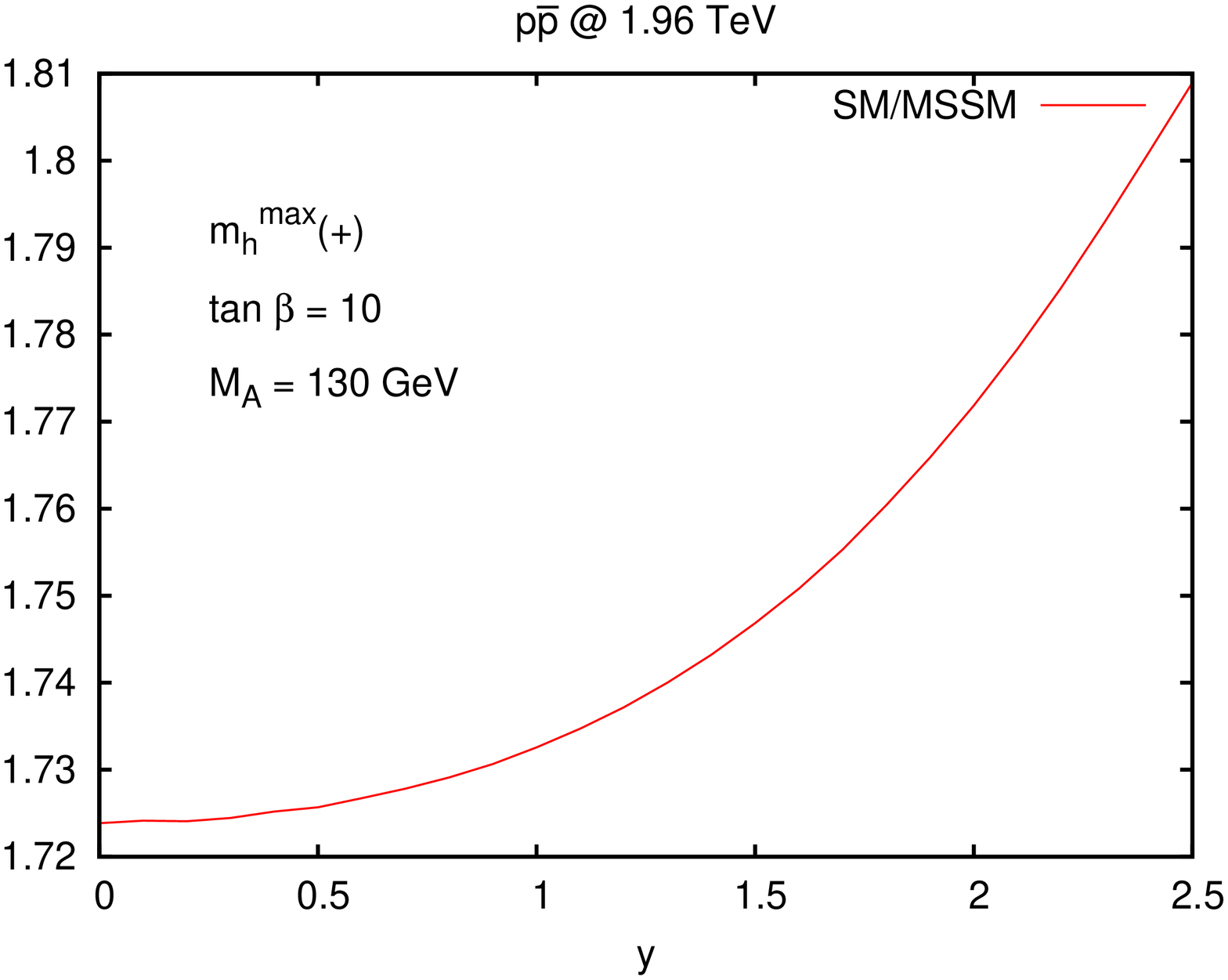} 
\\      (a) & (b)
    \end{tabular}
    \parbox{.9\textwidth}{
      \caption[]{\label{fig::ytev}\sloppy (a) Rapidity distribution of
        the Higgs boson at \nlo{} in the \sm{} and the \mssm{}, using the
        \mhmax{+} scenario at the Tevatron.  (b) Ratio of the \mssm{}
        and the \sm{} distribution.  }}
  \end{center}
\end{figure}

The same conclusions hold for the Tevatron, for which the results are
displayed in \fig{fig::ytev}.

\section{Fourth matter generation}\label{sec::4th}

It is well-known that a \sm{}-like \fourth{} generation of quarks
(denoted $(t_4,b_4)$ in what follows) would increase the Higgs
production cross section significantly, leading to a much larger
exclusion region from the Tevatron search
results~\cite{Aaltonen:2010sv}. On the other hand, the Higgs exclusion
from electro-weak precision data would soften considerably with a
\fourth{} generation~\cite{Kribs:2007nz}. Effects of a \fourth{}
generation on the inclusive \sm{}$_4$ Higgs cross section have been
evaluated through \nnlo{} \qcd{}~\cite{Anastasiou:2010bt,Li:2010fu}. At
\lo{}, also $H$+jet and $H$+2~jet production has been studied in this
model~\cite{Li:2010fu}.

In this section, we evaluate the effect of an \mssm{}-like \fourth{}
generation on the Higgs production cross section through \nlo{}.  Of
course, also the \susy{} constraints on the Higgs mass change
significantly with the presence of a \fourth{}
generation~\cite{Fok:2008yg,Litsey:2009rp,Dawson:2010jx}.  For this
first study, we will use the approximation
\begin{equation}
\begin{split}
\mhiggs^2 = (\mhiggs^{\mssm{}})^2 +
\sum_{q=t_4,b_4}\frac{3}{2\pi^2}\frac{M_4^4}{v^2}
\log\frac{M_S^2}{M_4^2}\,,\qquad
M_S^2 = M_{\rm SUSY}^2 + M_4^2\,,
\end{split}
\end{equation}
where $v=246$\,GeV, $M_{\rm SUSY}$ is given by the scenarios as defined
in \sct{sec::scenarios}, and $M_4$ is the mass of the fourth generation
quarks which are taken to be degenerate. The 3-generation result
$\mhiggs^{\mssm}$ is again taken from {\tt
  FeynHiggs}~\cite{Frank:2006yh,Degrassi:2002fi,Heinemeyer:1998np,
  Heinemeyer:1998yj}. Furthermore, for the two lighter \fourth{}
generation squark masses we also assume $\msquark{1}=M_4$
($q\in\{b_4,t_4\}$), while the heavier ones are set to
$\msquark{2}=(2M_{\rm SUSY}^2 + M_4^2)^{1/2}$.  The form of the coupling
constants for the \fourth{} generation (s)quarks is taken to be
identical to the first three generations, i.e., they are given by
Eqs.\,(\ref{eq::sbottomhiggs-1}) and (\ref{eq::ttob}) of
Appendix~\ref{app::couplings}, with the obvious replacements of the
masses and mixing angles. For recent \fourth{} generation search limits,
see Refs.\,\cite{Aaltonen:2009nr,Cox:2009mx}.

\begin{figure}
  \begin{center}
    \begin{tabular}{cc}
      \includegraphics[bb=60 100 560
        750,width=.36\textwidth,angle=-90]{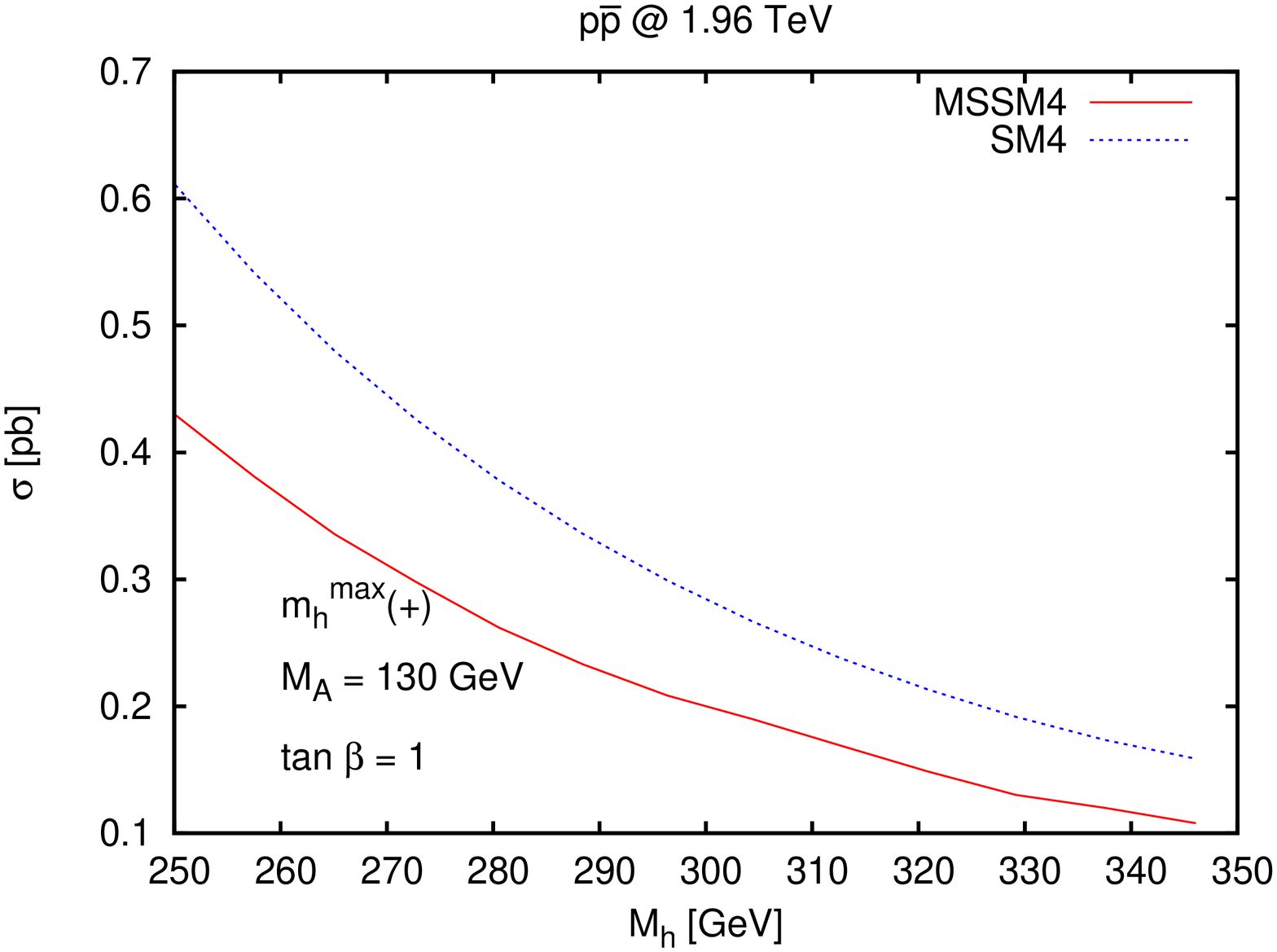} &
      \includegraphics[bb=60 100 560
        750,width=.36\textwidth,angle=-90]{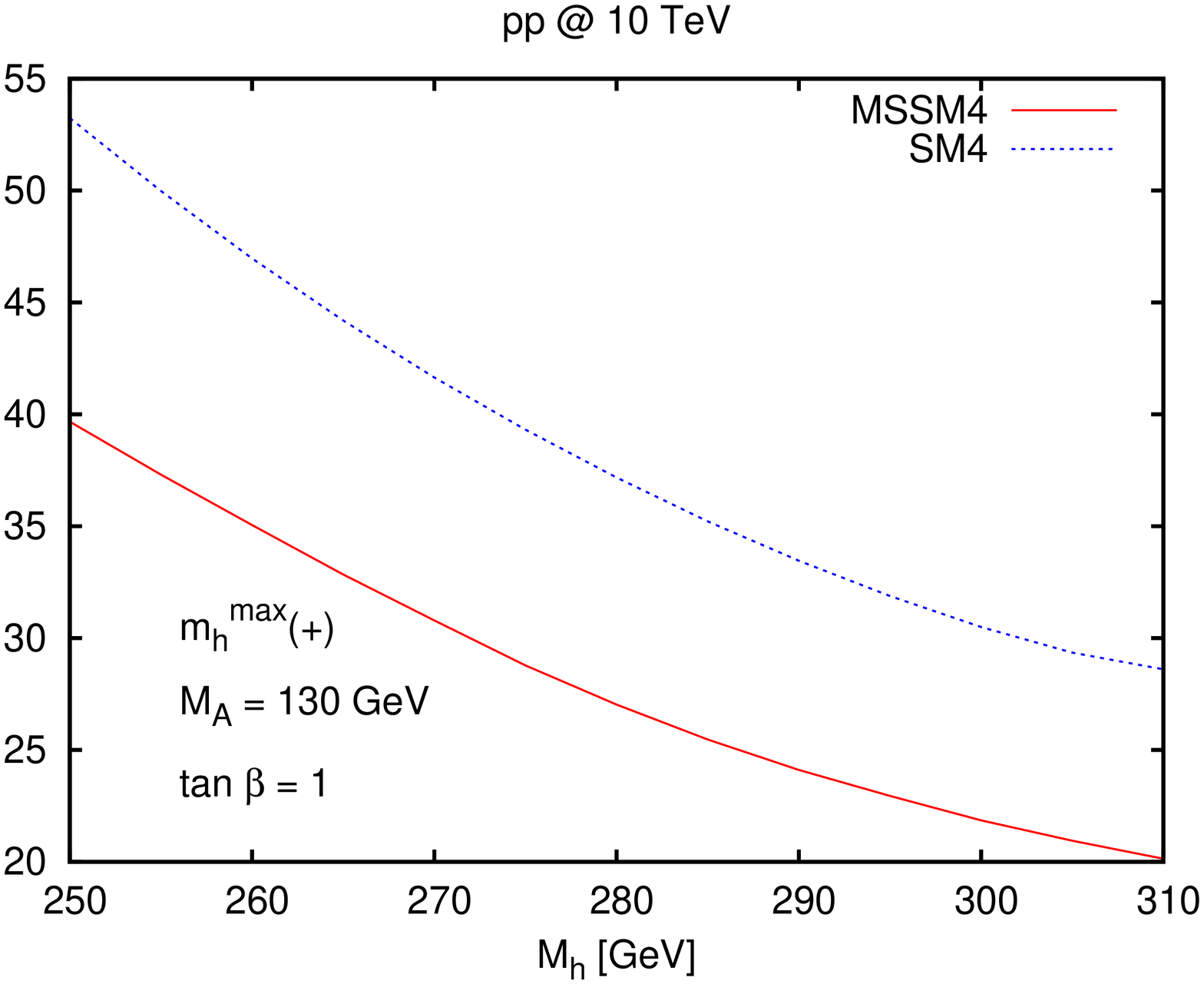}
\\      (a) & (b)
    \end{tabular}
    \parbox{.9\textwidth}{
      \caption[]{\label{fig::4gen}\sloppy Higgs production cross section
        as a function of the light Higgs boson mass (a)~at the Tevatron
        and (b)~at the \lhc{}, with a \fourth{} \sm{}- and \mssm{}-like
        generation of (s)fermions.}}
  \end{center}
\end{figure}

\fig{fig::4gen}\,(a) displays the Tevatron and \lhc{} results for the
\sm{}$_4$ and the \mssm{}$_4$ using the \susy{} parameters as in the
\mhmax{+} scenario. In order to avoid non-perturbativity of the
$hb_4\bar{b}_4$ coupling, we set $\tan\beta=1$. On top of the well-known
factor of $\sim 9$ from the \sm{}$\to$\sm$_4$ transition, we find
a 30\% decrease when switching to the \mssm{}$_4$. This value
depends quite sensitively on the actual values of the \fourth{}
generation squark masses and suggests further, more detailed
investigations.

\section{Conclusions}\label{sec::conclusions}

The various contributions of the \nlo{} \qcd{} corrections to the cross
section for Higgs production in gluon fusion have been combined in order
to evaluate it consistently within the \mssm{}. Both quark and squark
effects have been taken into account in the top and the bottom sector,
including the mixed quark/squark/gluino and all interference effects.
The numerical results have been presented for a set of selected \mssm{}
scenarios, but our implementation allows for any other reasonable set of
parameters.

For the total inclusive cross section, we constructed a ``best
approximation'' which includes the known \nnlo{} \qcd{} top-quark
induced corrections. We also provided results for Higgs distributions in
$\pt$ through \lo{} and in $y$ through \nlo{}. Finally, the effect of a
\fourth{} \mssm{} generation on the Higgs production rate has been
investigated.

In future works, we will study the effect of changing the
renormalization scheme, in particular for the bottom/sbottom sector,
include the recently evaluated \nnlo{} effects for the top/stop sector,
and consider also the production of the heavy Higgs (see also
Ref.\,\cite{Degrassi:2010eu}).

We hope that these results, plus similar ones provided
on-line~\cite{url}, will be useful input for experimental analyses.  We
will gladly provide numbers for other scenarios to the interested
reader, and are planning to release the code for public use in the near
future.

\paragraph{Acknowledgments.}
We would like to thank G.~Degrassi, S.~Heinemeyer, K.~Ozeren,
P.~Slavich, and G.~Weiglein for enlightening comments and
discussions. Furthermore, we thank M.~Steinhauser and N.~Zerf for
pointing out a number of typos in the original version of the
manuscript. This work was supported by {\abbrev DFG}, contract
HA~2990/3-1.

\begin{appendix}

\section{Feynman rules and coupling constants}\label{app::couplings}
The Feynman rules for the \susy-\qcd{} vertices can be found in
Refs.\,\cite{Harlander:2004tp,Harlander:2009mn}. Here, we give in
addition the ones for the bottom- and sbottom-Higgs vertices:
\begin{center}
\begin{tabular}{llll}
\raisebox{-4.5em}{
\includegraphics[bb=120 550 310 725,width=10em]{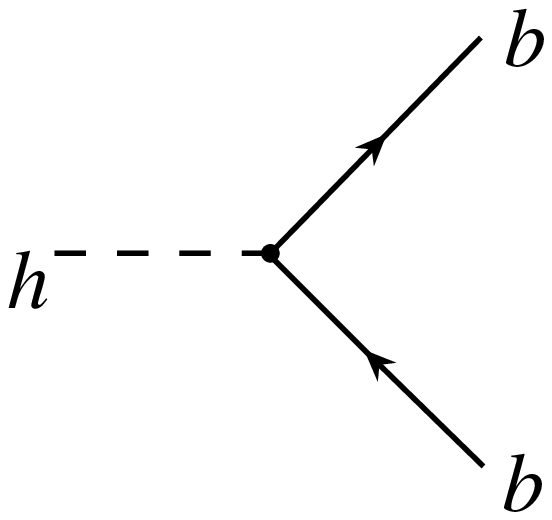}} &\!\!\!
$\displaystyle i \frac{m_b}{v}\,g^h_b$&\qquad\qquad
\raisebox{-4.5em}{
\includegraphics[bb=120 550 310 725,width=10em]{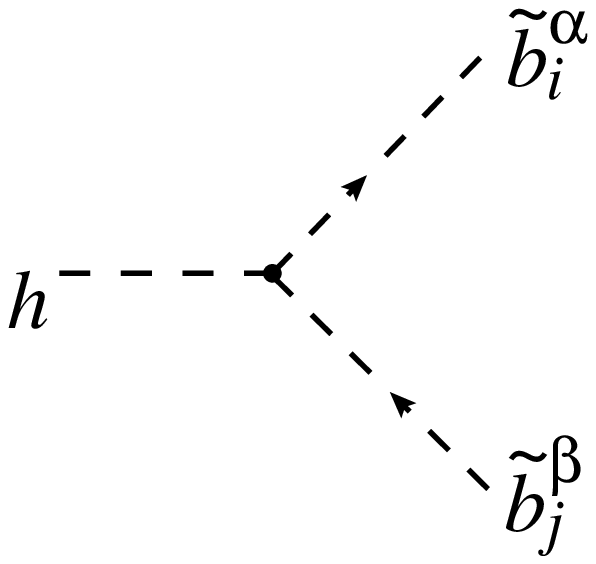}} &
$\displaystyle i \frac{m_b^2}{v}\,g^h_{b,ij}$
\end{tabular}
\end{center}

The bottom-Higgs coupling constant reads
\begin{equation}
\begin{split}
g_b^h = -\frac{\sin\alpha}{\cos\beta}\,,
\end{split}
\end{equation}
and the sbottom-Higgs couplings are
\newcommand{\cew}[1]{c_{#1}^{\rm EW}}
\begin{equation}
\begin{split}
g^h_{b,ij} &=g^{h,\rm EW}_{b,ij}
+ g_{b,ij}^{h,\mu} + g_{b,ij}^{h,\alpha}\,,
\label{eq::sbottomhiggs}
\end{split}
\end{equation}
with
\begin{equation}
\begin{split}
g_{b,11}^{h,{\rm EW}} &= \cew{1,b}\,\cos^2\theta_b +
\cew{2,b}\,\sin^2\theta_b\,,\\
g_{b,22}^{h,{\rm EW}} &= \cew{1,b}\,\sin^2\theta_b +
\cew{2,b}\,\cos^2\theta_b\,,\\
g_{b,12}^{h,{\rm EW}} &= 
g_{b,21}^{h,{\rm EW}} = 
\frac{1}{2}(\cew{2,b}-\cew{1,b})\,\sin 2\theta_b\,,\\
g_{b,11}^{h,\mu} &= - g_{b,22}^{h,\mu} = 
-\frac{\muSUSY}{\mbottom}\,\frac{\cos(\alpha-\beta)}{\cos^2\beta}
\sin2\theta_b\,,\\
g_{b,12}^{h,\mu} &= 
g_{b,21}^{h,\mu} = -
\frac{\muSUSY}{\mbottom}\,\frac{\cos(\alpha-\beta)}{\cos^2\beta}
\cos2\theta_b\,,\\
g_{b,11}^{h,\alpha} &= -\frac{\sin\alpha}{\cos\beta}\left[
  2 + \frac{\msbottom{1}^2 - \msbottom{2}^2}{2\mbottom^2}\,
\sin^22\theta_b\,\right],\\
g_{b,22}^{h,\alpha} &= -\frac{\sin\alpha}{\cos\beta}\left[
  2 - \frac{\msbottom{1}^2 - \msbottom{2}^2}{2\mbottom^2}\,
\sin^22\theta_b\,\right],\\
g_{b,12}^{h,\alpha} &= 
g_{b,21}^{h,\alpha} = 
-\frac{\sin\alpha}{\cos\beta}\,
  \frac{\msbottom{1}^2 - \msbottom{2}^2}{2\mbottom^2}\,
\sin2\theta_b\cos2\theta_b\,,\\
\label{eq::sbottomhiggs-1}
\end{split}
\end{equation}
where
\begin{equation}
\begin{split}
\cew{1,b} &= \frac{2M_Z^2}{\mquark^2}\left( |I_{3}| - |Q|\sin^2\theta_W
\right)\,\sin(\alpha+\beta)\,,\\
\cew{2,b} &= 
\frac{2M_Z^2}{\mquark^2}\,|Q|\,\sin^2\theta_W\sin(\alpha+\beta)\,,\\
\sin\theta_W &= \sqrt{1 - \frac{M_W^2}{M_Z^2}}\,,\qquad
I_{3} = -\frac{1}{2}\,,\quad
Q = -\frac{1}{3}\,,
\label{eq::cew}
\end{split}
\end{equation}
and 
\begin{equation}
\begin{split}
\tan\beta = \frac{v_2}{v_1}\,,\qquad
v = \frac{2M_W}{g} = \frac{1}{\sqrt{\sqrt{2}G_F}} =
\sqrt{v_1^2+v_2^2} \approx 246\,{\rm GeV}\,,
\end{split}
\end{equation}
with $v_1$, $v_2$ the vacuum expectation values of the two Higgs
doublets.  The angle $\theta_b$ rotates the super-partners of the left-
and right-handed bottom quarks into their mass eigenstates, while
$\alpha$ does the same for the neutral components of the Higgs
doublets. In \eqn{eq::sbottomhiggs-1} we have already expressed
the trilinear couplings of the soft \susy{} breaking terms through
independent parameters:
\begin{equation}
\begin{split}
A_b &= \frac{\msbottom{1}^2 - \msbottom{2}^2}{2\mbottom}\sin2\theta_b +
\muSUSY\,\tan\beta\,.
\label{eq::ab}
\end{split}
\end{equation}
The expressions in \eqn{eq::sbottomhiggs-1} and \neqn{eq::cew} are
completely analogous to the ones for the scalar top sector, given in
Ref.\,\cite{Harlander:2004tp}. They can be transformed into each other
by the replacements
\begin{equation}
\begin{split}
t\leftrightarrow b\,,\qquad
&\muSUSY\leftrightarrow -\muSUSY\,,\qquad
\alpha \leftrightarrow \alpha+\frac{\pi}{2}\,,\qquad
\beta \leftrightarrow \beta+\frac{\pi}{2}\,,\\
& Q\leftrightarrow Q-1\,,\qquad
 I_{3}\leftrightarrow I_{3}-1\,.
\label{eq::ttob}
\end{split}
\end{equation}

\section{Expansion of quark terms}\label{app::a1q}

Analytic expressions for the virtual quark loop contribution to the
gluon fusion process at \nlo{} have been presented in\footnote{For the
  sake of clarity, let us remark that in the notation of
  Ref.\,\cite{Harlander:2005rq}, it is $F_0^HB_1^H = 2\,a_q^{(1)}/g_q^h$.}
Refs\,\cite{Harlander:2005rq,Anastasiou:2006hc,Aglietti:2006tp}.  It may
be convenient for the reader to have the relevant limits of these
expressions. We quote them here for the large quark mass
limit,
\begin{equation}
\begin{split}
\frac{a_q^{(1)}}{g_q^h}=&\
\frac{11}{4}
+ \frac{1237}{1080} \,\tau_q^{-1}
+ \frac{17863}{28350} \,\tau_q^{-2}
+ \frac{157483}{396900} \,\tau^3
+ \frac{636694}{2338875} \,\tau_q^{-4}
\\&
+ \frac{48712384706}{245827456875} \,\tau_q^{-5}
+ \frac{96272051048}{639151387875} \,\tau_q^{-6}
+ \frac{6428929236304}{54327867969375} \,\tau_q^{-7}
\\&
+ \frac{11720258014074752}{122835309478756875} \,\tau_q^{-8}
+ \frac{24899957625820672}{316945428161236875} \,\tau_q^{-9}
+ \order{\tau_q^{-{10}}}
\end{split}
\end{equation}
and for the small quark mass limit:
\begin{equation}
\begin{split}
\frac{a_q^{(1)}}{g_q^h} =&\
\frac{\tau_q}{4}\bigg[
47 
- 31\,\zeta_3 
- 18\,\zeta_4
+ \Lhq\,(-12 - 4\,\zeta_2 + 11\,\zeta_3) 
+ \Lhq^2\,(-\frac{9}{4} - \frac{1}{2}\,\zeta_2) 
\\&
+ \Lhq^3 
- \frac{5}{48}\,\Lhq^4
\bigg]
+\left(\frac{\tau_q}{4}\right)^{2}\bigg[
1 
- 10\,\zeta_2 
+ 128\,\zeta_3 
+ 72\,\zeta_4
+ \Lhq\,(-2 - 44\,\zeta_3) 
\\&
+ \Lhq^2\,(-\frac{11}{2} + 2\,\zeta_2) 
- 9\,\Lhq^3 
+ \frac{5}{12} \,\Lhq^4
\bigg]
+\left(\frac{\tau_q}{4}\right)^{3}\bigg[
\frac{1073}{8}
+ \frac{109}{2} \,\zeta_2
- 80\,\zeta_3 
\\&
+ 36\,\zeta_4
+ \Lhq\,(\frac{543}{4} + 40\,\zeta_2 - 32\,\zeta_3) 
+ \Lhq^2\,(\frac{179}{2} + 8\,\zeta_2) 
+ \frac{119}{12} \,\Lhq^3
+ \frac{1}{6 }\,\Lhq^4
\bigg]
\\&
+ \order{\tau_q^4}\,,
\end{split}
\end{equation}
where $\zeta_n\equiv \zeta(n)$ is Riemann's zeta function evaluated at
$n$, with
\begin{equation}
\begin{split}
\zeta_2 = \frac{\pi^2}{6} = 1.64493\ldots\,,\qquad
\zeta_3 = 1.20206\ldots\,,\qquad
\zeta_4 = \frac{\pi^4}{90} = 1.08232\ldots\,.
\end{split}
\end{equation}
\end{appendix}

\section{Total cross section for other \susy{} scenarios}\label{app::res}

\begin{figure}
  \begin{center}
    \begin{tabular}{cc}
      \includegraphics[bb=60 100 560
        750,width=.36\textwidth,angle=-90]{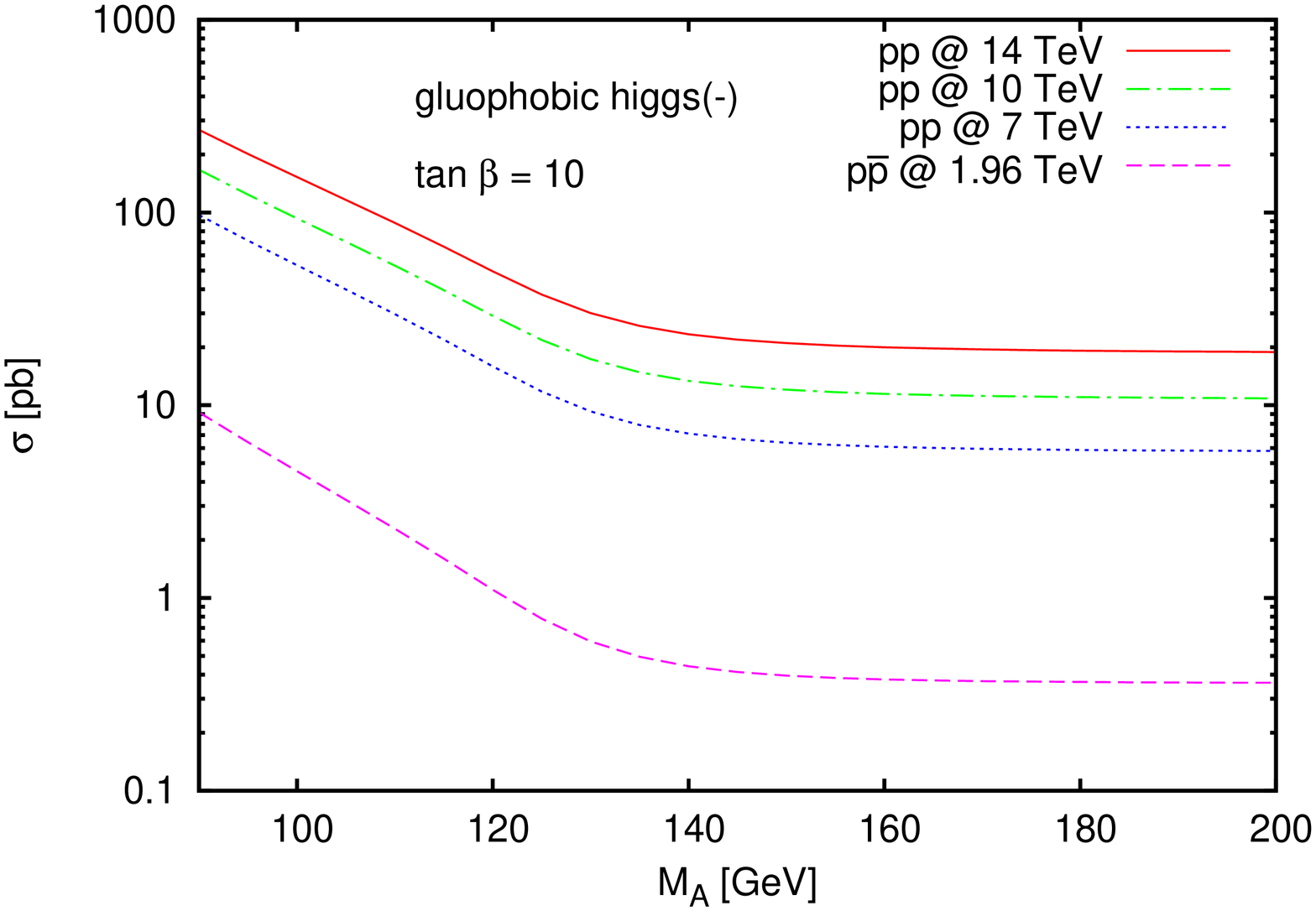} &
      \includegraphics[bb=60 100 560
        750,width=.36\textwidth,angle=-90]{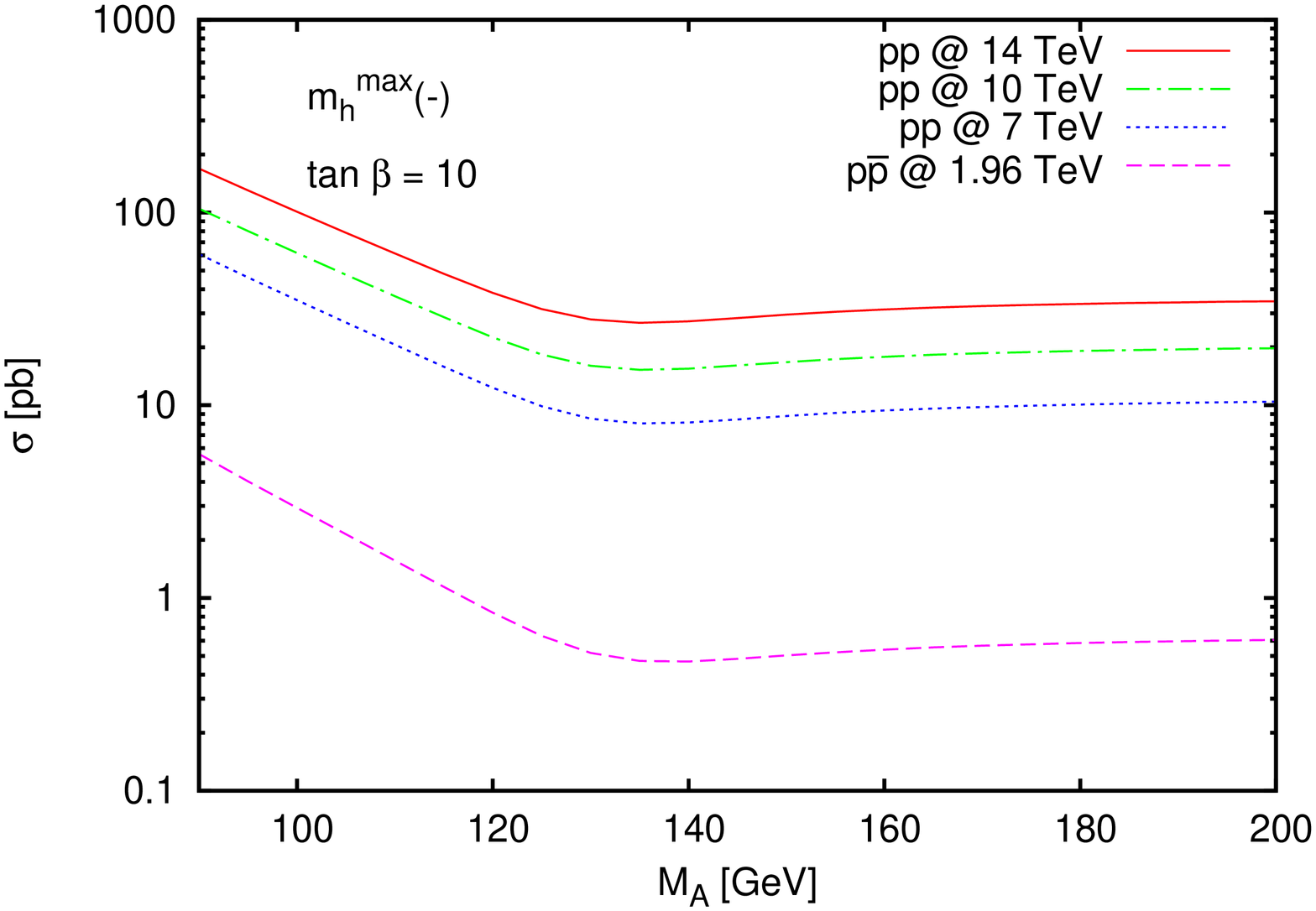} \\
      (a) & (b) \\
      \includegraphics[bb=60 100 560
        750,width=.36\textwidth,angle=-90]{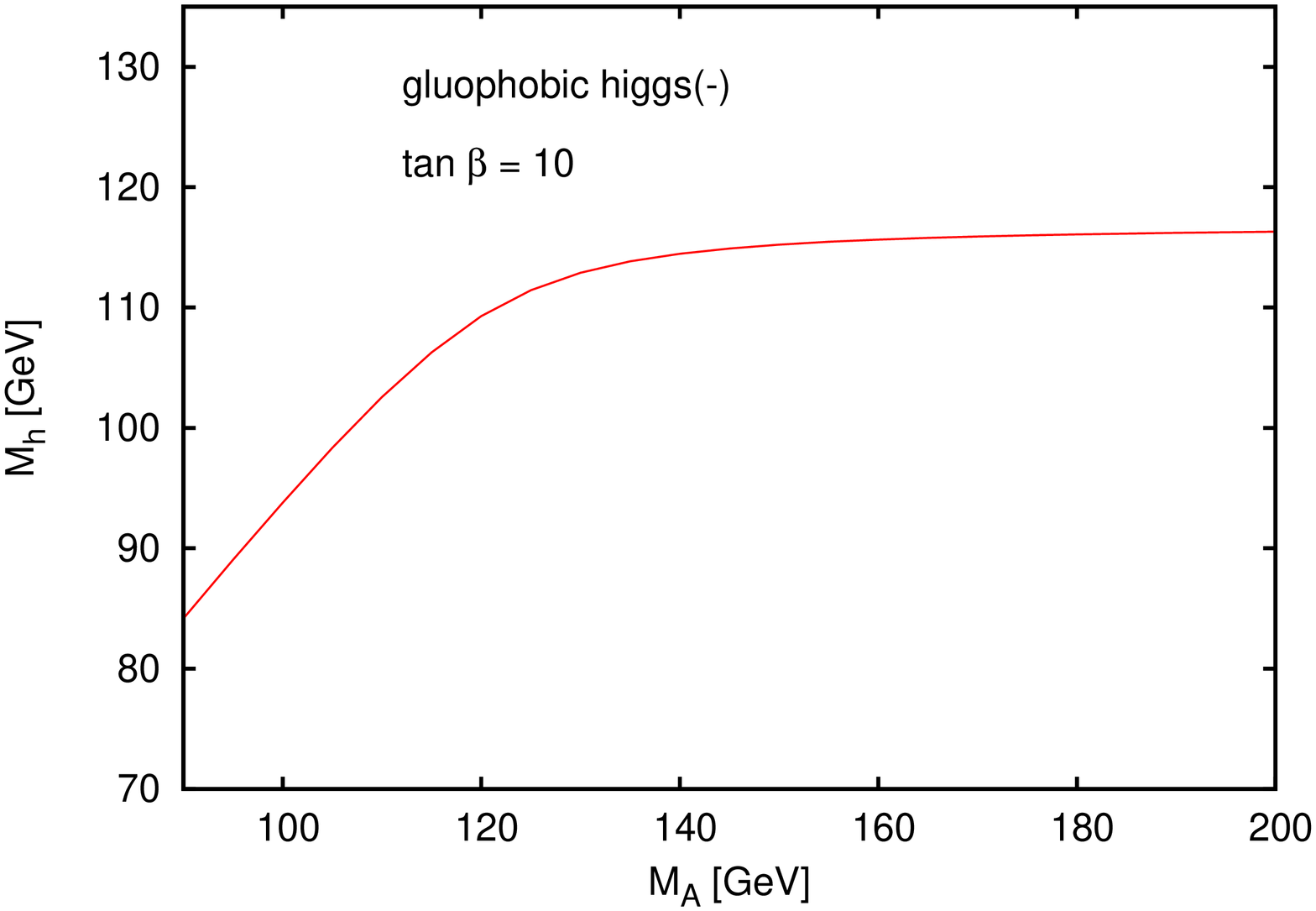} &
      \includegraphics[bb=60 100 560
        750,width=.36\textwidth,angle=-90]{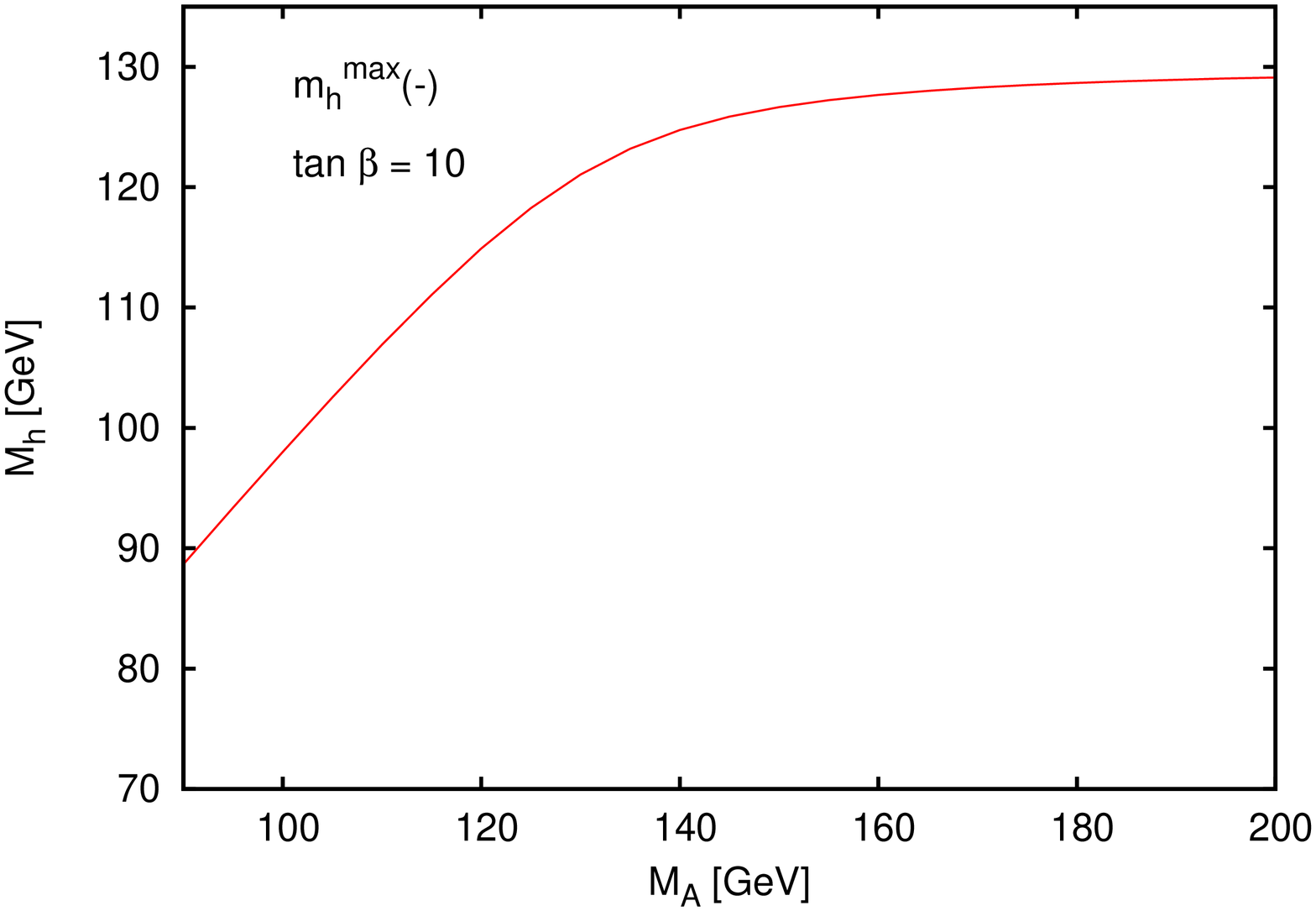} \\
      (c) & (d)
    \end{tabular}
    \parbox{.9\textwidth}{
      \caption[]{\label{fig::gluophobic.mhmax-}\sloppy Inclusive total
        cross section for gluon fusion in the \mssm{}.  (a)
        \gluophobic{-};\quad (b) \mhmax{-}.  Panels (c) and (d) show the
        corresponding light Higgs boson mass. Note that the scenarios
        \gluophobic{+} and \mhmax{+} can be found in
        \fig{fig::gluophobic.mhmax+}.  }}
  \end{center}
\end{figure}

\begin{figure}
  \begin{center}
    \begin{tabular}{cc}
      \includegraphics[bb=60 100 560
        750,width=.36\textwidth,angle=-90]{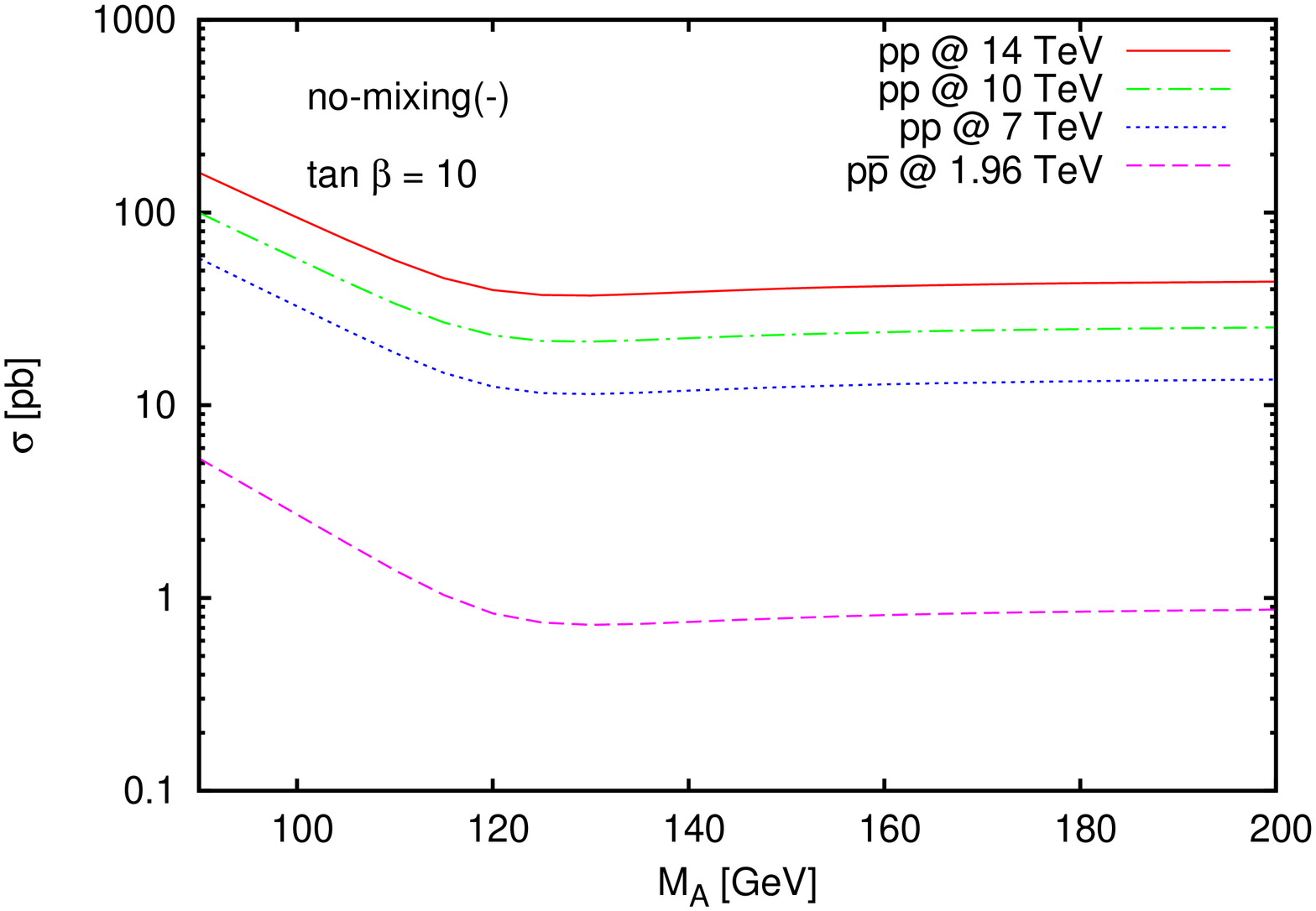} &
      \includegraphics[bb=60 100 560
        750,width=.36\textwidth,angle=-90]{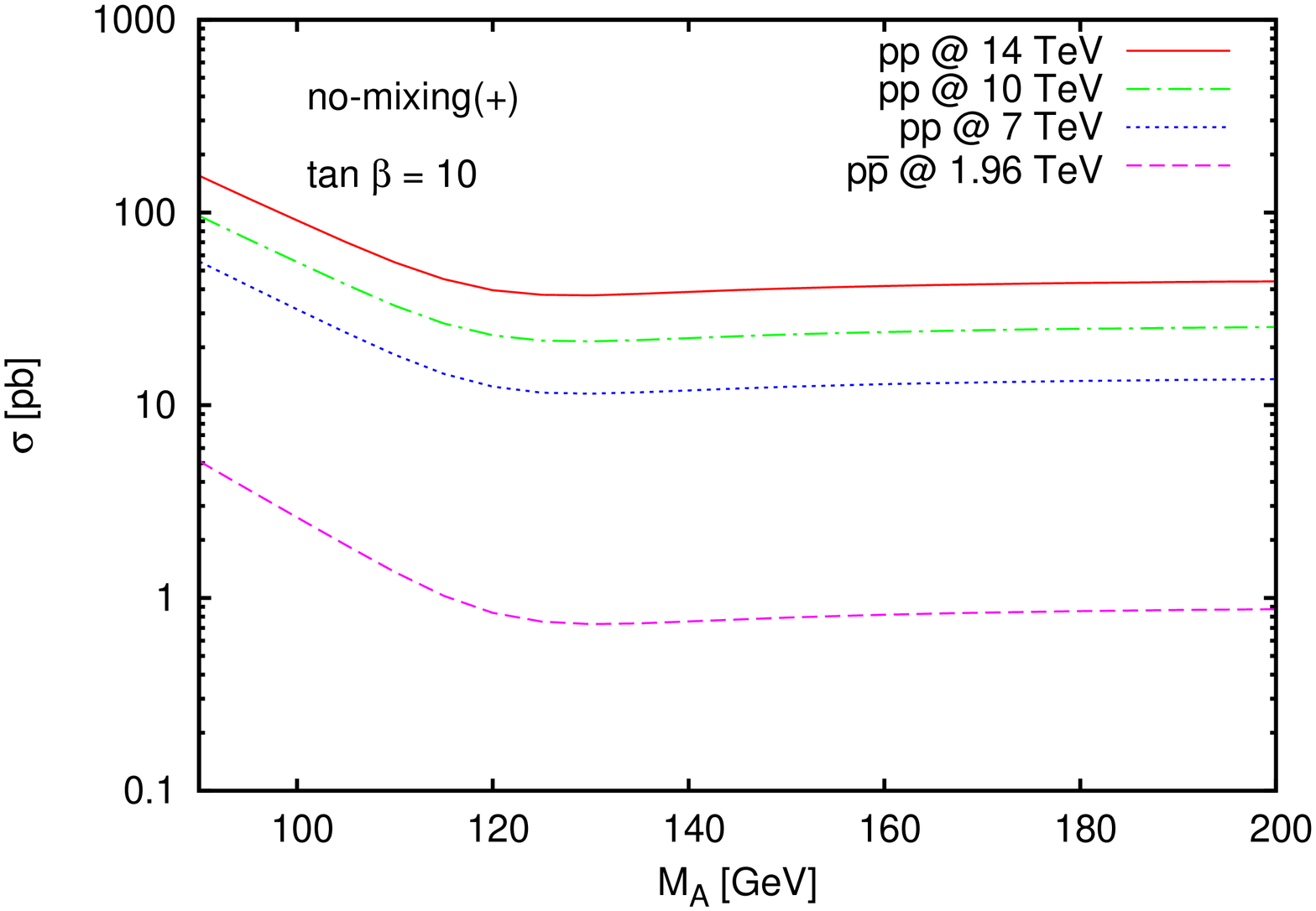} \\
      (a) & (b) \\
      \includegraphics[bb=60 100 560
        750,width=.36\textwidth,angle=-90]{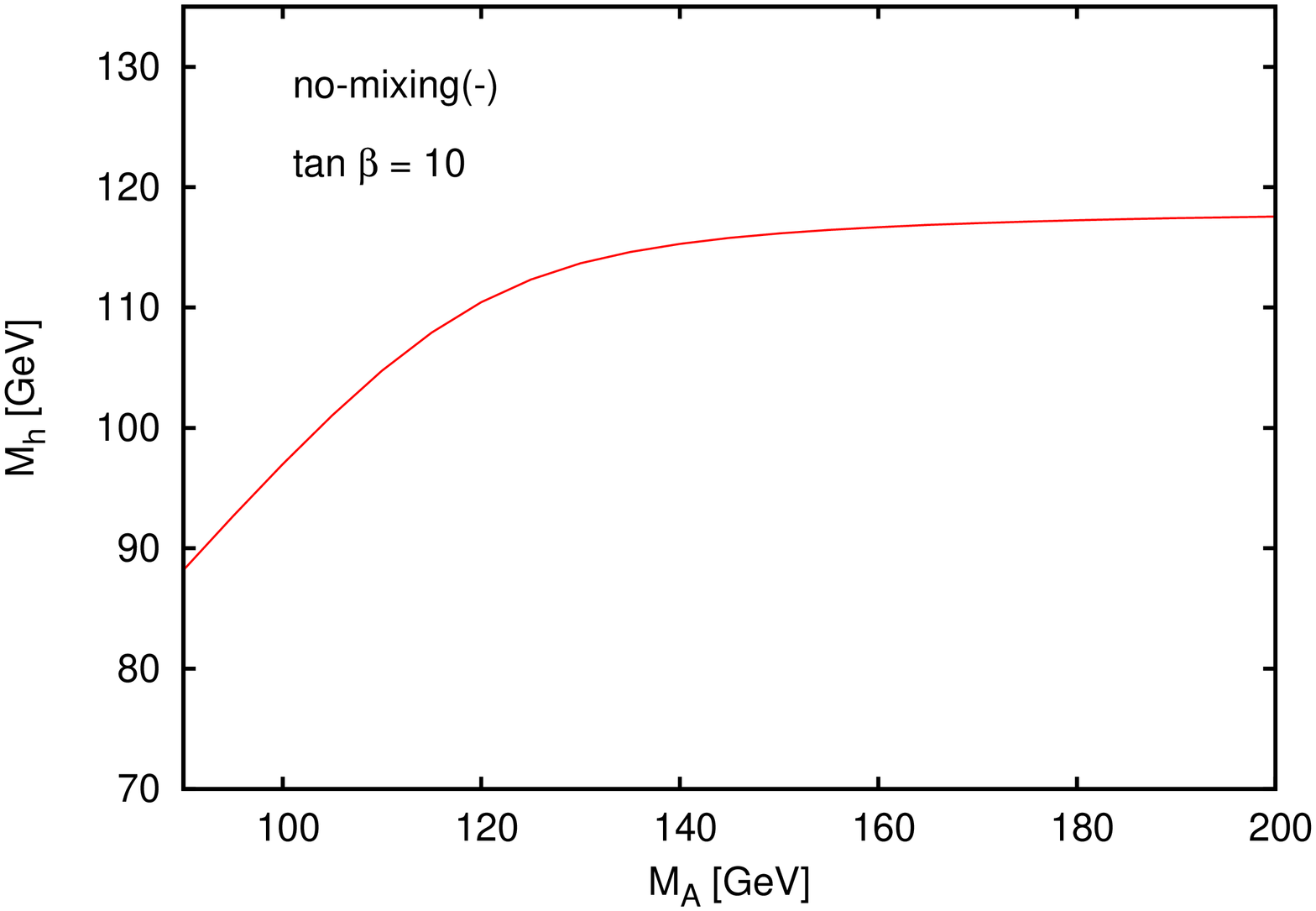} &
      \includegraphics[bb=60 100 560
        750,width=.36\textwidth,angle=-90]{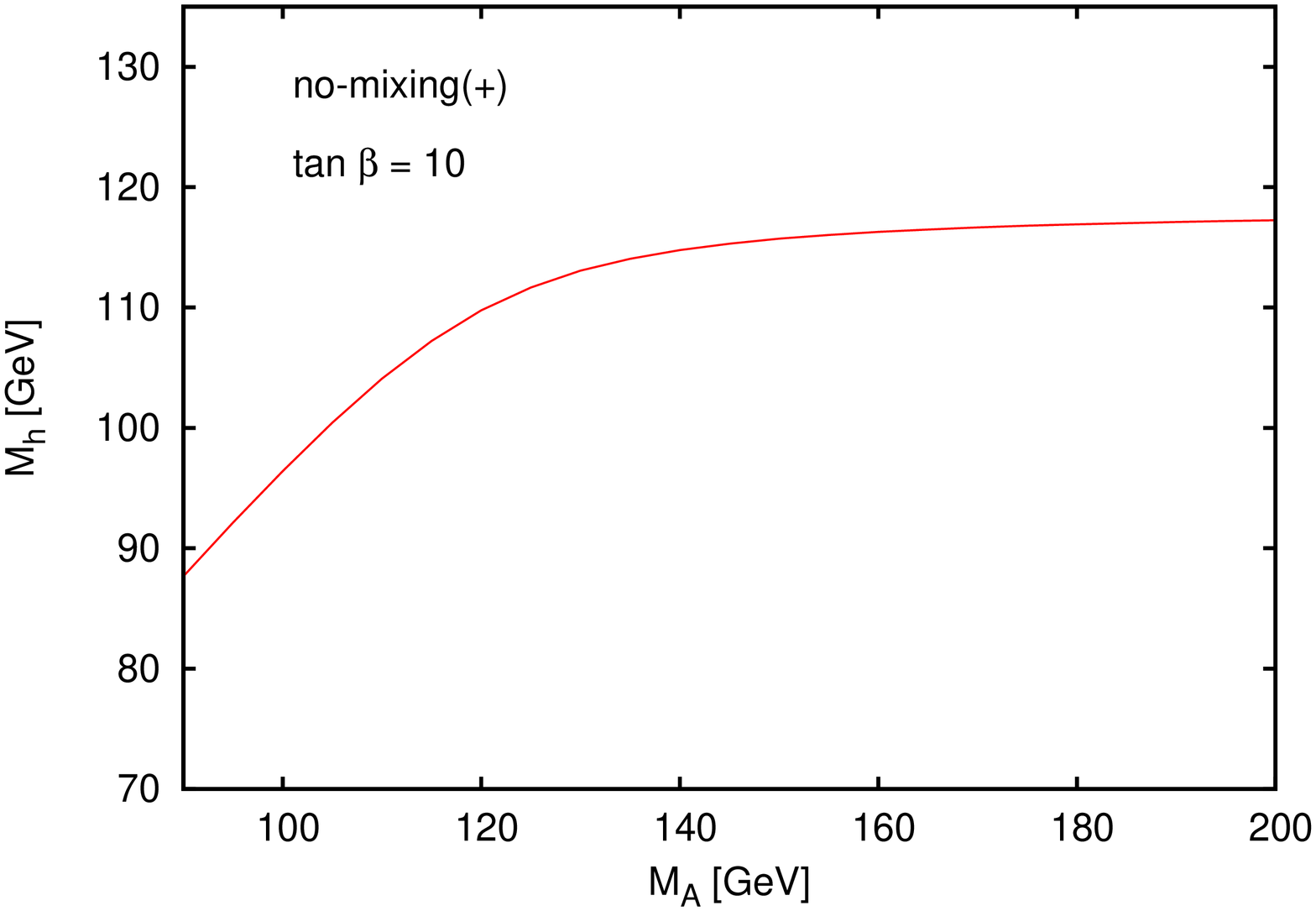} \\
      (c) & (d)
    \end{tabular}
    \parbox{.9\textwidth}{
      \caption[]{\label{fig::nomix}\sloppy Inclusive total cross section
        for gluon fusion in the \mssm{}.  (a) \nomixing{-};\quad (b)
        \nomixing{+}.  Panels (c) and (d) show the corresponding light
        Higgs boson mass.  }}
  \end{center}
\end{figure}

\begin{figure}
  \begin{center}
    \begin{tabular}{cc}
      \includegraphics[bb=60 100 560
        750,width=.36\textwidth,angle=-90]{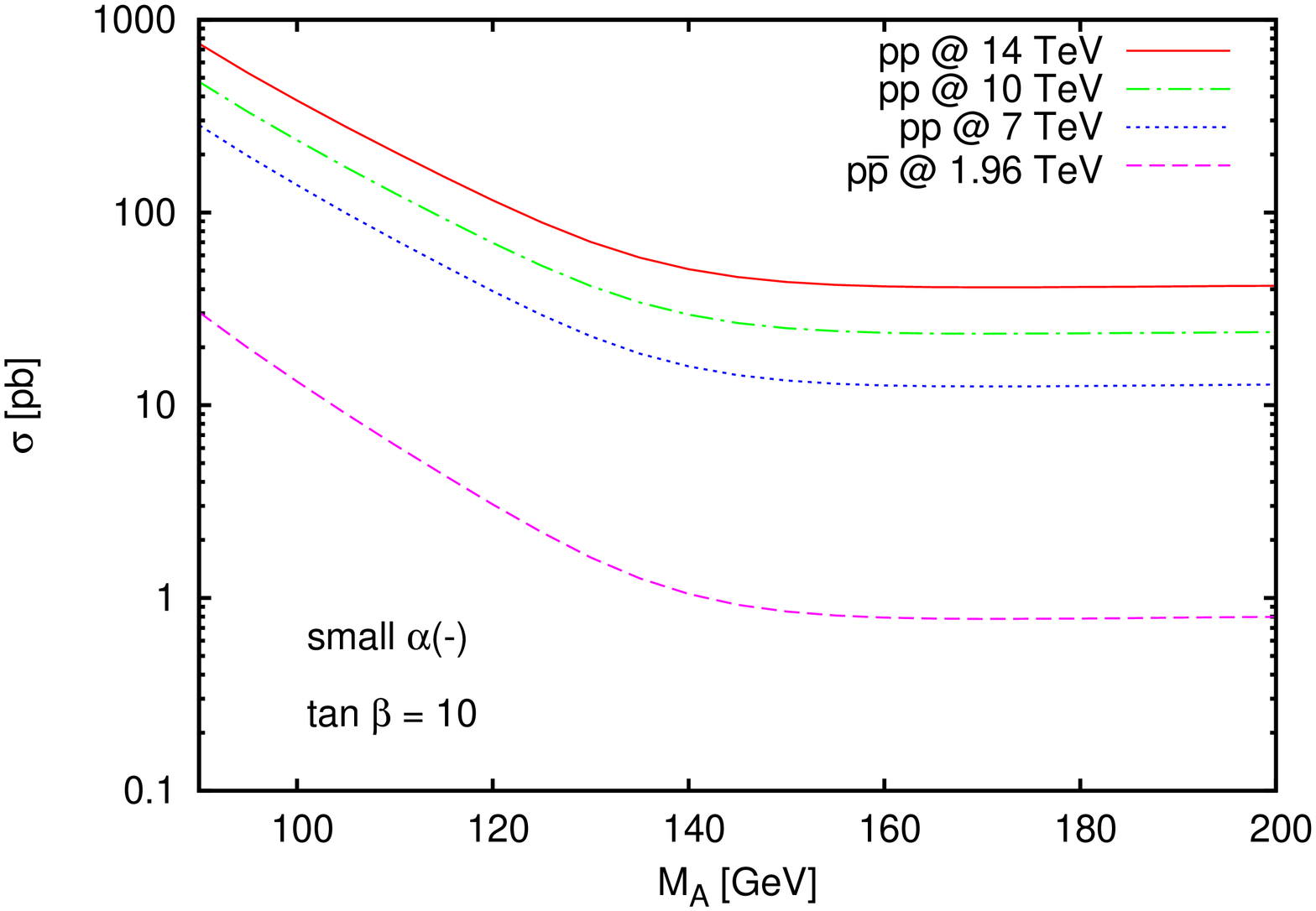} &
      \includegraphics[bb=60 100 560
        750,width=.36\textwidth,angle=-90]{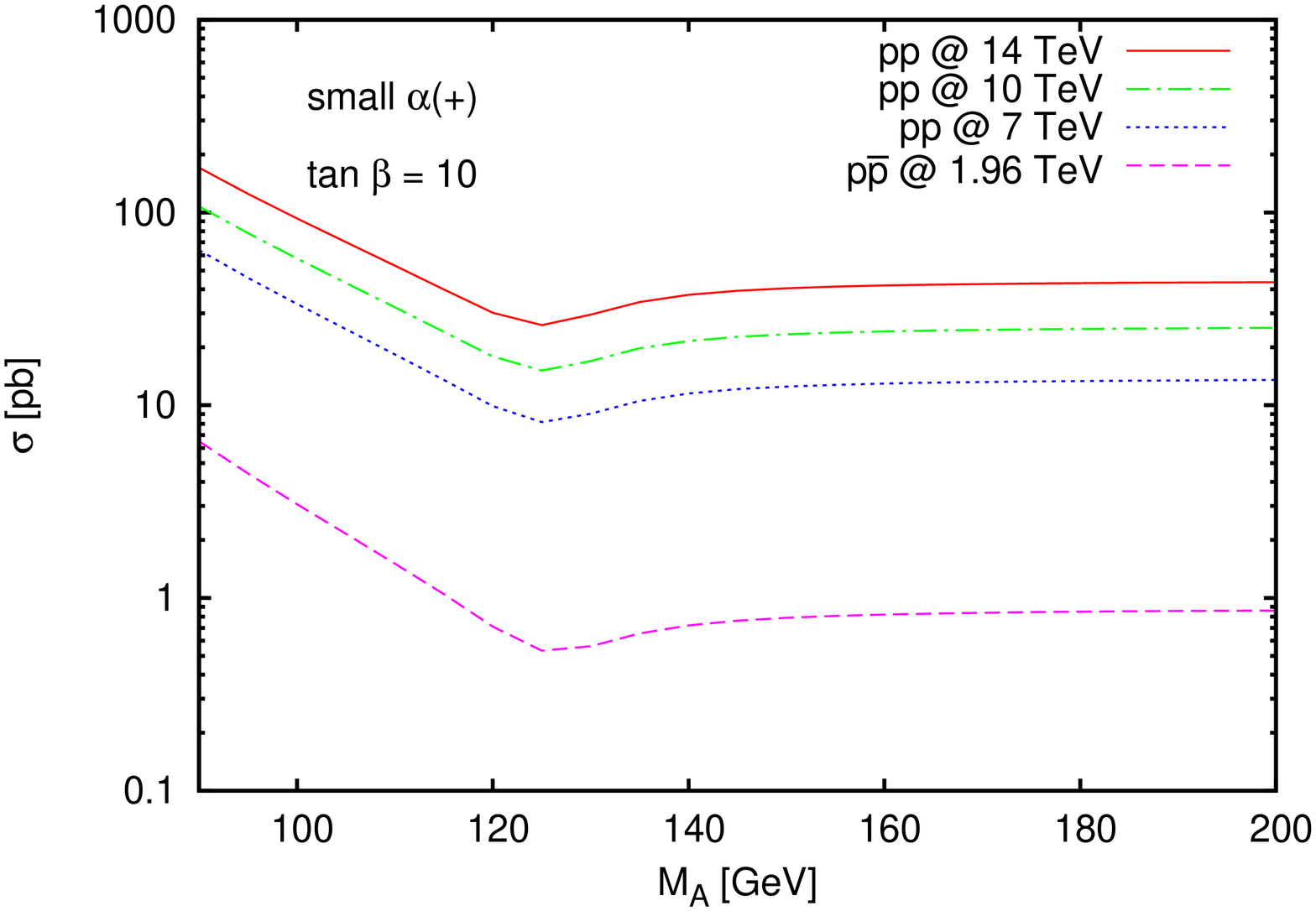} \\
      (a) & (b) \\
      \includegraphics[bb=60 100 560
        750,width=.36\textwidth,angle=-90]{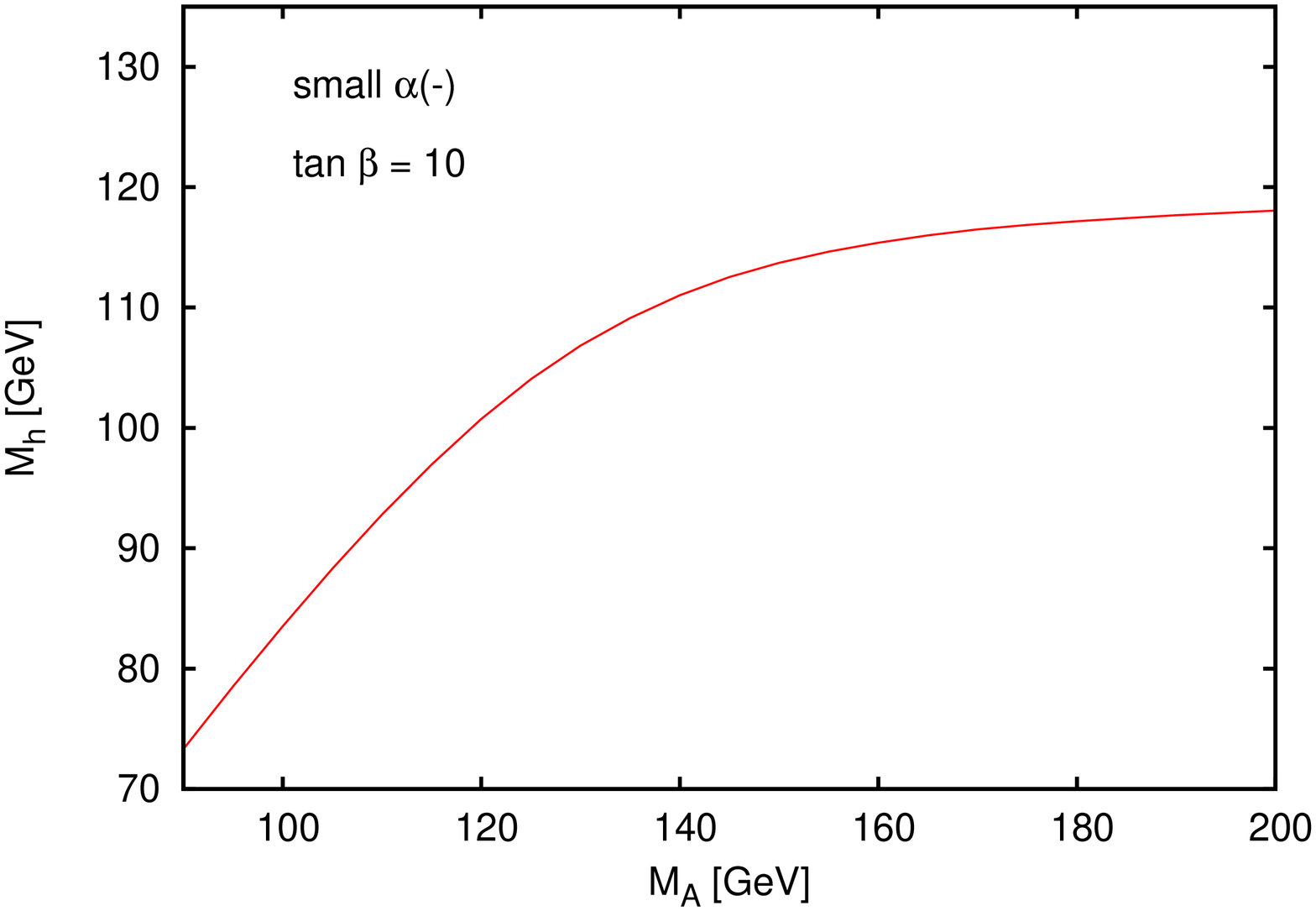} &
      \includegraphics[bb=60 100 560
        750,width=.36\textwidth,angle=-90]{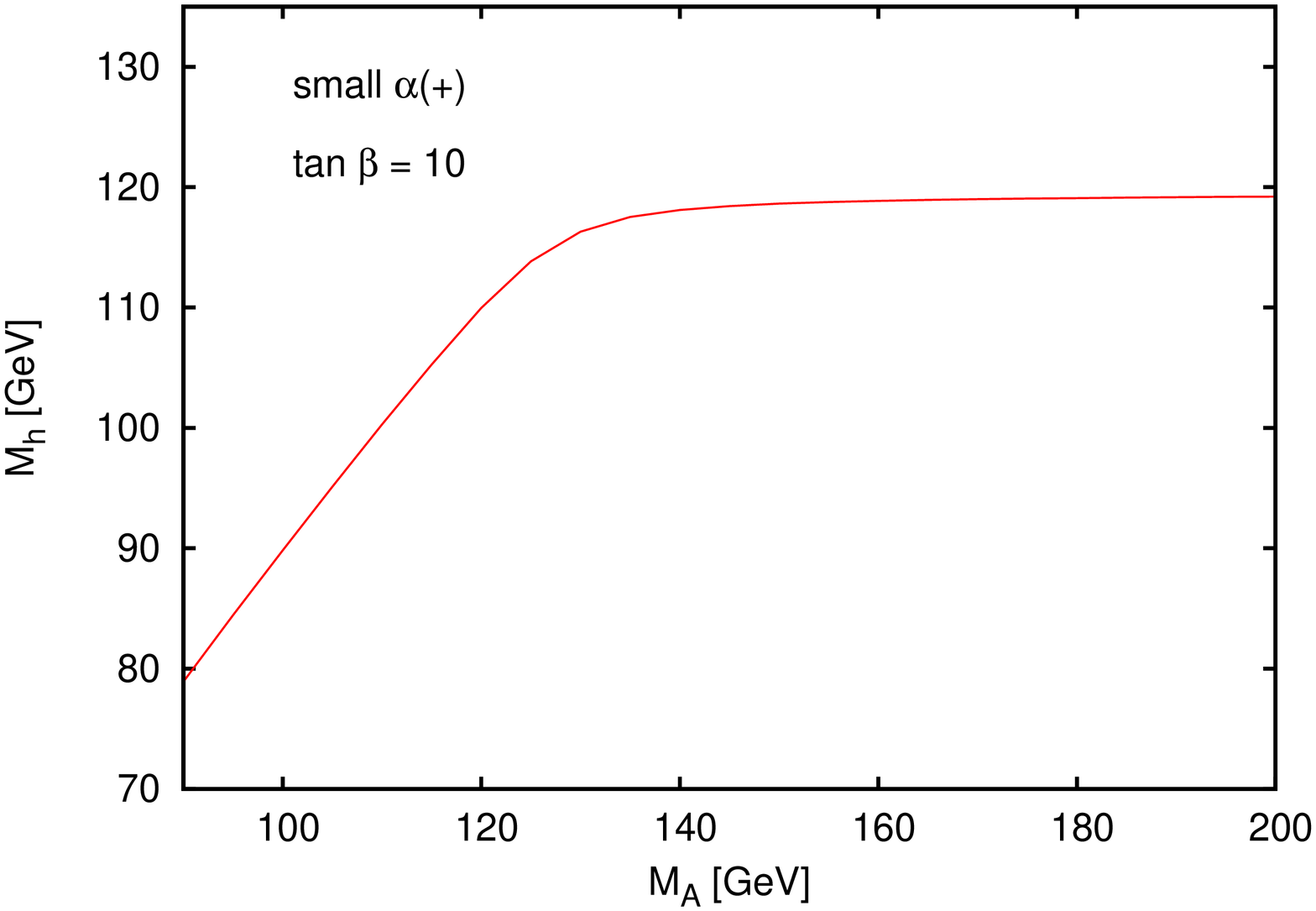} \\
      (c) & (d)
    \end{tabular}
    \parbox{.9\textwidth}{
      \caption[]{\label{fig::smalla}\sloppy Inclusive total cross
        section for gluon fusion in the \mssm{}.  (a)
        \smallalpha{-};\quad (b) \smallalpha{+}.  Panels (c) and (d)
        show the corresponding light Higgs boson mass.  }}
  \end{center}
\end{figure}

This section collects the results for the total inclusive cross section
in the \susy{} scenarios of
Eqs.\,(\ref{eq::gluophobic})--(\ref{eq::smalla}) which have not been
discussed in the main text. We restrict ourselves to graphical
representations; the corresponding data tables can be found at the
{\abbrev URL}
\cite{url}. Figs.\,\ref{fig::gluophobic.mhmax-}--\ref{fig::smalla} are
simply the analogues of \fig{fig::gluophobic.mhmax+} for these
scenarios.

\def\app#1#2#3{{\it Act.~Phys.~Pol.~}\jref{\bf B #1}{#2}{#3}}
\def\apa#1#2#3{{\it Act.~Phys.~Austr.~}\jref{\bf#1}{#2}{#3}}
\def\annphys#1#2#3{{\it Ann.~Phys.~}\jref{\bf #1}{#2}{#3}}
\def\cmp#1#2#3{{\it Comm.~Math.~Phys.~}\jref{\bf #1}{#2}{#3}}
\def\cpc#1#2#3{{\it Comp.~Phys.~Commun.~}\jref{\bf #1}{#2}{#3}}
\def\epjc#1#2#3{{\it Eur.\ Phys.\ J.\ }\jref{\bf C #1}{#2}{#3}}
\def\fortp#1#2#3{{\it Fortschr.~Phys.~}\jref{\bf#1}{#2}{#3}}
\def\ijmpc#1#2#3{{\it Int.~J.~Mod.~Phys.~}\jref{\bf C #1}{#2}{#3}}
\def\ijmpa#1#2#3{{\it Int.~J.~Mod.~Phys.~}\jref{\bf A #1}{#2}{#3}}
\def\jcp#1#2#3{{\it J.~Comp.~Phys.~}\jref{\bf #1}{#2}{#3}}
\def\jetp#1#2#3{{\it JETP~Lett.~}\jref{\bf #1}{#2}{#3}}
\def\jphysg#1#2#3{{\small\it J.~Phys.~G~}\jref{\bf #1}{#2}{#3}}
\def\jhep#1#2#3{{\small\it JHEP~}\jref{\bf #1}{#2}{#3}}
\def\mpl#1#2#3{{\it Mod.~Phys.~Lett.~}\jref{\bf A #1}{#2}{#3}}
\def\nima#1#2#3{{\it Nucl.~Inst.~Meth.~}\jref{\bf A #1}{#2}{#3}}
\def\npb#1#2#3{{\it Nucl.~Phys.~}\jref{\bf B #1}{#2}{#3}}
\def\nca#1#2#3{{\it Nuovo~Cim.~}\jref{\bf #1A}{#2}{#3}}
\def\plb#1#2#3{{\it Phys.~Lett.~}\jref{\bf B #1}{#2}{#3}}
\def\prc#1#2#3{{\it Phys.~Reports }\jref{\bf #1}{#2}{#3}}
\def\prd#1#2#3{{\it Phys.~Rev.~}\jref{\bf D #1}{#2}{#3}}
\def\pR#1#2#3{{\it Phys.~Rev.~}\jref{\bf #1}{#2}{#3}}
\def\prl#1#2#3{{\it Phys.~Rev.~Lett.~}\jref{\bf #1}{#2}{#3}}
\def\pr#1#2#3{{\it Phys.~Reports }\jref{\bf #1}{#2}{#3}}
\def\ptp#1#2#3{{\it Prog.~Theor.~Phys.~}\jref{\bf #1}{#2}{#3}}
\def\ppnp#1#2#3{{\it Prog.~Part.~Nucl.~Phys.~}\jref{\bf #1}{#2}{#3}}
\def\rmp#1#2#3{{\it Rev.~Mod.~Phys.~}\jref{\bf #1}{#2}{#3}}
\def\sovnp#1#2#3{{\it Sov.~J.~Nucl.~Phys.~}\jref{\bf #1}{#2}{#3}}
\def\sovus#1#2#3{{\it Sov.~Phys.~Usp.~}\jref{\bf #1}{#2}{#3}}
\def\tmf#1#2#3{{\it Teor.~Mat.~Fiz.~}\jref{\bf #1}{#2}{#3}}
\def\tmp#1#2#3{{\it Theor.~Math.~Phys.~}\jref{\bf #1}{#2}{#3}}
\def\yadfiz#1#2#3{{\it Yad.~Fiz.~}\jref{\bf #1}{#2}{#3}}
\def\zpc#1#2#3{{\it Z.~Phys.~}\jref{\bf C #1}{#2}{#3}}
\def\ibid#1#2#3{{ibid.~}\jref{\bf #1}{#2}{#3}}
\def\otherjournal#1#2#3#4{{\it #1}\jref{\bf #2}{#3}{#4}}
\newcommand{\jref}[3]{{\bf #1} (#2) #3}
\newcommand{\bibentry}[4]{#1, {\it #2}, #3\ifthenelse{\equal{#4}{}}{}{, }#4.}
\newcommand{\hepph}[1]{{\tt hep-ph/#1}}
\newcommand{\mathph}[1]{[math-ph/#1]}
\newcommand{\arxiv}[2]{{\tt arXiv:#1}}

\end{document}